\documentclass[a4paper,11pt]{article}
\pdfoutput=1 
						
\usepackage[T1]{fontenc} 
\usepackage{tikz}
\usepackage{graphics}
\usepackage{graphicx}
\usepackage{float}
\usepackage{caption}
\usepackage{subcaption}
\usepackage{bbding}
\usepackage{jheppub} 
\usepackage{color}		
\usepackage{slashed}		
\usepackage{multirow}
\usepackage{adjustbox}

\title{Radiative Two-Loop Neutrino Masses with Dark Matter}

\author[a]{C.~Simoes}
\author[a]{and D. Wegman}
 \affiliation[a]{IFPA, D\'ep. AGO, Universit\'e de
      Li\`ege, B\^at B5, Sart Tilman B-4000 Li\`ege 1,
      Belgium}

\emailAdd{csimoes@ulg.ac.be}
\emailAdd{dwegman@ulg.ac.be}

\abstract{Using the Weinberg operator, we present a full collection of genuine two-loop models for neutrino mass generation, which contain a dark matter particle as one of the internal messengers. These models can be constructed simply by adding new fields that are singlets or doublets of $\mathsf{SU(2)_L}$. 
We ensure the stability of the dark matter candidate by the addition of a $\mathsf{Z_2}$ symmetry that will also be used to forbid tree level or one-loop diagrams. Thus we only present models where the main contribution for neutrinos masses is generated from the corresponding two-loop diagram. We also discuss a short outline corresponding to some phenomenological characteristics of these models.}


\begin{document} 
\maketitle

\section{Introduction}
\label{sec:intro}
The existence of neutrino masses and dark matter~(DM) have corroborated the need for physics beyond the standard model~(BSM). While neutrino experiments~\cite{Forero:2014bxa} have continuously supplied us with better information about parameters of the neutrino sector, direct DM detection experiments~\cite{Ahmed:2009zw,Aprile:2012nq,Akerib:2015rjg}, given the lack in the direct detection signals, have only provided with bounds on the DM mass. Nevertheless, cosmological evidence~\cite{Ade:2013zuv} has shown arguably enough proof of its existence. 

With the addition of each new field (or any source of new physics), there will be a growing number of parameters added to the Standard Model~(SM). Therefore, connecting this expanding number of parameters via physical arguments (that should arise from concrete models) helps reduce the freedom of these variables by tracing them to a single origin. With this idea in mind, many proposed models have attempted to connect neutrino masses to DM~\cite{Krauss:2002px,Hirsch:2004he,Aoki:2008av,Sierra:2008wj,Asaka:2005pn,Restrepo:2015ura,Klasen:2013jpa,Wang:2015saa,Merle:2016scw}, one of the most appealing of which is the Ma-Scotogenic model~\cite{Ma:2006km}. Radiative neutrino models as the above mentioned  are very attractive, they not only give a natural explanation to the smallness of neutrino masses, but also given the new fields added, a DM candidate may be used as a messenger, ensuring a relation between neutrinos and DM. 

The exact nature of neutrinos is still unclear; they may behave similarly to the charged leptons, i.e., Dirac particles, or entirely different by being their own antiparticles, referred to as Majorana particles. The existence of heavy Majorana neutrinos is a very appealing scenario to explain the smallnesses of SM neutrino masses ~\cite{Ma:1998dn}.

If one wants to write an effective theory for Majorana neutrino masses, one can start with the SM Lagrangian and add higher order non-normalizable operators ~\cite{Angel:2012ug}, 
\begin{equation}
\mathcal{L}= \mathcal{L_\text{SM}} +  \sum_{n>4} \frac{C^n}{\Lambda^{n-4}}\, \mathcal{O}_n\,,
\label{eq:higer}
\end{equation}
where $\Lambda$ is the energy scale, $C^n$ is a constant and $\mathcal{O}_n$ is an operator of order $n$. 

It can be shown~\cite{Weinberg:1979sa} that there is only one possibility for $n=5$, commonly referred in literature as the Weinberg dim=5 operator, 
\begin{equation}
\mathcal{O}_5 \sim \overline{L^\texttt{c}}\,L\,H\,H\,,
\label{eq:weinberg}
\end{equation} 
where $H$ is the Higgs field, $L$ the SM lepton field, and the $\mathsf{SU(2)_L}$ contractions have not been written for simplicity.

This operator violates lepton number by two units and after the Higgs gets a vacuum expectation value~(VEV), $\left\langle H\right\rangle = v$, one can use this operator to calculate Majorana neutrino masses. Given the smallness of neutrino masses, assuming a mass of the order $m_{\nu} \sim O(10^{-1})$~eV \cite{Forero:2014bxa}, it is possible to conclude that for the dim=5 operator, the constant $C$ on eq.\eqref{eq:higer} is very small and proportional to the mass scale $C \sim  10^{-12}  \times(\Lambda/\text{TeV})$. This equation shows that is possible to have small neutrino masses at tree level with a very high scale, $\Lambda \sim (10^{15}-10^{16})\,\text{GeV}$, where new physics is required. Conversely, one can reproduce small neutrino masses by using a radiative method, with $C \sim (y^2/ 4\pi)^m$ and $m \geq 2$, where $m=2$ corresponds to one-loop, $m=3$ to two-loop and so on. 
An approximate calculation  shows that for $m=5$ (i.e., four-loop diagrams) the value of the $\Lambda$ scale is too low ($\sim \text{eV}$), in other words, within the dim=5 Weinberg operator, it is only possible to reproduce the scale of neutrino masses up to models with three-loops.
Also  higher dimensional  operators are possible~\cite{Lehman:2014jma,Babu:2001ex,Bonnet:2009ej,Liao:2016qyd}, although the Weinberg operator remains a more elegant and simple way to generate the masses.

At tree level Majorana neutrinos can acquire mass via three mechanisms \cite{delAguila:2008cj}, called seesaw type-I~\cite{Minkowski:1977sc,Yanagida:1979as,GellMann:1980vs,Mohapatra:1979ia}, -II~\cite{Schechter:1980gr,Magg:1980ut,Mohapatra:1980yp,Cheng:1980qt,Perez:2008ha} and -III~\cite{Foot:1988aq,Franceschini:2008pz}.  To be able to recreate neutrino experimental data type-I needs the addition of (at least) two fermion singlets, type-II requires one scalar triplet and type-III needs (at least) two fermion triplets.
At one-loop there are four possible diagrams for the Weinberg operator~\cite{Bonnet:2012kz}, a study of these one-loop radiative neutrino mass models with viable dark matter candidates was done in ref.~\cite{Restrepo:2013aga}.
At the two-loop level the Weinberg operator has twenty genuine different diagrams (by "genuine" two-loop models we mean the ones for which, given a specific set of fields, there are no contributions to the neutrino masses either from tree level (seesaw) or from one-loop.); a systematic classification of these two-loop realizations was presented in ref.~\cite{Sierra:2014rxa}.

As mentioned above, another hint for the need of new physics is the cosmological evidence of a new type of matter. If DM is a particle, present constrains dictate that it should be color neutral, without electric charge (or a very small one~\cite{Dolgov:2013una,DelNobile:2015bqo,Agrawal:2016quu,Foot:2014uba}), and fit experimental constraints, as the ones coming from relic abundance and direct detection experiments \cite{Abe:2014gua}. Also the particle must be stable in the sense that must be long-lived (experimental bounds set this limit around seven orders of magnitude higher than the lifetime of the universe~\cite{Ibarra:2013zia,Rott:2014kfa,Ando:2015qda,Giesen:2015ufa}). But given the lacking evidence of DM via direct detection, indirect detection or colliders, its basic nature has not been determined yet~\cite{Agrawal:2010fh}, in that sense DM could be scalar~\cite{Burgess:2000yq,Barger:2007im,Hambye:2009pw,Cline:2013gha,Brdar:2013iea,Aoki:2014cja}, fermionic~\cite{Kim:2006af,Kusenko:2009up,Kumericki:2012bh}, vectorial~\cite{McDonald:1993ex,Birkedal:2006fz,Bhattacharya:2011tr,Baek:2012se} or even have multiple components~\cite{Boehm:2003ha,Cao:2007fy,Profumo:2009tb,Bhattacharya:2013hva,Bhattacharya:2013nya,Kajiyama:2013rla}.

The electric charge of a  particle inside a multiplet is calculated using its isospin $I_3$ and hypercharge $Y$. If we require a DM candidate to fit the restriction of non-existing electrical charge,  $Q=I_3 +Y/2=0$, we arrive to the condition that only a particle where $Y=-2\,I_3$ can be DM. 

If one wants a particle to be stable, the most common mechanism is to include an extra unbroken symmetry that will forbid Lagrangian terms that would induce decay. Particularly, the cyclic Abelian group $\mathsf{Z_n}$ is most commonly used, $\mathsf{Z_2}$ being the simplest of these symmetries. Ten years ago, a new method of stabilizing DM was introduced~\cite{Cirelli:2005uq,Cirelli:2007xd,Cirelli:2009uv}: in this minimal dark matter scenarios the high $\mathsf{SU(2)_L}$ representation of the new fields does not permit the construction of renormalizable operators that allow the decay of the neutral component of the new field. Nevertheless it was proven that this scenario is not compatible with neutrino masses at one-loop, although higher loops might still be possible~\cite{Ahriche:2015wha,Sierra:2016qfa,Sierra:2016rcz}.

Electroweak~(EW) multiplets that contain a DM candidate will have tree level interactions with quarks via Z bosons, giving a direct detection cross section via nucleon recoil that are proportional to $Y^2$ (the hypercharge squared)~\cite{Essig:2007az}. Experimental results have eliminated these particles as DM, if their mass is bigger than 10 GeV, they would produce scattering cross sections big enough, that would have been seen given current limits. This condition eliminates multiplets with even number of fields (i.e., doublets, quartets etc.), except those with $Y=0$ ~\cite{Cirelli:2005uq,Cirelli:2007xd,Cirelli:2009uv}. In models with multiple added fields it is possible to have other kind of candidates if they mix with the above-mentioned particles. Also, there is an exception in the case of the scalar doublet with hypercharge $Y=1$. For these models, that have two scalar doublets (SM Higgs plus the one added) a mass splitting can be enforced between the scalar and pseudo-scalar~\cite{Farzan:2012ev}, this can eliminate the coupling with the Z at tree level, leaving only loop induced detection, allowing the neutral component of the doublet to have a detection rate that is still lower than experimental bounds.

In this article we will present multiple models with two-loop neutrino masses and a DM candidate that helps to generate the mass of the neutrinos, all of the candidates being either fermionic or scalar singlets, or a scalar doublet (or a mixture of the latter two). All of these models are unique and present interesting phenomenological features consistent with current experimental bounds. We will follow closely the results presented in~\cite{Sierra:2014rxa}, although we have two main differences. The first difference is that while they have shown that there are twenty independent ways to create genuine two-loop diagrams, we only present models for nineteen diagrams, the reason for this is that one of the diagrams does not give any viable model that fit our conditions. The second difference is that restrictions for genuine two-loop models presented in~\cite{Sierra:2014rxa} get relaxed with the implementation of the $\mathsf{Z_2}$ symmetry; one can see that some fields and interactions that where forbidden since they allow tree and one-loop contributions, are now allowed in our models.

In section~\ref{sec:results} we present the models, including for all of them the corresponding Lagrangian interactions and the neutrino mass equations, while in section~\ref{sec:pheno} we present a small review on phenomenological applications of some of the fields, that can be used to discriminate some models. In Appendix~\ref{sec:app} we show all the integrals that are needed to calculate the neutrino masses., and in Appendix~\ref{sec:appB} we show two examples of the models to clarify some of the aspects of their construction.

\section{The Models}
\label{sec:results}

In this section we present all genuine two-loop neutrino models that are possible to construct by only adding to the SM colorless $\mathsf{SU(2)_L}$ singlets and doublets.  We do not have a specific rule to explicitly forbid the one-loop models from appearing, instead, we construct all possible models that generate a two-loop diagram and eliminate the ones with particle content that can be used to build any tree level or one-loop diagram. It is important to keep in mind, that for all of the models that will be presented, all BSM fields contain only colorless particles, we will not discuss particles that have $\mathsf{SU(3)_c}$ charges as possible messenger inside the loops, doing so would increase the number of possible models greatly. Nevertheless, the extension of the models to include colored particles is quite simple: one just needs to assign the same  $\mathsf{SU(3)_c}$ charge to all particles around one of the loops, while simultaneously avoiding charging the particle that would be the DM candidate. This is always possible for all models that will be presented. These two-loop setups do allow for a portion of the particles to be charged under color but still have other colorless particles, including the DM candidate. Although it is not the focus of this article, we want to remark that this is one of the more interesting aspects about these models and it embodies one of the important differences between one-loop and two-loop scenarios.

The results will be divided into two main sections according to the number of fields needed to complete the diagrams: those that require seven particles~(section~\ref{sub:7particles}) and those with six particles~(section~\ref{sub:6particles}). The theoretical difference between models in sections~\ref{sub:7particles} and~\ref{sub:6particles} is that models in the former have solely Yukawa and cubic interactions, while in the latter a scalar quartic coupling is always present. 

While some of the fields used to complete the diagrams can be SM, there is an exception: the Higgs, a scalar with quantum numbers $(\mathbf{2},1)$ under $\mathsf{SU(2)_L \times U(1)_Y}$, cannot be present in the internal lines since it will fracture the two-loop diagram and generate a one-loop mass contribution after EW symmetry breaking, this effect has been carefully explained in section~3.3 of ref.~\cite{Farzan:2012ev}. Furthermore, all new scalars are taken to be inert, i.e., vanishing VEV, to avoid as well one-loop diagrams from appearing after $\mathsf{SU(2)_L}$ breaking. The SM fields that might be used to complete the loop are then the fermions $L=(\mathbf{2},1)$ and $e_R=(\mathbf{1},2)$. 
The field $L$ appears explicitly in the Weinberg operator  and therefore is used as two of the external legs in the diagrams, it has a left-handed chiral structure encoded in the projector $P_L = (1-\gamma_5)/2$. BSM fermions needed to construct the models presented  lack this chiral structure and are vector-like in nature.

There are multiple ways of drawing the two-loop diagrams. We will choose a particular way for which the figure does not have any crossing lines in a plane (i.e., any vertex in the drawing will represent a term in the Lagrangian). Sketching them in this fashion will make the two-loops clear and visually explicit. Based on the drawings, we will divide each section into two categories and we will label the models in each of them by the category's number and a lowercase letter, e.g., for Category 1, with eight possible diagrams, the diagrams will be numbered from 1.a to 1.h. The diagrams are traced in the standard way, the dashed lines refer to scalars and the solid lines to fermions. The particles in the loops are generically labeled\footnote{For simplicity, we dropped the labels of the legs -- they should be understood as the SM lepton doublet $L$ and the Higgs doublet $H$. Furthermore, when writing the Lagrangians we will make explicit the fermionic or scalar nature of the fields, substituting $X$ by $F$ or $S$, respectively.} as $X_i$ with $i=1$ to 7 in section~\ref{sub:7particles} and $i=1$ to 6 in section~\ref{sub:6particles}.

We assume the presence of a $\mathsf{Z_2}$ symmetry in all models. This symmetry, introduced with the purpose of ensuring the stability of the DM candidate, will forbid (when necessary) Lagrangian terms that can be used to construct type-I seesaw masses and, in most cases, forbids one-loop diagrams. Since we are considering only singlets and doublets of $\mathsf{SU(2)_L}$, we do not need to be concerned about scenarios with seesaw type-II and type-III. Additionally, since we do not allow fermion singlets with zero hypercharge and uncharged under $\mathsf{Z_2}$  (i.e., right-handed neutrinos) it is not possible to write neutrino Dirac masses. 

Following what was explained in the introduction, we are only interested in models that explicitly have an DM particle functioning as an internal messenger in the loops. Therefore we have excluded all of the models that, while theoretically allow for a  genuine two-loop neutrino mass and have a field that contains a DM candidate, do not have an electrically neutral particle in the loop. This can happen, for example, when one decomposes the fields into their corresponding components and the only contribution to the neutrino mass comes from diagrams where all of the particles inside the loops have an electric charge.

In order to build the two-loop diagrams we need to specify the $\mathsf{SU(2)_L \times U(1)_Y \times Z_2}$ quantum numbers of each particle $X_i$. Models of Categories 1, 2 and 3 have four different $\mathsf{SU(2)_L}$ possibilities while models of Category 4 have just two, we give these possibilities labeled by roman numerals (\textit{i}-\textit{iv}). The hypercharge $Y$ (with normalization $Q=I_3+Y/2$)  is encoded in the parameters $\alpha$ and $\beta$, and has integer values $-4<\alpha,\beta<4$. While models can be built using higher hypercharges, we will not be using those since they would involve particles with ``exotic'' electric charge such as $Q=3$ or particles with higher charges that spoil the perturbativity of $\alpha_\text{EW}$.

In what $\mathsf{Z_2}$ is concerned, the SM particles transform as even ($\boldsymbol{+}$) while the BSM particles transform as either even ($\boldsymbol{+}$) or odd ($\boldsymbol{-}$). This ambivalence results from multiple possibilities to assign the $\mathsf{Z_2}$ charges while allowing for all the interactions in the loop. Ignoring the trivial assignments we have a total of 3 different combinations for the $\mathsf{Z_2}$ charges, identifiable by capital letters (A, B, C).

In each section we present the relevant Lagrangian terms for the vertices of the diagram as well as the neutrino mass matrix that can be calculated after integration. We will not be concerned either with the exact composition of the Yukawa matrices or with the possibility to recreate neutrino mixing since one can always use the Casas-Ibarra parametrization~\cite{Casas:2001sr} to compute it. 

The models are presented in the form of tables; for each Class (diagram) we give the hypercharge parameters $\alpha$ and $\beta$, the DM candidate appearing in the model, the minimum number of BSM fields needed to complete the diagram (considering that fields with the same quantum numbers are the same and  taking into account possible SM fields inside the loop). We also present the number of BSM scalar doublets, with hypercharge $Y=1,3$ ($\mathbf{2}^1_S$, $\mathbf{2}^3_S$), and finally we mark if the models contain a doubly charged fermion and/or scalar without distinguishing their nature, i.e., singlets with hypercharge $Y=4$ or doublets with $Y=3$.

\subsection{Seven-particle models}
\label{sub:7particles}

In this section we discuss all two-loop models for which seven particles are needed. The Lagrangian for these models will only have Yukawa and scalar cubic interactions (although some of these will only have the first type).

The diagrams can be split in two ways: those for which all external legs couple to outside particles in the loop - the planar diagrams or Category~1 (Figure~\ref{fig:category1}) and those in which two external legs couple to an internal line in the loop - the non-planar diagrams or Category~2 (Figure~\ref{fig:category2}). One has in total 15 different diagrams built with seven particles, 8 in Category~1 and 7 in Category~2.

There are three different $\mathsf{Z_2}$ charge assignments for each category which are given in the upper right corner of Tables~\ref{tab:assignmentscategory1} and~\ref{tab:assignmentscategory2}.

\subsubsection{Category 1}
\label{sub:cat1} 

\begin{figure}[H]
	\centering
	\captionsetup[subfigure]{labelformat=empty}
	\begin{subfigure}[b]{0.25\textwidth}
		\includegraphics[scale=0.2]{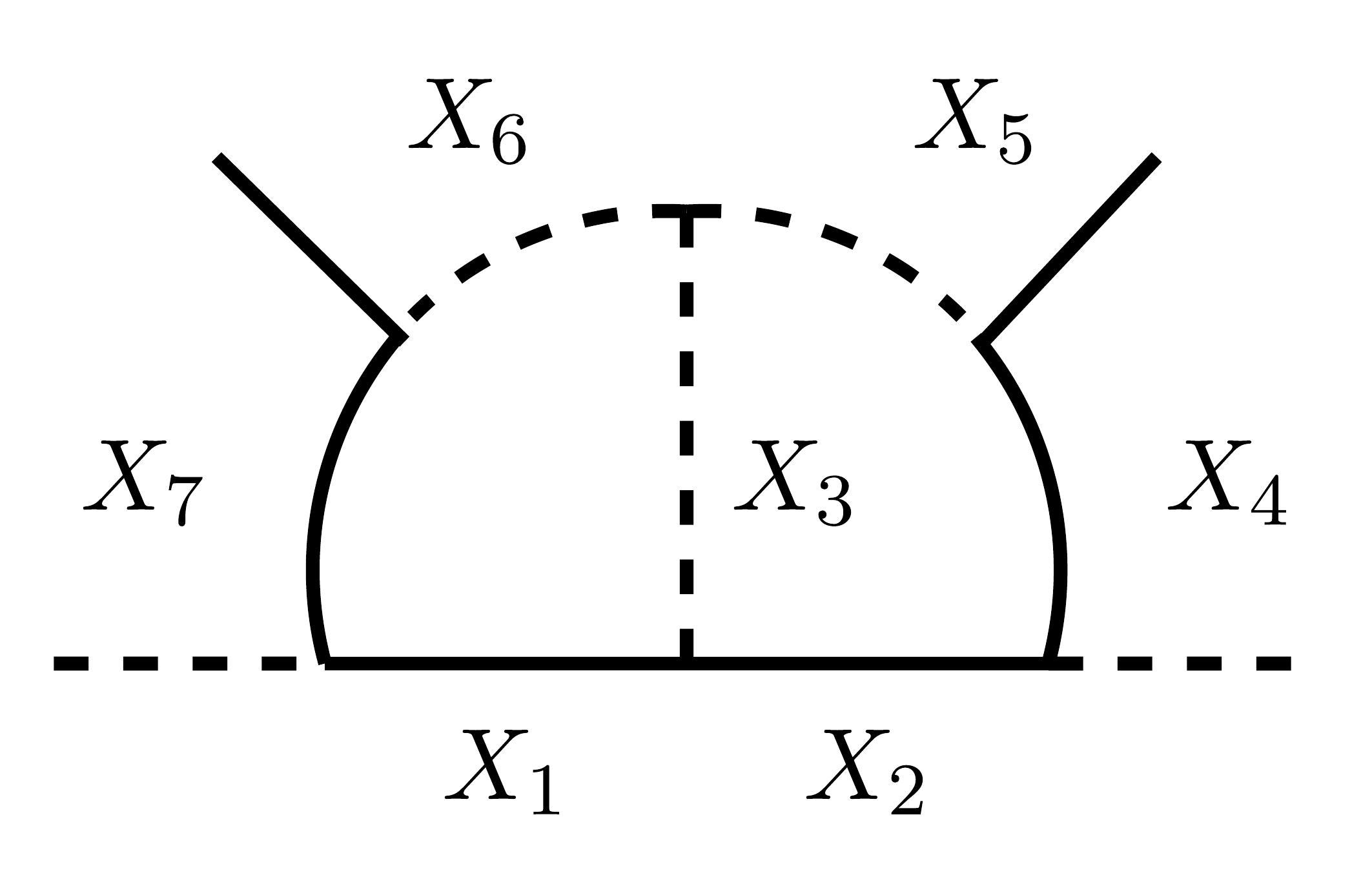}		
		\caption{\hspace{0.5cm} 1.a}
	\end{subfigure}\hspace{0.8cm}
	\begin{subfigure}[b]{0.25\textwidth}
		\includegraphics[scale=0.2]{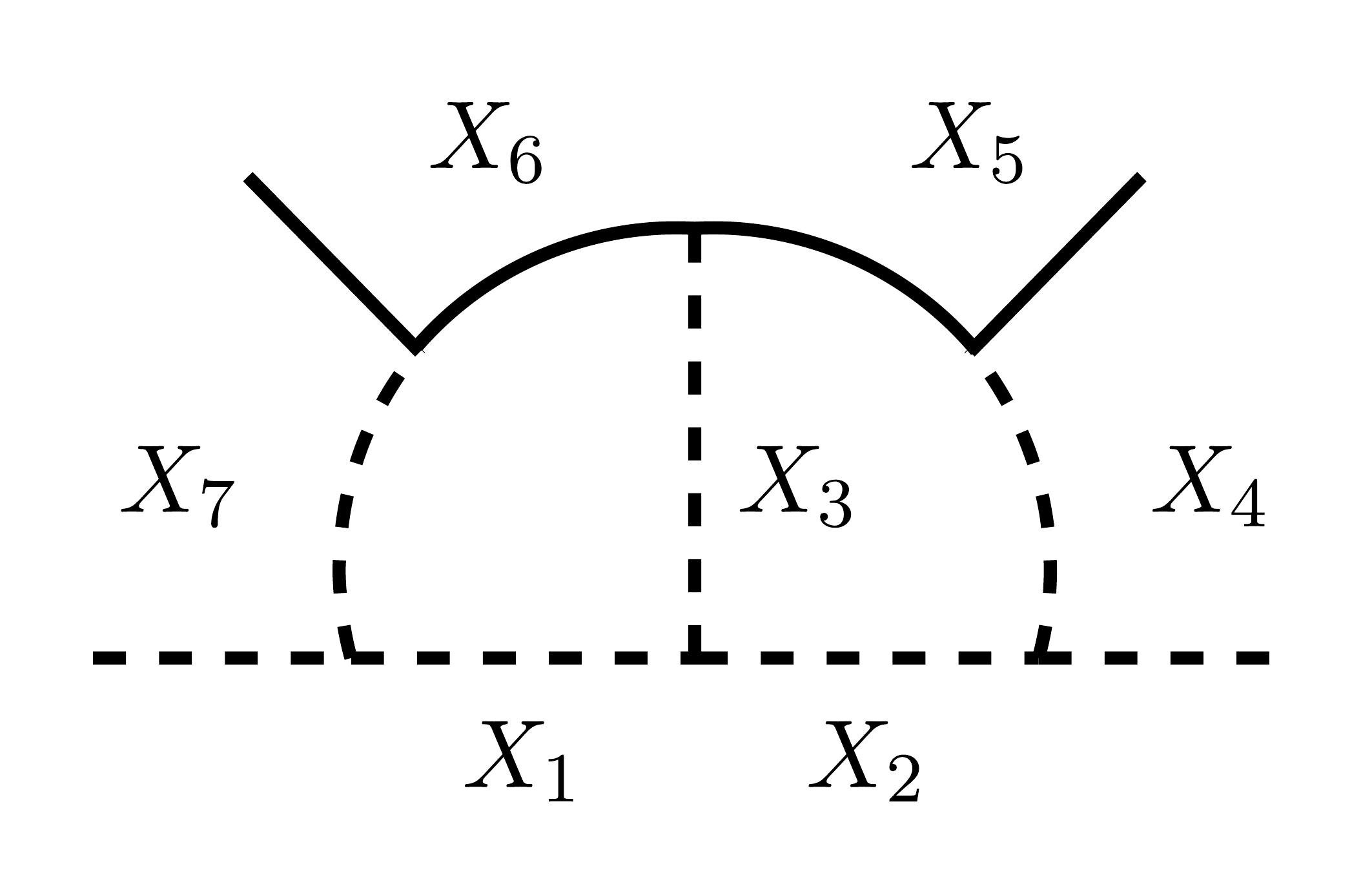}		
		\caption{\hspace{0.5cm} 1.b}
	\end{subfigure} \hspace{0.8cm}
	\begin{subfigure}[b]{0.25\textwidth}
		\includegraphics[scale=0.2]{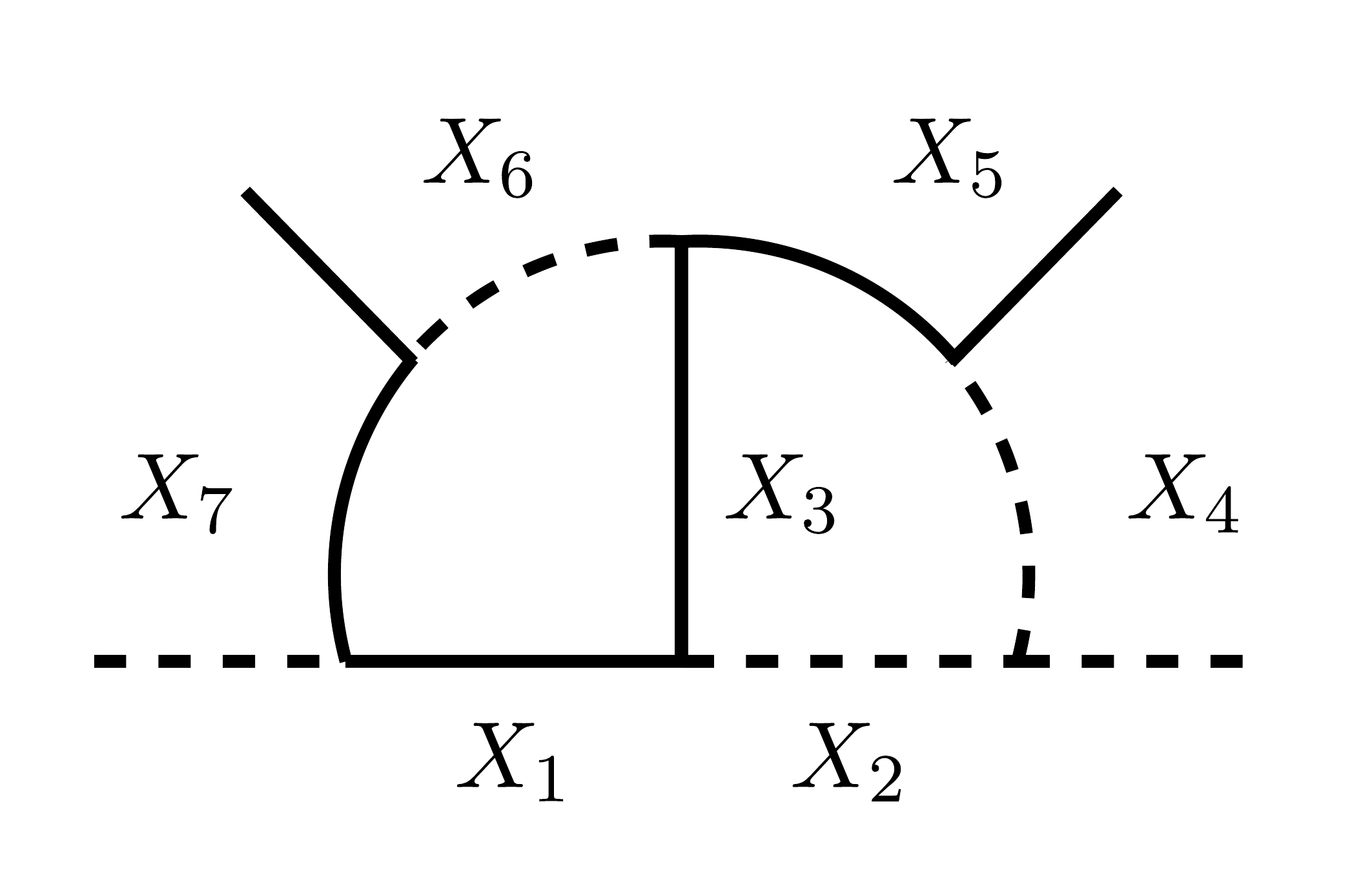}		
		\caption{\hspace{0.5cm} 1.c}
	\end{subfigure} \vspace{0.5cm}
	
	\begin{subfigure}[b]{0.25\textwidth}
		\includegraphics[scale=0.2]{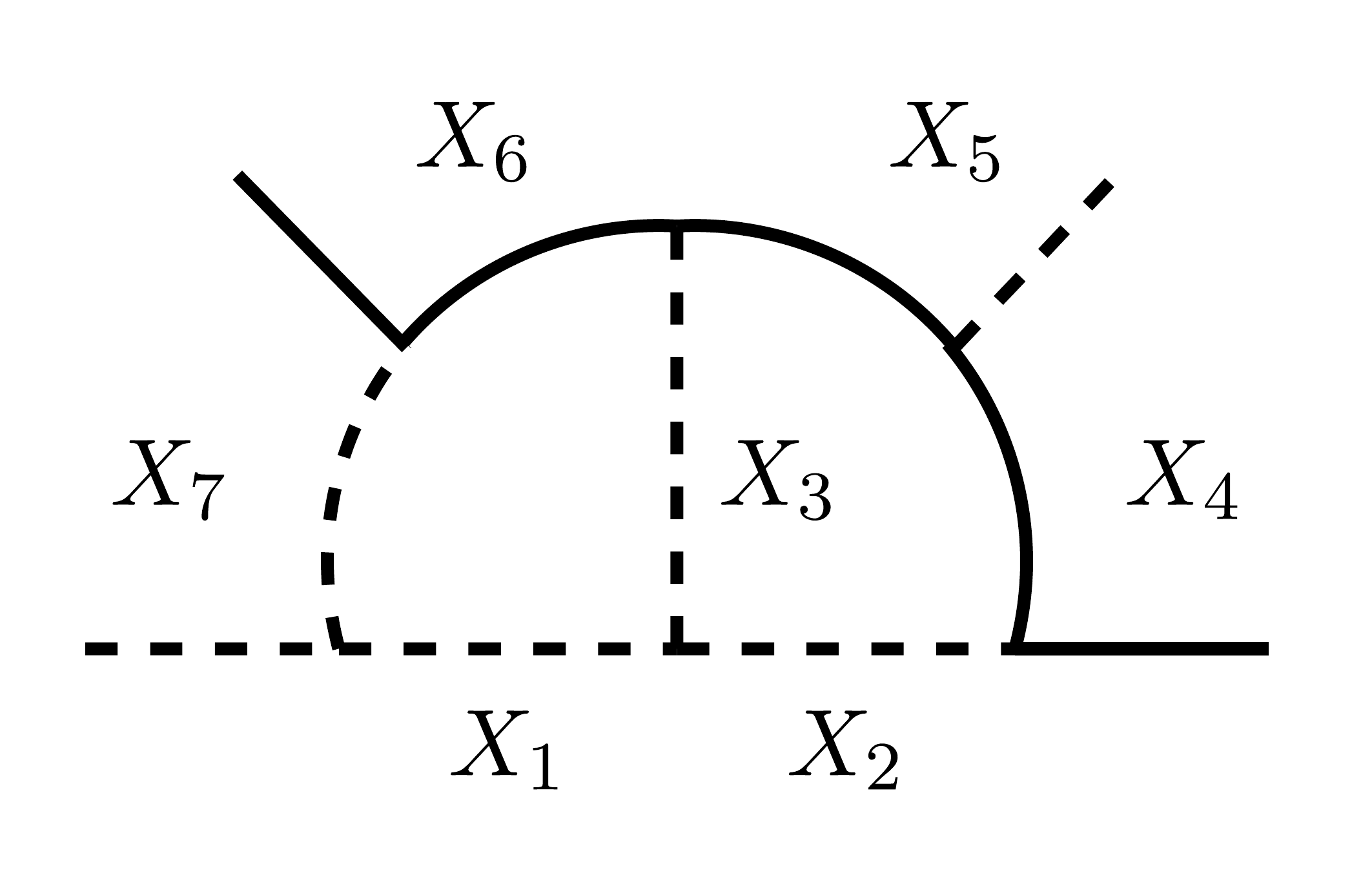}		
		\caption{\hspace{0.5cm}  1.d}
	\end{subfigure} \hspace{0.8cm}
	\begin{subfigure}[b]{0.25\textwidth}
		\includegraphics[scale=0.2]{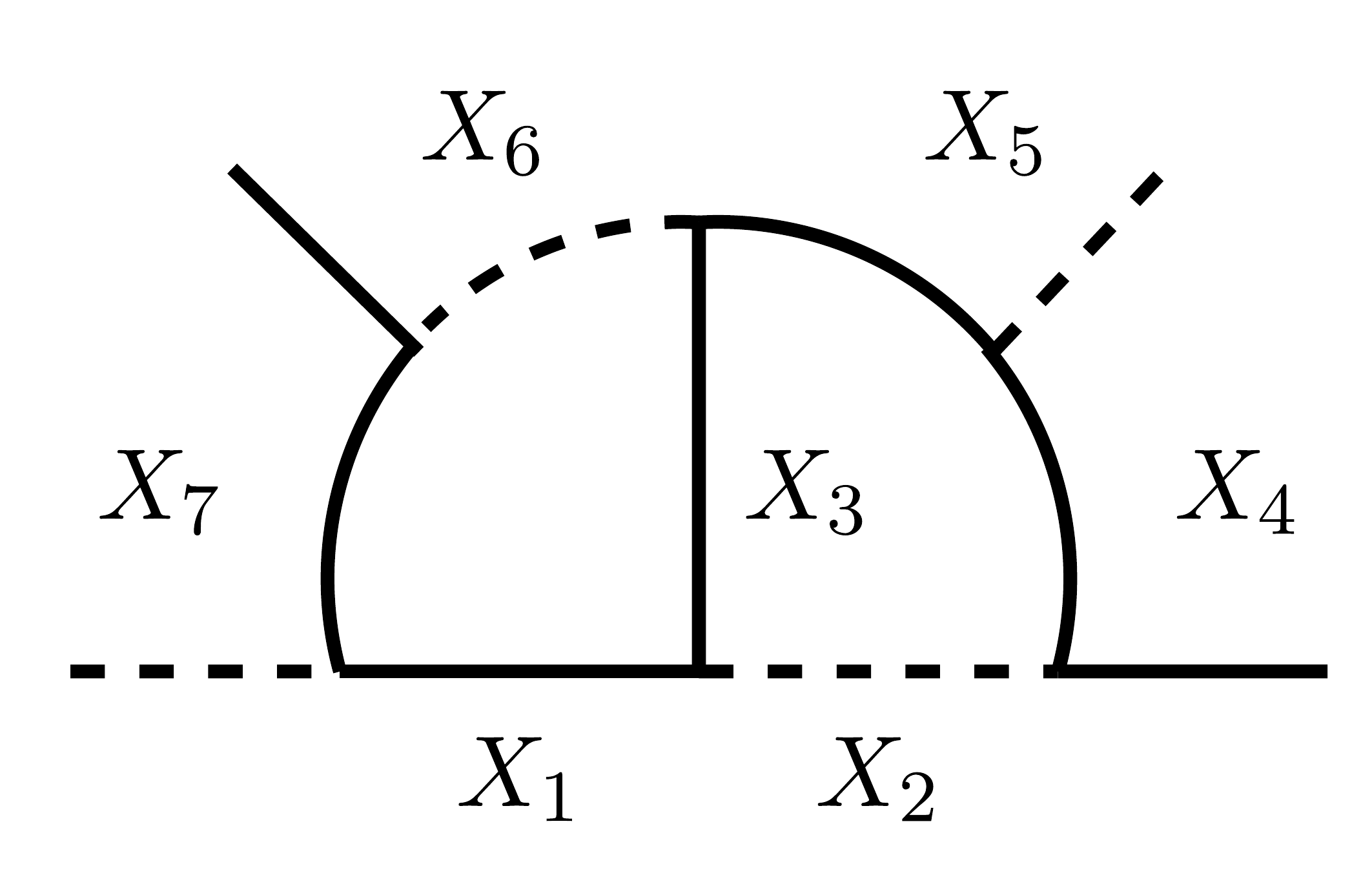}		
		\caption{\hspace{0.5cm}  1.e}
	\end{subfigure}\hspace{0.8cm}
	\begin{subfigure}[b]{0.25\textwidth}
		\includegraphics[scale=0.2]{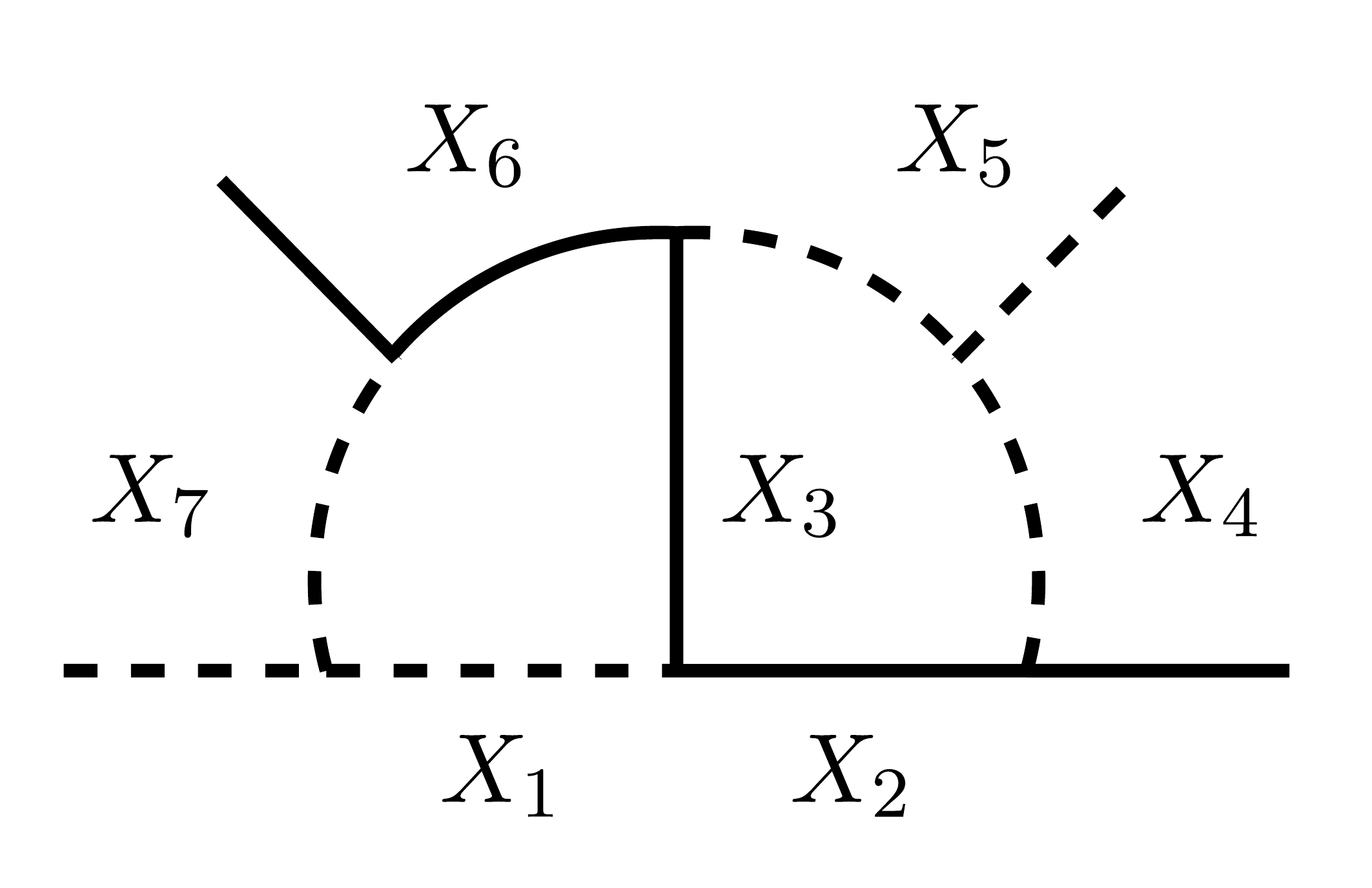}		
		\caption{\hspace{0.5cm}  1.f}
	\end{subfigure} \vspace{0.5cm}
	
	\begin{subfigure}[b]{0.25\textwidth}
		\includegraphics[scale=0.2]{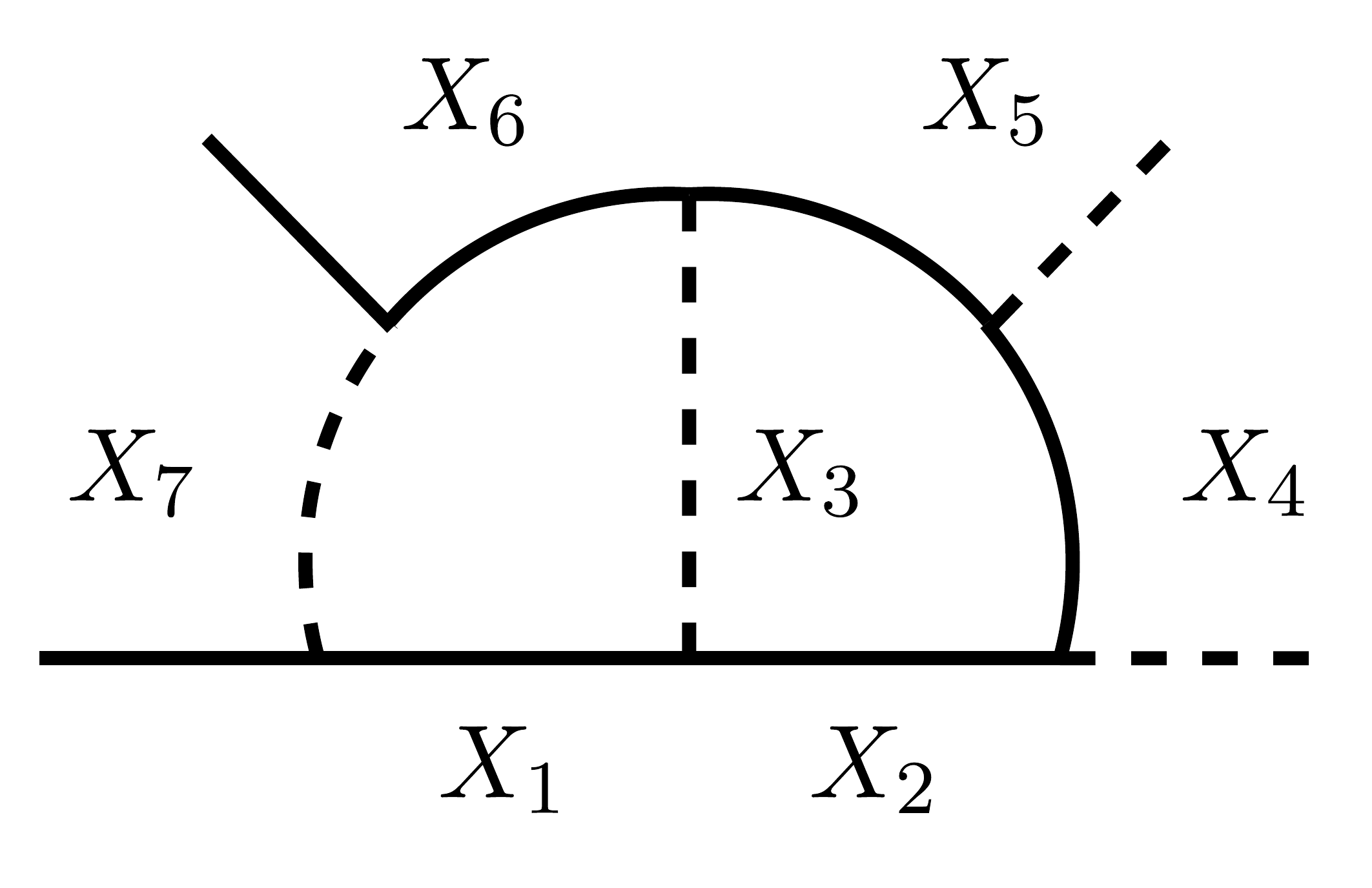}		
		\caption{\hspace{0.5cm}  1.g}
	\end{subfigure} \hspace{0.8cm}
	\begin{subfigure}[b]{0.25\textwidth}
		\includegraphics[scale=0.2]{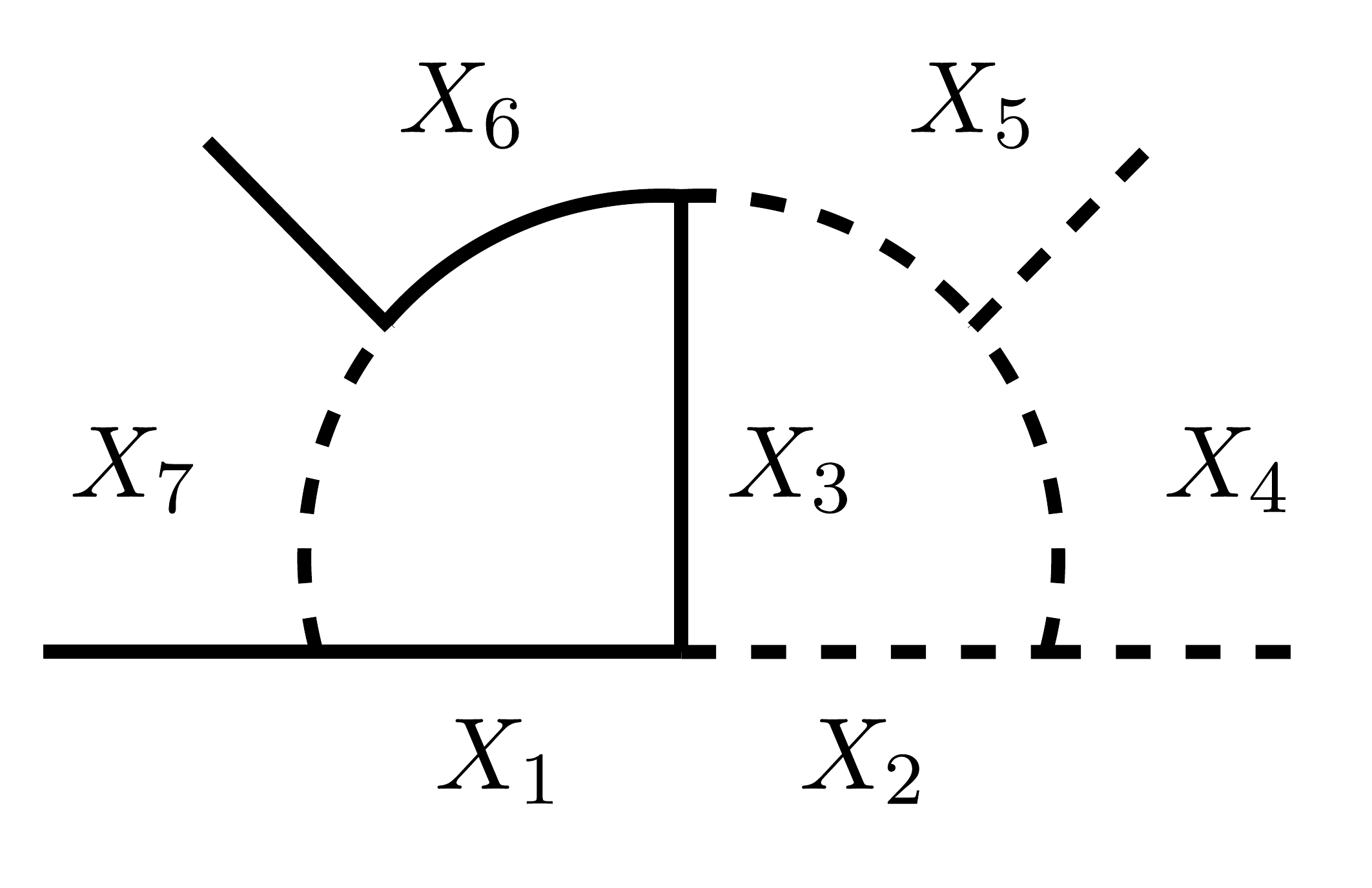}		
		\caption{\hspace{0.5cm} 1.h}
	\end{subfigure}
	\caption{\label{fig:category1} Two-loop diagrams for Class 1.a - 1.h in Category~1.}
\end{figure}

Category 1 is composed of 8 diagrams, shown in Figure~\ref{fig:category1}. Each diagram defines a Class; there are two classes with six Yukawas and no cubic interactions (1.d and 1.g), two diagrams for which there are five Yukawas and one cubic interaction (1.a and 1.e), three diagrams with four Yukawas and two cubic interactions (1.b, 1.f and 1.h) and just one diagram that presents three Yukawas and three cubic interactions (1.c).

In Table~\ref{tab:assignmentscategory1} we show the different assignments (\textit{i}-\textit{iv}) for the $\mathsf{SU(2)_L}$ (upper left), the hypercharge (bottom) and the $\mathsf{Z_2}$ charge (upper right) assignments. Notice that the hypercharge assignment is diagram dependent, i.e., depends on the geometry of the diagram, and hence one has three different assignments (one for each line of diagrams in Figure~\ref{fig:category1})

\begin{table}[H]
	\setlength{\tabcolsep}{2pt}
	\centering
	\begin{tabular}{|c||c|c|c|c|c|c|c|}
		\hline 
		& $X_{1}$ & $X_{2}$ & $X_{3}$ & $X_{4}$ & $X_{5}$ & $X_{6}$ & $X_{7}$\\
		\hline 
		\hline 
		\textit{i} & \textbf{1} & \textbf{1} & \textbf{1} & \textbf{2} & \textbf{1} & \textbf{1} & \textbf{2}\\
		\hline 
		\textit{ii} & \textbf{1} & \textbf{2} & \textbf{2} & \textbf{1} & \textbf{2} & \textbf{1} & \textbf{2}\\
		\hline 
		\textit{iii} & \textbf{2} & \textbf{1} & \textbf{2} & \textbf{2} & \textbf{1} & \textbf{2} & \textbf{1}\\
		\hline 
		\textit{iv} & \textbf{2} & \textbf{2} & \textbf{1} & \textbf{1} & \textbf{2} & \textbf{2} & \textbf{1}\\
		\hline 
	\end{tabular} \hspace{1cm}
	\begin{tabular}{|c||c|c|c|c|c|c|c|}
		\hline 
		& $X_{1}$ & $X_{2}$ & $X_{3}$ & $X_{4}$ & $X_{5}$ & $X_{6}$ & $X_{7}$\\
		\hline 
		\hline 
		A & $\boldsymbol{+}$ & $\boldsymbol{-}$ &$\boldsymbol{-}$ & $\boldsymbol{-}$ & $\boldsymbol{-}$ &$\boldsymbol{+}$ & $\boldsymbol{+}$\\
		\hline 
		B & $\boldsymbol{-}$ & $\boldsymbol{+}$ & $\boldsymbol{-}$ & $\boldsymbol{+}$ & $\boldsymbol{+}$ & $\boldsymbol{-}$ & $\boldsymbol{-}$\\
		\hline 
		C & $\boldsymbol{-}$ & $\boldsymbol{-}$ & $\boldsymbol{+}$ & $\boldsymbol{-}$ & $\boldsymbol{-}$ & $\boldsymbol{-}$ & $\boldsymbol{-}$\\
		\hline 
	\end{tabular} \vspace{0.3cm}
	
	\begin{tabular}{|c||c|c|c|c|c|c|c|}
		\hline 
		& $X_{1}$ & $X_{2}$ & $X_{3}$ & $X_{4}$ & $X_{5}$ & $X_{6}$ & $X_{7}$\\
		\hline 
		\hline 
		a, b, c & $\alpha$  & $\beta$ & $\alpha-\beta$ & $\beta+1$ & $\beta$ & $\alpha$ & $\alpha-1$\\
		\hline 
		d, e, f & $\alpha$ & $\beta$ & $\alpha-\beta$ & $\beta-1$ & $\beta$ & $\alpha$ & $\alpha-1$\\
		\hline 
		g, h & $\alpha$ & $\beta$ & $\alpha-\beta$ & $\beta+1$ & $\beta+2$ & $\alpha+2$ & $\alpha+1$\\
		\hline 
	\end{tabular}
	\caption{\label{tab:assignmentscategory1}  
		$\mathsf{SU(2)_L}$ (upper left), $\mathsf{Z_2}$ (upper right) and  $\mathsf{U(1)_Y}$ (bottom) assignments for particles $X_1$ to $X_7$ of Class 1.a - 1.h in Figure~\ref{fig:category1}.}
\end{table}

\paragraph{Class 1.a}

There are four models in Class 1.a for which it is possible to generate neutrino masses and have stable DM candidates. In all the cases the DM candidate is a scalar: $\mathbf{1}^0_{S}$ in one model and $\mathbf{2}^1_{S}$ in the remaining ones. There is an interesting model for which only four BSM particles are required however this model does not have any doubly charged particle possible testing at LHC. The results are presented in Table~\ref{tab:res1a}. 

The most generic Lagrangian is given by,
\begin{equation}
\begin{aligned}
\mathcal{L}^{}_\text{1.a}&= Y^{ia}_1\, (\overline{L^c}_i\,P_L) {F_4}_a \,S_5^{\dagger}\,+\,Y^{ab}_2\,\overline{F_4}_a\, {F_2}_b\, H \, +\, Y^{bc}_3\,\overline{F_2}_b {F_1}_c\, S_3^{\dagger}\,+\,Y^{cd}_4\,\overline{F_1}_c\,{F_7}_d\,H \\
&\,+\,  Y^{dj}_5\, \overline{F_7}_d (P_L\, L_j)\,S_6 \,+\,\mu S_3 S_5 S_6^{\dagger}\,+\, \text{h.c.}\,,
\end{aligned}
\label{eq:L1a}
\end{equation}
where $i,\,j=1,\,2,\,3$ are the generation indices of the SM fields and $a$, $b$, $c$, $d$ are generation indices of the BSM fermions. 

The mass matrix is then given as,

\begin{equation}
\label{eq:M1a}
\begin{aligned}
\left(M^\text{1.a}_{\nu}\right)_{ij}&=\frac{\mu\, (Y^{ia}_5\,Y^{ab}_3\,Y^{bj}_1+Y^{ib}_1\,Y^{ba}_3\,Y^{aj}_5)}{4\,(2 \pi)^8}\,\sin (2\theta_{17})\,\sin (2 \theta_{24}) \\
& \times  \sum_{\alpha,\,\beta\,=\,1}^{2} (-1)^\alpha\, (-1)^\beta \,\left(2\,{m_{17}}^{}_{\alpha} \, {m_{24}}^{}_{\beta} \, 
I^1_\chi\,+\, 
I^{k^2}_\chi\, +\, 
I^{q^2}_\chi\, -\, 
I^{(k+q)^2}_\chi \right)\,.
\end{aligned}
\end{equation}
The masses $m_\text{AB}$ and the angles $\theta_\text{AB}$ refer to the mass eigenstate of a particle, $X_{AB}$, that got mixed between particles $X_A$ and $X_B$ and its mixing angle, respectively. The indices $\alpha$ and $\beta$ indicate the different components of the mass eigenstates. Hereinafter we will use this terminology for all the classes. The integrals of $I^{1,\,q^2,\,k^2,\,(q+k)^2}_\chi$ type are shown in the appendix and are computed explicitly in~\cite{Sierra:2014rxa}. We should consider for this class $\chi\equiv \{S_6,\,{F_{17}}^{}_\alpha,\,S_3,\,S_5,\,{F_{24}}_\beta\}$.

\begin{table}[H]
	\centering
	\begin{tabular}{|c|c|c|c|c|}
		\hline 
		& \multicolumn{4}{c|}{A}\\
		\hline 
		\hline 
		& \textit{i} & \multicolumn{2}{c|}{\textit{ii}} & \textit{iii} \\
		\hline 
		$\alpha$ & 2 & 2 & 4 & 3\\
		\hline 
		$\beta$ & 2 & 1 & 3 & 2\\
		\hline 
		DM & $\mathbf{1}_{S}$ & $\mathbf{2}_{S}$ & $\mathbf{2}_{S}$ & $\mathbf{2}_{S}$\\
		\hline 
		\# & 5 & 4 & 7 & 6\\
		\hline 
		$\mathbf{2}^1_{S}/\mathbf{2}^3_{S}$ & 1/$\times$ & 2/$\times$ & $\times$/$\times$ & $\times$/$\times$\\
		\hline 
		$F^{++}/S^{++}$ & \checkmark/$\times$ & $\times$/$\times$ & \checkmark/\checkmark{} & \checkmark/\checkmark{}\\
		\hline 
	\end{tabular} 
	\caption{\label{tab:res1a} Realizations of Class 1.a of Figure~\ref{fig:category1}. For each set of $\mathsf{SU(2)_L}$ quantum numbers (roman numerals) and $\mathsf{Z_2}$ assignments (capital letters A, B, C) we give the hypercharge in terms of $\alpha$ and $\beta$ parameters, the DM candidates, the number of BSM particles (\#), the number of $\mathbf{2}^1_{S}$ and $\mathbf{2}^3_{S}$ present as well as the existence (or not) of doubly-charged fermions ($F^{++}$) and doubly-charged scalars ($S^{++}$). The symbol $\checkmark$($\times$) means present (absent) while the symbol $\boldsymbol{\ast}$, if present in the DM row, means mixing between $\mathbf{1}^0_{S}$ and $\mathbf{2}^1_{S}$. The $\mathsf{SU(2)_L}$, $\mathsf{Z_2}$ and $\mathsf{U(1)_Y}$ assignments are given in Table~\ref{tab:assignmentscategory1}.}
\end{table}

\paragraph{Class 1.b}

This class has five models for which one can generate neutrino masses. All models have scalar DM candidates stabilized by $\mathsf{Z_2}$ symmetry whose assignment is given in Table~\ref{tab:assignmentscategory1}. The most appealing model needs five BSM fields to be drawn and presents a DM candidate that is a mixture between $\mathbf{1}^0_{S}$ and $\mathbf{2}^1_{S}$, referred as $\ast$ in the table of results, Table~\ref{tab:res1b}.
The Lagrangian for this class is
\begin{equation}
\begin{aligned}
\mathcal{L}^{}_\text{1.b}&= 
Y^{ia}_1\,(\overline{L^\texttt{C}}_i\,P_L) {F^\texttt{C}_5}_a \,S_4\,+\,
Y^{ab}_2\,\overline{F^\texttt{C}_5}_a\, {F^\texttt{C}_6}_b\, S_3 \, +\, 
Y^{bj}_3\,\overline{F^\texttt{C}_6}_b\,(P_L\, L_j)\,S^\dagger_7 \,+\,
\mu_1\,H\,S^\dagger_1\,S_7\\
&\,+\,\mu_2\,S_1\,S^\dagger_2\,S^\dagger_3\,+\,
\mu_3\,H\,S_2\,S_4^\dagger\,+\,\text{h.c.}\,,
\end{aligned}
\label{eq:L1b}
\end{equation}
for which the neutrino mass matrix is given by
\begin{equation}
\label{eq:M1b}
\begin{aligned}
\left(M^\text{1.b}_{\nu}\right)_{ij}&=\frac{\mu_2\, (Y^{ia}_3\,Y^{ab}_2\,Y^{bj}_1+Y^{ib}_1\,Y^{ba}_2\,Y^{aj}_3)}{4(2 \pi)^8}\,\sin(2\theta_{17})\,\sin (2\theta_{24})  \\
&\times\sum_{\alpha,\beta = 1}^{2} (-1)^\alpha (-1)^\beta \left(2\,m_{5}\,m_{6}\,I_\chi^1\,+\, I_\chi^{k^2}\, +\,I_\chi^{q^2}\, -\, I^{(k+q)^2}_\chi\right)\,,
\end{aligned}
\end{equation}
with $\chi\equiv \{F_6,\,{S_{17}}_\alpha, S_3, F_5,\,{S_{24}}_\beta\}$.

\begin{table}[H]
	\centering
	\begin{tabular}{|c|c|c|c|c|c|}
		\hline 
		& \multicolumn{5}{c|}{A}\\
		\hline 
		\hline 
		& \textit{i} & \multicolumn{3}{c|}{\textit{ii}} & \textit{iii}\\
		\hline 
		$\alpha$ & -2 & 4 & -2 & -2 & -3\\
		\hline 
		$\beta$ & -2 & 3 & -1 & -3 & -4\\
		\hline 
		DM & $\boldsymbol{\ast}$ & $\mathbf{2}_{S}$ & $\boldsymbol{\ast}$ & $\mathbf{2}_{S}$ & $\mathbf{2}_{S}$\\
		\hline 
		\# & 6 & 7 & 5 & 6 & 7\\
		\hline 
		$\mathbf{2}^1_{S}/\mathbf{2}^3_{S}$ & 1/1 & 1/2 & 2/1 & 1/2 & 1/2 \\
		\hline 
		$F^{++}/S^{++}$ & $\times$/\checkmark{} & \checkmark/\checkmark{} & $\times$/\checkmark{} & \checkmark/\checkmark{} & \checkmark/\checkmark{}\\
		\hline 
	\end{tabular}
	\caption{\label{tab:res1b} The same as in Table~\ref{tab:res1a} for Class 1.b of Figure~\ref{fig:category1}.}
\end{table}

\paragraph{Class 1.c}

There are four models generating neutrino masses via diagram 1.c: two of them lead to scalar DM and the other two to fermion singlet DM candidates. The models in the first and last column of Table~\ref{tab:res1c} are similar in all respects other than the $\mathsf{Z_2}$ assignment.
The Lagrangian for this class is given by 
\begin{equation}
\begin{aligned}
\mathcal{L}^{}_\text{1.c}&= 
Y^{ia}_1\,(\overline{L^\texttt{C}}_i\,P_L) {F^\texttt{C}_5}_a \,S_4\,+\,
Y^{ab}_2\,\overline{F^\texttt{C}_5}_a\, {F_3}_b\, S^\dagger_6 \, +\, 
Y^{bc}_3\,\overline{F_3}_b\,{F_1}_c\, S^\dagger_2\,+\,  
Y^{cd}_4\,\overline{F_1}_c\,{F_7}_d\,H \\
&+\,Y^{dj}_5\,\overline{F_7}_d\,(P_L\, L_j)\,S_6 \,+\,
\mu\,H\,S_2\,S_4^{\dagger}\,+\,\text{h.c.}\,,
\end{aligned}
\label{eq:L1c}
\end{equation}
where

\begin{equation}
\label{eq:M1c}
\begin{aligned}
\left(M^\text{1.c}_{\nu}\right)_{ij}&=\frac{ (Y^{ia}_5\,Y^{ab}_3\,Y^{bc}_2\,Y^{cj}_1+Y^{ic}_1\,Y^{cb}_2\,Y^{ba}_3\,Y^{aj}_5)}{4(2 \pi)^8}\sin(2\theta_{17}) \sin(2\theta_{24}) \\
&\times \sum_{\alpha,\,\beta\,=\,1}^{2} (-1)^\alpha\,(-1)^\beta \,\left(2\,{m_{17}}_{\alpha}\,m_{3}\,m_{5}\,I_\chi^1 \,+\, \left({m_{17}}^{}_{\alpha}\,-\,m_{3}\,-\,m_{5}\right)\,I_\chi^{k^2}\right. \\
& \left. + \left(-{m_{17}}^{}_{\alpha}\,-\,m_{3}\,+\,m_{5}\right)\, I_\chi^{q^2}\, +\, \left({m_{17}}^{}_{\alpha}\, +\, m_{3}\,+\,m_{5}\right)\,I_\chi^{(k+q)^2}\right)\,,
\end{aligned}
\end{equation}
with $\chi\equiv \{S_6,\,{S_{17}}_\alpha, F_3, F_5,\,{S_{24}}_\beta\}$.

\begin{table}[H]
	\centering
	\begin{tabular}{|c|c|c|c|c|}
		\hline 
		& \multicolumn{3}{c|}{A} & B\\
		\hline 
		\hline 
		& \textit{i} & \multicolumn{2}{c|}{\textit{ii}} & \textit{i}\\
		\hline 
		$\alpha$ & 2 & 2 & 2 & 2\\
		\hline 
		$\beta$ & 2 & 1 & -1 & 2\\
		\hline 
		DM & $\mathbf{1}_{F}$ & $\mathbf{2}_{S}$ & $\boldsymbol\ast$ & $\mathbf{1}_{F}$\\
		\hline 
		\# & 5 & 4 & 5 & 6\\
		\hline 
		$\mathbf{2}^1_{S}/\mathbf{2}^3_{S}$  & $\times$/1 & 2/$\times$ & 1/$\times$ & $\times$/1 \\
		\hline 
		$F^{++}/S^{++}$ & $\times$/\checkmark{} & $\times/\times$ & \checkmark/$\times$  & $\times$/\checkmark{}\\
		\hline 
	\end{tabular}
	\caption{\label{tab:res1c} The same as in Table~\ref{tab:res1a} for diagram 1.c of Figure~\ref{fig:category1}.}
\end{table}

\paragraph{Class 1.d}
This class presents fourteen realizations~(see results in Table~\ref{tab:res1d}) for the two-loop neutrino mass generation. The three $\mathsf{Z_2}$ charge assignments as well as the four possible $\mathsf{SU(2)_L}$ assignments are present. There is no fermion DM candidate coming from this class and the most appealing model, a three Higgs doublet model, needs only four BSM fields to be drawn. The Langrangian is
\begin{equation}
\begin{aligned}
\mathcal{L}^{}_\text{1.d}&= 
Y^{ia}_1\,(\overline{L^\texttt{C}}_i\,P_L) {F^\texttt{C}_4}_a \,S_2\,+\,
Y^{ab}_2\,\overline{F^\texttt{C}_4}_a\, {F^\texttt{C}_5}_b\, H \, +\, 
Y^{bc}_3\,\overline{F^\texttt{C}_5}_b\,{F^\texttt{C}_6}_c\, S_3\,+\,  
\,Y^{cj}_4\,\overline{F^\texttt{C}_6}_c\,(P_L\, L_j)\,S^\dagger_7 \\
&\,+\,\mu_1\,H\,S_1^{\dagger}\,S_7\,+\,
\mu_2\,S_1\,S_2^{\dagger}\,S^\dagger_3\,+\,\text{h.c.}\,,
\end{aligned}
\label{eq:L1d}
\end{equation}
and the mass matrix 
\begin{equation}
\label{eq:M1d}
\begin{aligned}
\left(M^\text{1.d}_{\nu}\right)_{ij}&=\frac{\mu_2\, (Y^{ia}_1\,Y^{ab}_3\,Y^{bj}_4+Y^{ic}_4\,Y^{ba}_3\,Y^{aj}_1)}{4(2 \pi)^8}\,\sin(2 \theta_{17})\,\sin(2\theta_{45})\\
& \times \sum_{\alpha,\beta\,=\,1}^{2} (-1)^\alpha\, (-1)^\beta\,\left(2\,{m_{45}}^{}_{\alpha}\,m_{6}\, I_\chi^1\, +\, I_\chi^{k^2}\, +\, I_\chi^{q^2}\,-\, I_\chi^{(k+q)^2} \right)\,,
\end{aligned}
\end{equation}
with $\chi\equiv \{S_2,\,{F_{45}}_\alpha, S_3, F_6,\,{S_{17}}_\beta\}$.

\begin{table}[H]
	\setlength{\tabcolsep}{1pt}
	\centering
	\begin{tabular}{|c|c|c|c|c|c|c|c|c|c|c|c|c|c|c|}
		\hline 
		& \multicolumn{4}{c|}{A} & \multicolumn{7}{c|}{B} & \multicolumn{3}{c|}{C}\\
		\hline 
		\hline 
		& \multicolumn{2}{c|}{\textit{i}} & \textit{ii} & \textit{iv} & \multicolumn{2}{c|}{\textit{i}} & \textit{ii} & \multicolumn{3}{c|}{\textit{iii}} & \textit{iv} & \textit{iii} & \multicolumn{2}{c|}{\textit{iv}}\\
		\hline 
		$\alpha$ & -2 & 4 & -2 & -3 & 2 & 4 & 2 & 1 & 3 & 1 & 3 & 1 & 1 & 1\\
		\hline 
		$\beta$ & -2 & 4 & -1 & -3 & 2 & 4 & 3 & 2 & 2 & 4 & 3 & 4 & -3 & -1\\
		\hline 
		DM & $\mathbf{1}_{S}$ & $\mathbf{1}_{S}$ & $\mathbf{2}_{S}$ & $\mathbf{1}_{S}$ & $\boldsymbol \ast$ & $\mathbf{1}_{S}$ & $\mathbf{2}_{S}$ & $\boldsymbol \ast$ & $\mathbf{2}_{S}$ & $\boldsymbol \ast$ & $\mathbf{1}_{S}$ & $\boldsymbol \ast$ & $\boldsymbol \ast$ & $\boldsymbol \ast$\\
		\hline 
		\# & 6 & 7 & 5 & 7 & 5 & 7 & 5 & 4 & 5 & 7 & 6 & 7 & 7 & 5\\
		\hline 
		$\mathbf{2}^1_{S}/\mathbf{2}^3_{S}$ & $\times$/1 & $\times$/1 & 2/1 & $\times$/2 & 1/$\times$ & $\times$/1 & 2/1 & 2/$\times$ & 1/1 & 1/1 & $\times$/2 & 1/1 & 1/1 & 2/$\times$\\
		\hline 
		$F^{++}/S^{++}$ & \checkmark/\checkmark{} & \checkmark/\checkmark{} & $\times$/\checkmark{} & \checkmark / \checkmark{} & $\times /\times$  & \checkmark /\checkmark{} & \checkmark /\checkmark{} & $\times /\times$ & \checkmark/\checkmark{} & \checkmark /\checkmark{} & \checkmark /\checkmark{} & \checkmark/\checkmark{} & \checkmark /\checkmark{} & $\times /\times$\\
		\hline 
	\end{tabular}
	\caption{\label{tab:res1d} The same as in Table~\ref{tab:res1a} for diagram 1.d of Figure~\ref{fig:category1}.}
\end{table}

\paragraph{Class 1.e}

There are three models that can generate neutrino masses via Class 1.e diagrams, solutions in Table~\ref{tab:res1e} below. In all the models, the DM candidate is a fermionic singlet. The Lagrangian is written as follows
\begin{equation}
\begin{aligned}
\mathcal{L}^{}_\text{1.e}&= 
Y^{ia}_1\,(\overline{L^\texttt{C}}_i\,P_L) {F^\texttt{C}_4}_a \,S_2\,+\,
Y^{ab}_2\,\overline{F^\texttt{C}_4}_a\, {F^\texttt{C}_5}_b\, H \, +\, 
Y^{bc}_3\,\overline{F^\texttt{C}_5}_b\,{F_3}_c\, S^\dagger_6\,+\,  
Y^{cd}_4\,\overline{F_3}_c\,{F_1}_d\,S^\dagger_2\\
&\,+\,\,Y^{de}_5\,\overline{F_1}_d\,{F_7}_e\,H\,+\,
Y^{ej}_6\,\overline{F_7}_e\,(P_L\, L_j)\,S_6\,+\,\text{h.c.}\,,
\end{aligned}
\label{eq:L1e}
\end{equation}
and the neutrino mass matrix is given by
\begin{equation}
\label{eq:M1e}
\begin{aligned}
\left(M^\text{1.e}_{\nu}\right)_{ij}&=\frac{(Y^{ia}_1\,Y^{ab}_3\,Y^{bc}_4\,Y^{cj}_6+Y^{ic}_6\,Y^{cb}_4\,Y^{ba}_3\,Y^{aj}_1)}{4(2 \pi)^8}\,\sin (2 \theta_{17})\,\sin (2 \theta_{45}) \\
& \times \sum_{\alpha,\beta\,=\,1}^{2} (-1)^\alpha \,(-1)^\beta\,\left(2\,{m_{45}}^{}_{\alpha}\,m_{3}\,{m_{17}}^{}_{\beta}\,I_\chi^1\,+\,\left({m_{45}}^{}_{\alpha}\,-\,m_{3}\,-\,{m_{17}}^{}_{\beta}\right)\, I_\chi^{k^2} \right. \\
& \left. + \left(-{m_{45}}^{}_{\alpha}\,-\, m_{3}\, +\,{m_{17}}^{}_{\beta}\right)\,I_\chi^{q^2}\,+\,\left({m_{45}}_{\alpha}\,+\, m_{3}\, +\,{m_{17}}_{\beta}\right)\, I_\chi^{(k+q)^2} \right) \,,
\end{aligned}
\end{equation}
where $\chi\equiv \{S_2,\,{F_{45}}_\alpha, F_3, S_6,\,{F_{17}}_\beta\}$.

\begin{table}[H]
	\centering
	\begin{tabular}{|c|c|c|c|}
		\hline 
		& \multicolumn{3}{c|}{A}\\
		\hline 
		\hline 
		& \multicolumn{2}{c|}{\textit{i}} & \textit{iv}\\
		\hline 
		$\alpha$ & 2 & 4 & 3\\
		\hline 
		$\beta$ & 2 & 4 & 3\\
		\hline 
		DM & $\mathbf{1}_{F}$ & $\mathbf{1}_{F}$ & $\mathbf{1}_{F}$\\
		\hline 
		\# & 5 & 7 & 6\\
		\hline 
		$\mathbf{2}^1_{S}/\mathbf{2}^3_{S}$ & $\times$/$\times$ & $\times/\times$ & $\times$/2\\
		\hline 
		$F^{++}/S^{++}$ & $\times/\times$ & \checkmark/$\times$ & \checkmark/\checkmark\\
		\hline 
	\end{tabular}
	\caption{\label{tab:res1e} The same as in Table~\ref{tab:res1a} for Class 1.e of Figure~\ref{fig:category1}.}
\end{table}

\paragraph{Class 1.f} 

In this class one has nine models with both fermionic and scalar DM candidates. In the best case scenario, we need to have five BSM fields to complete the two-loop diagram. The solutions are presented in Table~\ref{tab:res1f}. The generic Lagrangian of this class is
\begin{equation}
\begin{aligned}
\mathcal{L}^{}_\text{1.f}&= 
Y^{ia}_1\,(\overline{L^\texttt{C}}_i\,P_L) {F_2}_a \,S^\dagger_4\,+\,
Y^{ab}_2\,\overline{F_2}_a\, {F^\texttt{C}_3}_b\,S_1 \, +\, 
Y^{bc}_3\,\overline{F^\texttt{C}_3}_b\,{F^\texttt{C}_6}_c\, S_5\,+\,
Y^{cj}_4\,\overline{F^\texttt{C}_6}_c\,(P_L\, L_j)\,S^\dagger_7 \\
&\,+\,\mu_1\,H\,S_4\,S^\dagger_5\,+\,
\mu_2\,H\,S^{\dagger}_1\,S_7\,+\,\text{h.c.}\,.
\end{aligned}
\label{eq:L1f}
\end{equation}

The neutrino mass matrix is given by
\begin{equation}
\label{eq:M1f}
\begin{aligned}
\left(M^\text{1.f}_{\nu}\right)_{ij}&=\frac{ (Y^{ia}_1\,Y^{ab}_2\,Y^{bc}_3\,Y^{cj}_4+Y^{ic}_4\,Y^{cb}_3\,Y^{ba}_2\,Y^{aj}_1)}{4(2 \pi)^8}\sin (2 \theta_{17}) \sin (2 \theta_{45}) \\
& \times \sum_{\alpha,\,\beta\,=\,1}^{2} (-1)^\alpha\, (-1)^\beta \left(2\,m_{2} \, m_{3}\, m_{6} \, I_\chi^1\,+\,(m_{2}\, -\, m_{3}\, -\,m_{6}) \,I_\chi^{k^2}  \right. \\
& \hspace{20mm}\left. + \,(- m_{2} - m_{3} + m_{6})\, I_\chi^{q^2}\,+\,(m_{2}\,+\,m_{3}\,+\,m_{6})\,I_\chi^{(k+q)^2} \right)\,,
\end{aligned}
\end{equation}
with $\chi\equiv \{F_2,\,{S_{45}}_\alpha, F_3, F_6,\,{S_{17}}_\beta\}$.

\begin{table}[H]
	\centering
	\begin{tabular}{|c|c|c|c|c|c|c|c|c|c|}
		\hline 
		& \multicolumn{7}{c|}{A} & \multicolumn{2}{c|}{C}\\
		\hline 
		\hline 
		& \multicolumn{2}{c|}{\textit{i}} & \multicolumn{3}{c|}{\textit{ii}} & \multicolumn{2}{c|}{\textit{iv}} & \textit{ii} & \textit{iv}\\
		\hline 
		$\alpha$ & -2 & 4 & -2 & -2 & 4 & -3 & -3 & 4 & 1\\
		\hline 
		$\beta$ & -2 & 4 & -1 & 1 & 1 & -3 & 1 & 1 & -3\\
		\hline 
		DM & $\mathbf{1}_{F}$ & $\mathbf{1}_{F}$ & $\mathbf{2}_{S}$ & $\boldsymbol \ast$ & $\boldsymbol \ast$ & $\mathbf{1}_{F}$ & $\boldsymbol \ast$ & $\boldsymbol \ast$ & $\boldsymbol \ast$\\
		\hline 
		\# & 6 & 7 & 5 & 6 & 7 & 7 & 7 & 7 & 7\\
		\hline 
		$\mathbf{2}^1_{S}/\mathbf{2}^3_{S}$ & $\times$/2 & $\times$/2 & 1/1 & 1/1 & 1/1 & $\times$/2 & 1/1 & 1/1 & 1/1\\
		\hline 
		$F^{++}/S^{++}$ & $\times$/\checkmark & \checkmark/\checkmark & $\times$/\checkmark & \checkmark/\checkmark & \checkmark/\checkmark & \checkmark/\checkmark & \checkmark/\checkmark & \checkmark/\checkmark & \checkmark/\checkmark\\
		\hline 
	\end{tabular}
	\caption{\label{tab:res1f} The same as in Table~\ref{tab:res1a} for Class 1.f of Figure~\ref{fig:category1}.}
\end{table}

\paragraph{Class 1.g}

Diagram 1.g leads to four models divided by two $\mathsf{Z_2}$ charge assignments (B and C) and the $\mathsf{SU(2)_L}$ assignment \textit{iv}. The DM can be both fermion or scalar singlet. The solutions are shown in Table~\ref{tab:res1g}. The Lagrangian is given by	
\begin{equation}
\begin{aligned}
\mathcal{L}^{}_\text{1.g}&= 
Y^{ia}_1\,(\overline{L^\texttt{C}}_i\,P_L) {F_6}_a \,S^\dagger_7\,+\,
Y^{ab}_2\,\overline{F_6}_a\, {F_5}_b\, S_3 \, +\, 
Y^{bc}_3\,\overline{F_5}_b\,{F_4}_c\, H\,+\,  
Y^{cd}_4\,\overline{F_4}_c\,{F_2}_d\,H\\
&\,+\,Y^{de}_5\,\overline{F_2}_d\,{F_1}_e\,S^\dagger_3\,+\,
Y^{ej}_6\,\overline{F_1}_e\,(P_L\, L_j)\,S_7\,+\,\text{h.c.}\,,
\end{aligned}
\label{eq:L1g}
\end{equation}
while the neutrino mass matrix by
\begin{equation}
\begin{aligned}
(M^\text{1.g}_{\nu})_{ij} &= \frac{(Y^{ia}_1\,Y^{ab}_2\,Y^{bc}_5\,Y^{cj}_6+Y^{ic}_6\,Y^{cb}_5\,Y^{ba}_2\,Y^{aj}_1)}{(2 \pi)^8}  \\
&  \times \sum_{\alpha=1}^3 R_{1 \alpha}R_{2 \alpha} \left(2\,m_{1} \, m_{245 \alpha}\, m_{6} \, I_\gamma^1\,+\,(2m_{245 \alpha}\, -\, m_{1}\, -\,m_{6}) \,I_\gamma^{k^2}  \right. \\
& \hspace{20mm}\left. + \,(- m_{1} - m_{6} )\, I_\gamma^{q^2}\,+\,(m_{1}\,+\,m_{6})\,I_\gamma^{(k+q)^2} \right)\,,
\end{aligned}
\label{eq:M1g}
\end{equation}
defining $\gamma\equiv \{ F_{234 \alpha},\, S_3,\, F_1,\, F_6,\, S_7\, \}$.

\begin{table}[H]
	\centering
	\begin{tabular}{|c|c|c|c|c|}
		\hline 
		& \multicolumn{2}{c|}{B} & \multicolumn{2}{c|}{C}\\
		\hline 
		\hline 
		& \multicolumn{2}{c|}{\textit{iv}} & \multicolumn{2}{c|}{\textit{iv}}\\
		\hline 
		$\alpha$ & -1 & -3 & -3 & -1\\
		\hline 
		$\beta$ & -3 & -3 & -1 & -3\\
		\hline 
		DM & $\mathbf{1}_{S}$ & $\mathbf{1}_{S}$ & $\mathbf{1}_{F}$ & $\mathbf{1}_{S}$\\
		\hline 
		\# & 4 & 5 & 5 & 5\\
		\hline 
		$\mathbf{2}^1_{S}/\mathbf{2}^3_{S}$ & $\times$/$\times$ & $\times$/$\times$ & $\times$/$\times$ & $\times$/$\times$\\
		\hline 
		$F^{++}/S^{++}$ & \checkmark/$\times$ & \checkmark/$\times$ & \checkmark/$\times$ & \checkmark/$\times$\\
		\hline 
	\end{tabular}
	\caption{\label{tab:res1g} The same as in Table~\ref{tab:res1a} for Class 1.g of Figure~\ref{fig:category1}.}
\end{table}

\paragraph{Class 1.h}

This class has no models that allow for a DM messenger inside the two-loop diagram with the discrete symmetry considered. We give the Lagrangian and the mass matrix just for completeness:
\begin{equation}
\begin{aligned}
\mathcal{L}^{}_\text{1.h}&= 
Y^{ia}_1\,(\overline{L^\texttt{C}}_i\,P_L) {F_6}_a \,S^\dagger_7\,+\,
Y^{ab}_2\,\overline{F_6}_a\, {F_3}_b\,S_5 \, +\, 
Y^{bc}_3\,\overline{F_3}_b\,{F_1}_c\, S^\dagger_2\,+\,
Y^{cj}_4\,\overline{F_1}_c\,(P_L\, L_j)\,S_7 \\
&\,+\,\mu_1\,H\,S_4\,S^\dagger_5\,+\,
\mu_2\,H\,S_2\,S^\dagger_4\,+\,\text{h.c.}\,.
\end{aligned}
\label{eq:L1h}
\end{equation}

\begin{equation}
\begin{aligned}
(M^\text{1.h}_{\nu})_{ij} &= \frac{(Y^{ia}_1\,Y^{ab}_2\,Y^{bc}_5\,Y^{cj}_6+Y^{ic}_6\,Y^{cb}_5\,Y^{ba}_2\,Y^{aj}_1)}{(2 \pi)^8}  \\
&  \times \sum_{\alpha=1}^3 R_{1 \alpha}R_{3 \alpha} \left(2\,m_{1} \, m_{3}\, m_{6} \, I_\gamma^1\,+\,(m_{1}\, +2\, m_{3}\, +\,m_{6}) \,I_\gamma^{k^2}  \right. \\
& \hspace{20mm}\left. + \,(- m_{1}\, -m_{6} )\, I_\gamma^{q^2}\,+\,(m_{1}\,+\,m_{3}+\,m_{6})\,I_\gamma^{(k+q)^2} \right)\,,
\end{aligned}
\label{eq:M1h}
\end{equation}
where $\gamma\equiv \{ S_{234 \alpha},\, F_3,\, F_1,\, F_6,\, S_7\, \}$.

\subsubsection{Category 2}
\label{sub:second}

\begin{figure}[H]
	\captionsetup[subfigure]{labelformat=empty}
	\centering
	\begin{subfigure}[b]{0.25\textwidth}
		\includegraphics[scale=0.2]{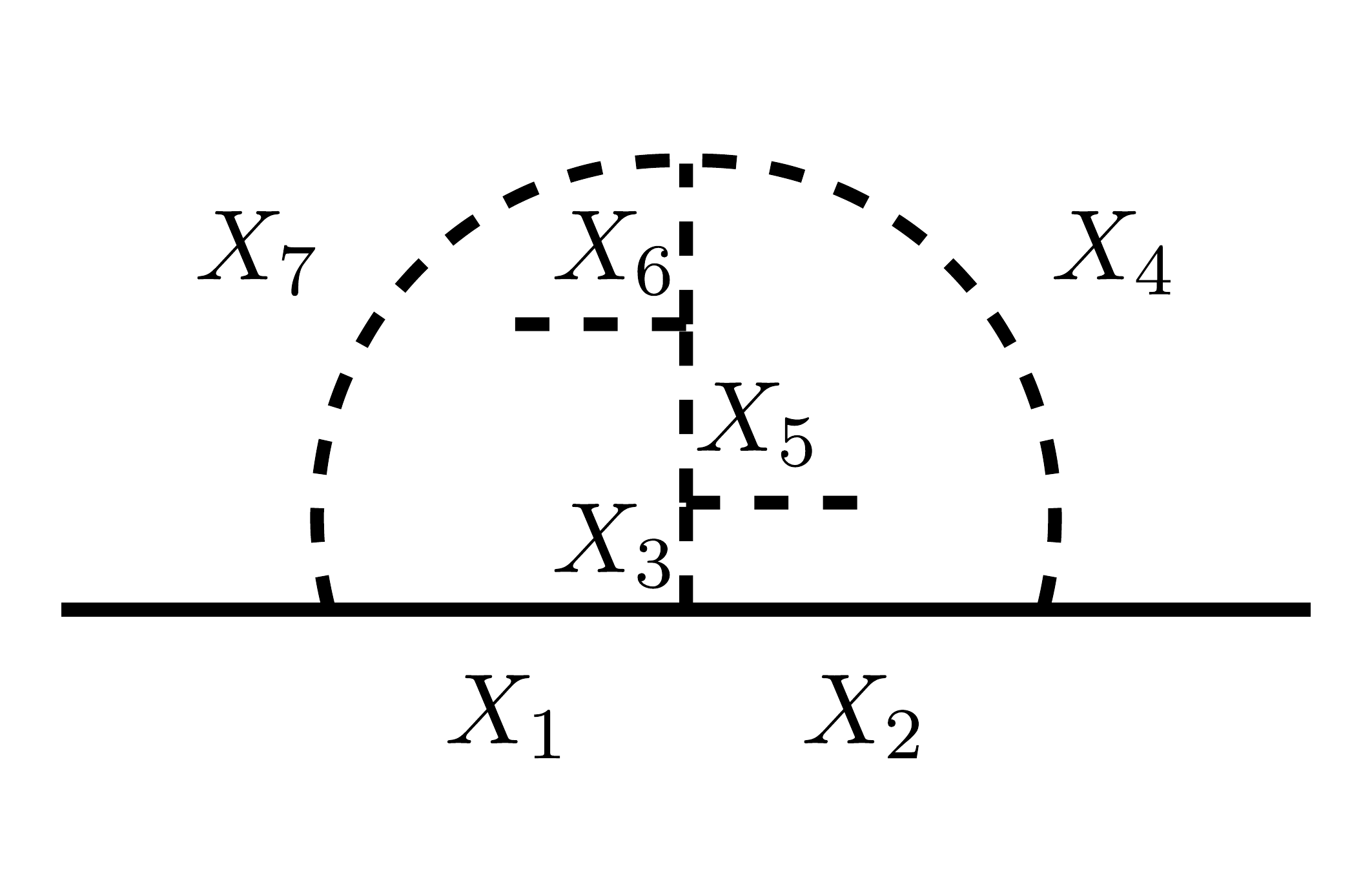}		
		\caption{\hspace{0.5cm} 2.a}
	\end{subfigure} \hspace{0.8cm}
	\begin{subfigure}[b]{0.25\textwidth}
		\includegraphics[scale=0.2]{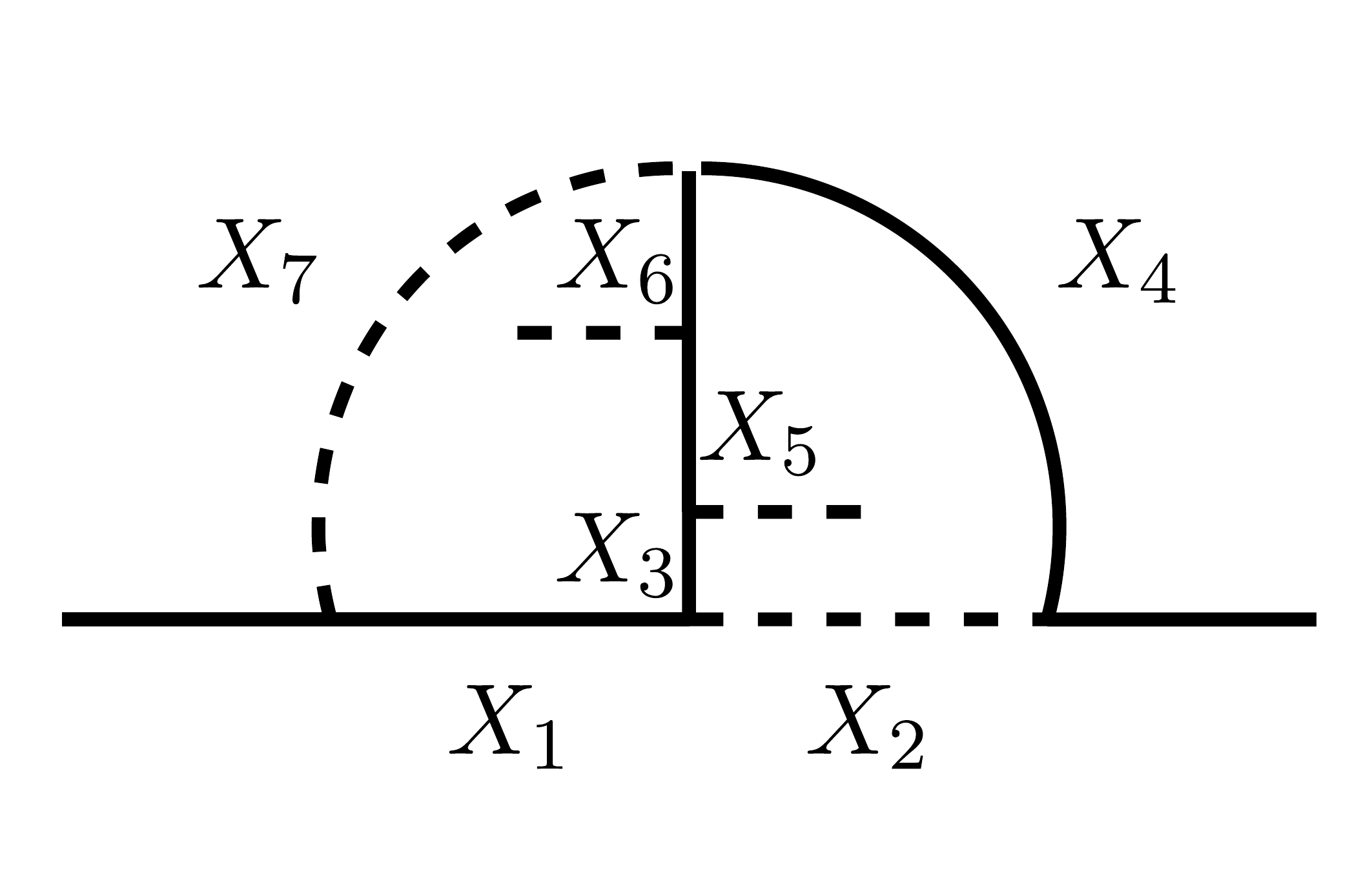}		
		\caption{\hspace{0.5cm}  2.b}
	\end{subfigure} \hspace{0.8cm}
	\begin{subfigure}[b]{0.25\textwidth}
		\includegraphics[scale=0.2]{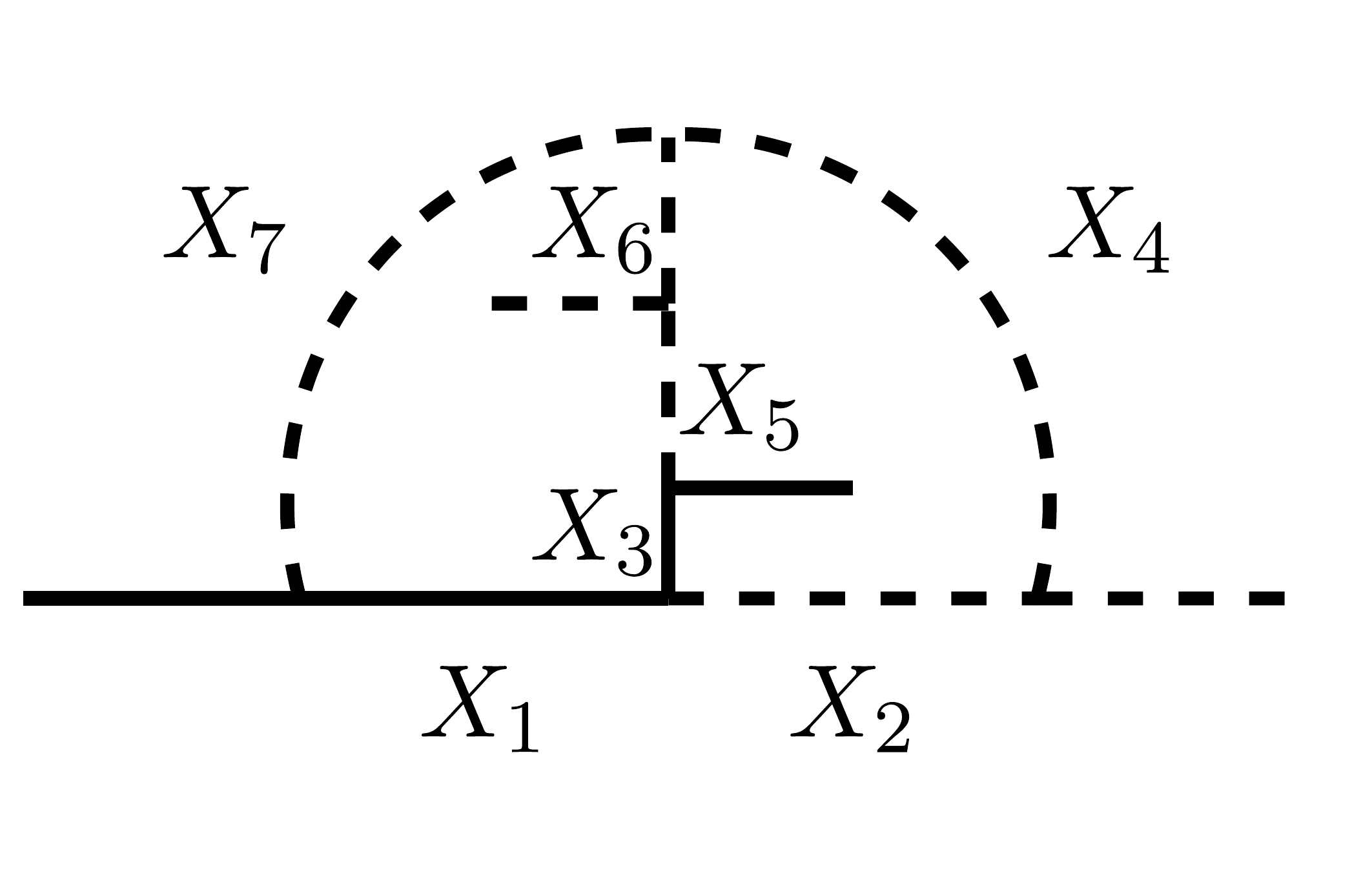}		
		\caption{\hspace{0.5cm}  2.c}
	\end{subfigure}
	\vspace{0.5cm}
	
	\begin{subfigure}[b]{0.25\textwidth}
		\includegraphics[scale=0.2]{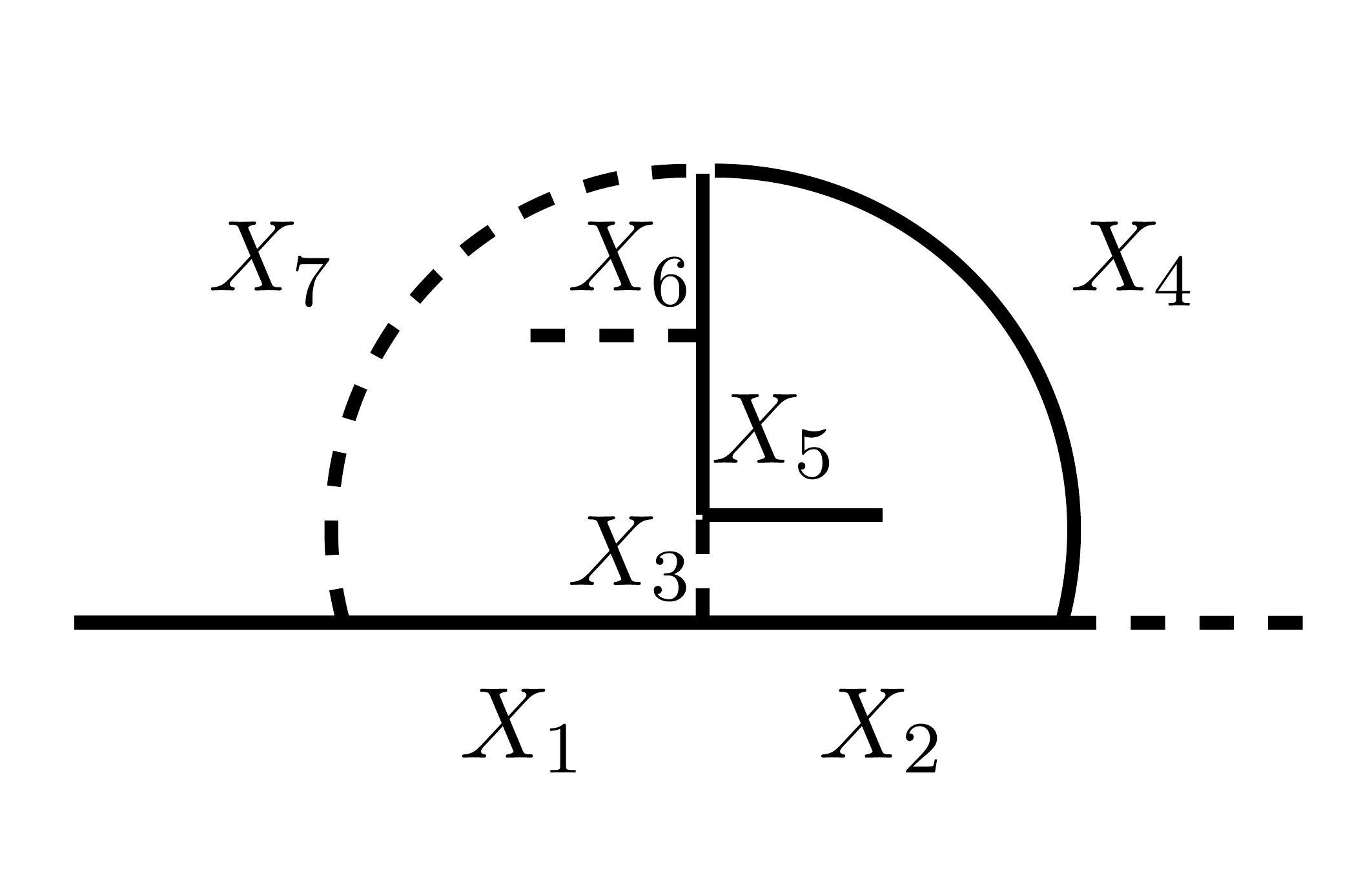}		
		\caption{\hspace{0.5cm} 2.d}
	\end{subfigure} \hspace{0.8cm}
	\begin{subfigure}[b]{0.25\textwidth}
		\includegraphics[scale=0.2]{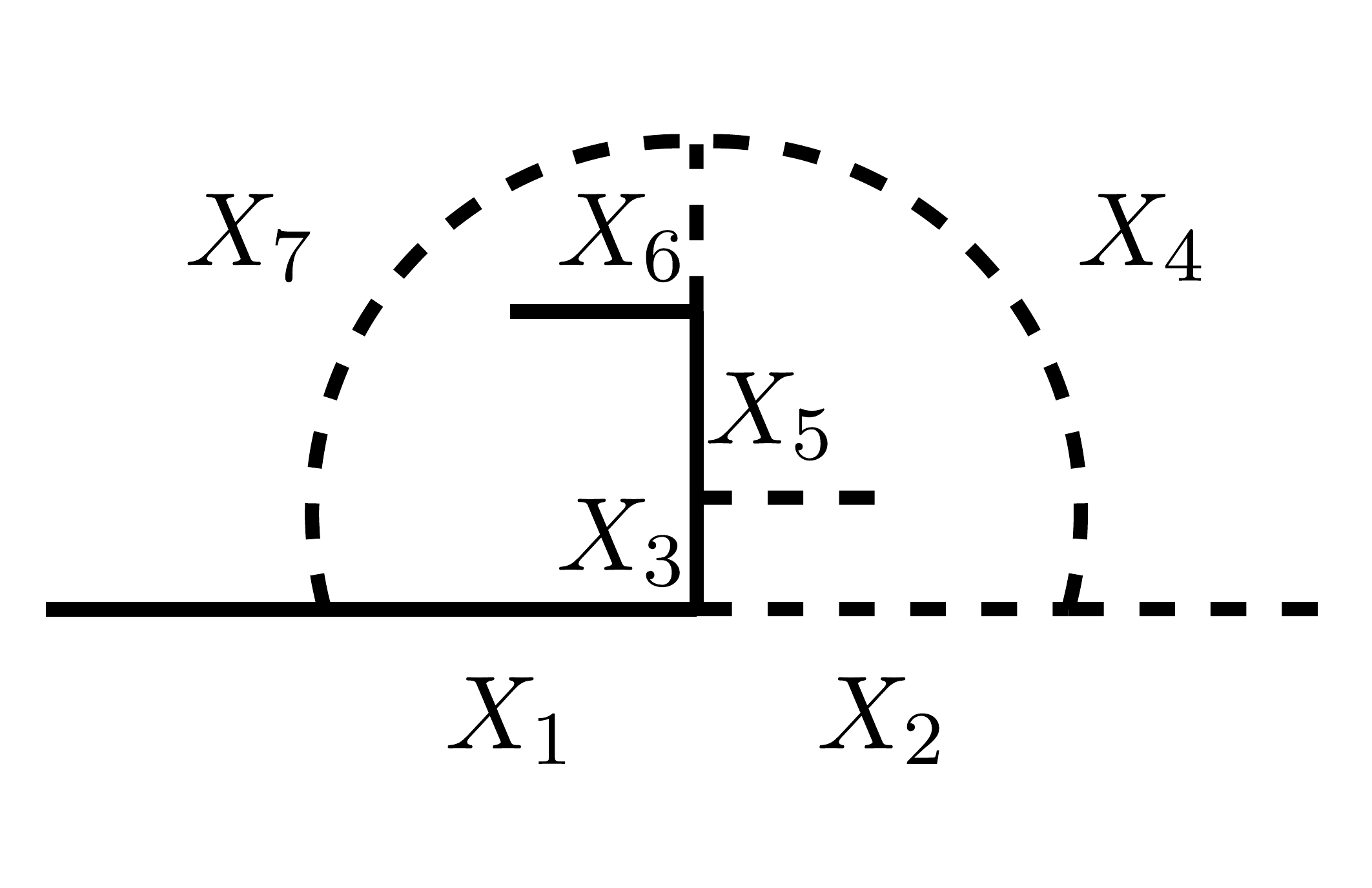}		
		\caption{\hspace{0.5cm} 2.e}
	\end{subfigure}         \hspace{0.8cm}
	\begin{subfigure}[b]{0.25\textwidth}
		\includegraphics[scale=0.2]{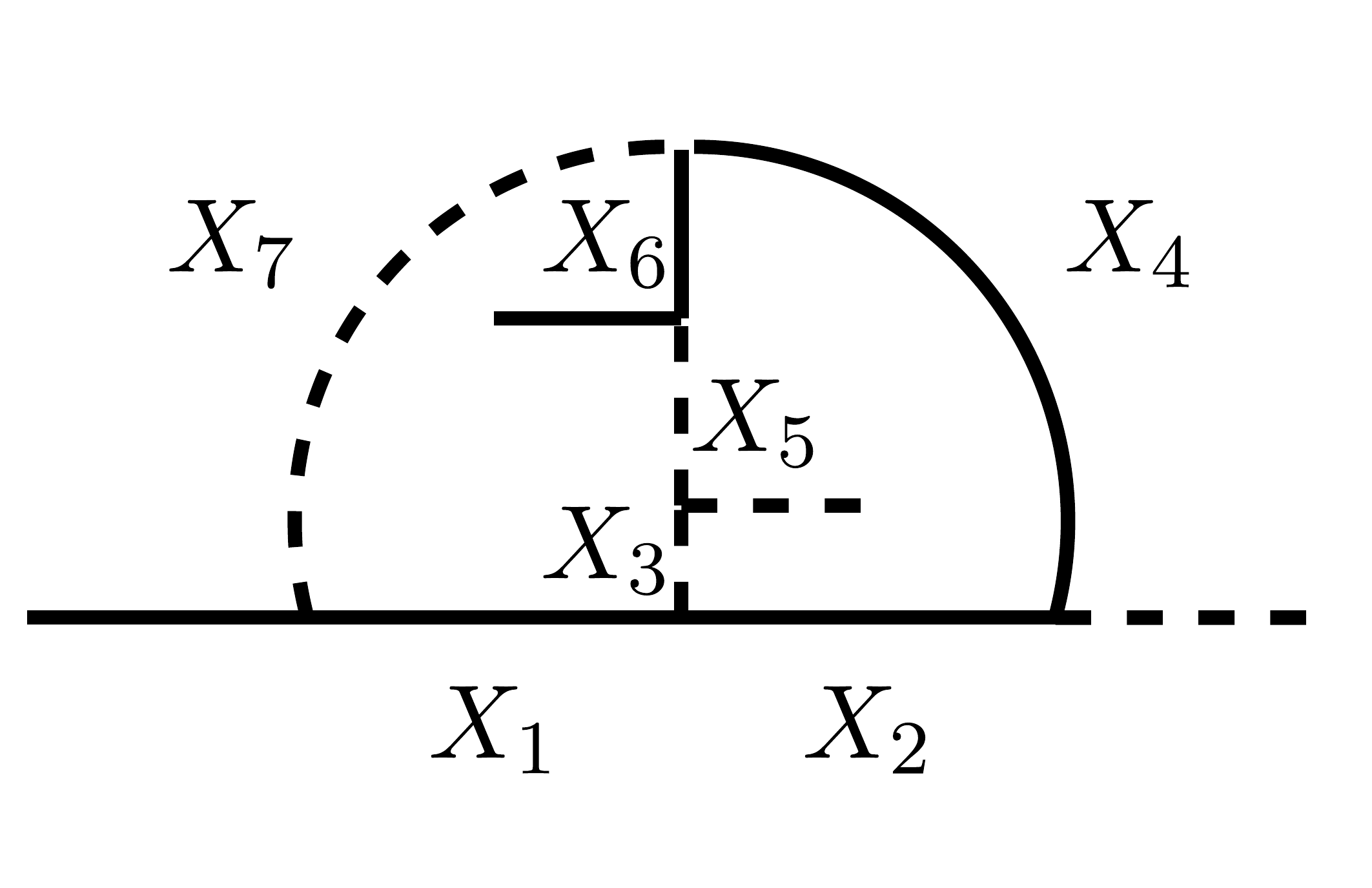}		
		\caption{\hspace{0.5cm}  2.f}
	\end{subfigure} \vspace{0.5cm}
	
	\begin{subfigure}[b]{0.25\textwidth}
		\includegraphics[scale=0.2]{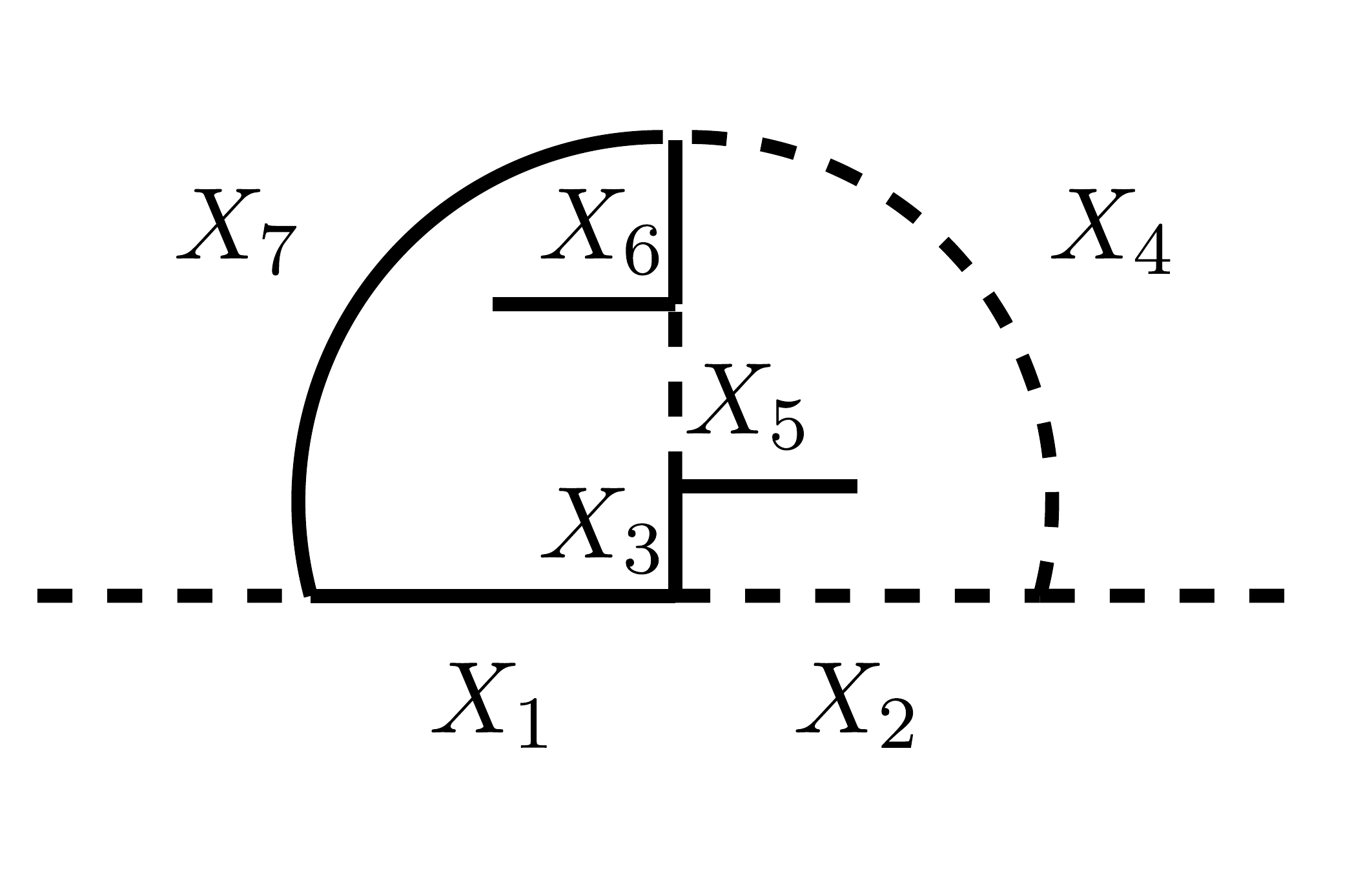}		
		\caption{\hspace{0.5cm} 2.g}
	\end{subfigure}
	\caption{\label{fig:category2} Two-loop diagrams for Class 2.a - 2.g of Category~2.}
\end{figure}

This category is composed of 7 diagrams, shown in Figure~\ref{fig:category2}. As in the previous category, the diagrams define a class and they have the following number of Yukawa/cubic scalar interactions: six/zero~(diagrams 2.b, 2.d and 2.f), five/one~(diagram 2.g), four/two~(diagram 2.e) and three/three~(diagrams 2.a and 2.c). The main difference between Category~1 and Category~2 is the geometry: in Category~2 the two external legs are placed inside the loops making these non-planar diagrams.

The $\mathsf{SU(2)_L}$, $\mathsf{Z_2}$ and $\mathsf{U(1)_Y}$ assignments are given, respectively, in the upper left, upper right and bottom of Table~\ref{tab:assignmentscategory2}. This category has four different hypercharge assignments which are related with the way how the SM fields are organized in the loop.

\begin{table}[H]
	\centering
	\begin{tabular}{|c||c|c|c|c|c|c|c|}
		\hline 
		& $X_{1}$ & $X_{2}$ & $X_{3}$ & $X_{4}$ & $X_{5}$ & $X_{6}$ & $X_{7}$\\
		\hline 
		\hline 
		\textit{i} & \textbf{1} & \textbf{1} & \textbf{1} & \textbf{2} & \textbf{2} & \textbf{1} & \textbf{2}\\
		\hline 
		\textit{ii} & \textbf{1} & \textbf{2} & \textbf{2} & \textbf{1} & \textbf{1} & \textbf{2} & \textbf{2}\\
		\hline 
		\textit{iii} & \textbf{2} & \textbf{1} & \textbf{2} & \textbf{2} & \textbf{1} & \textbf{2} & \textbf{1}\\
		\hline 
		\textit{iv} & \textbf{2} & \textbf{2} & \textbf{1} & \textbf{1} & \textbf{2} & \textbf{1} & \textbf{1}\\
		\hline 
	\end{tabular} \hspace{0.5cm}
	\begin{tabular}{|c||c|c|c|c|c|c|c|}
		\hline 
		& $X_{1}$ & $X_{2}$ & $X_{3}$ & $X_{4}$ & $X_{5}$ & $X_{6}$ & $X_{7}$\\
		\hline 
		\hline 
		A & $\boldsymbol{+}$ & $\boldsymbol{-}$ & $\boldsymbol{-}$ & $\boldsymbol{-}$ &$\boldsymbol{-}$ & $\boldsymbol{-}$ & $\boldsymbol{+}$\\
		\hline 
		B & $\boldsymbol{-}$ & $\boldsymbol{+}$ & $\boldsymbol{-}$ & $\boldsymbol{+}$ & $\boldsymbol{-}$ & $\boldsymbol{-}$ & $\boldsymbol{-}$\\
		\hline 
		C & $\boldsymbol{-}$ & $\boldsymbol{-}$ & $\boldsymbol{+}$ & $\boldsymbol{-}$ &$\boldsymbol{+}$ & $\boldsymbol{+}$ & $\boldsymbol{-}$\\
		\hline 
	\end{tabular}
	\vspace{0.5cm}	
	
	\begin{tabular}{|c||c|c|c|c|c|c|c|}
		\hline 
		& $X_{1}$ & $X_{2}$ & $X_{3}$ & $X_{4}$ & $X_{5}$ & $X_{6}$ & $X_{7}$\\
		\hline 
		\hline 
		a, b & $\alpha$ & $\beta$ & $\alpha-\beta$ & $\beta-1$ & $\alpha-\beta+1$ & $\alpha-\beta+2$ & $\alpha+1$\\
		\hline 
		c, d & $\alpha$ & $\beta$ & $\alpha-\beta$ & $\beta+1$ & $\alpha-\beta-1$ & $\alpha-\beta$ & $\alpha+1$\\
		\hline 
		e, f & $\alpha$ & $\beta$ & $\alpha-\beta$ & $\beta+1$ & $\alpha-\beta+1$ & $\alpha-\beta$ & $\alpha+1$\\
		\hline 
		g & $\alpha$ & $\beta$ & $\alpha-\beta$ & $\beta+1$ & $\alpha-\beta-1$ & $\alpha-\beta-2$ & $\alpha-1$\\
		\hline 
	\end{tabular}
	\caption{\label{tab:assignmentscategory2}  
		$\mathsf{SU(2)_L}$ (upper left), $\mathsf{Z_2}$ (upper right) and  $\mathsf{U(1)_Y}$ (bottom) assignments for particles $X_1$ to $X_7$ of diagrams 2.a - 2.g in Figure~\ref{fig:category2}.}
\end{table}

\paragraph{Class 2.a}

This class presents four solutions and the DM candidate is a mixture of the neutral states in $\mathbf{1}^0_{S}$ and $\mathbf{2}^1_{S}$ in all of them. The most appealing model requires four BSM fields but has no doubly-charged particles in it. The results are summed up in Table~\ref{tab:res2a}. The Lagrangian for this class is given by
\begin{equation}
\begin{aligned}
\mathcal{L}^{}_\text{2.a}&= 
Y^{ia}_1\,(\overline{L^\texttt{C}}_i\,P_L) {F_2}_a \,S^\dagger_4\,+\,
Y^{ab}_2\,\overline{F_2}_a\, {F_1}_b\,S^\dagger_3 \, +\, 
Y^{bj}_3\,\overline{F_1}_b\,(P_L\, L_j)\,S_7 \,+\,
\mu_1\,S_4\,S_6\,S^\dagger_7\\
&\,+\,\mu_2\,H\,S_5\,S^\dagger_6\,+\,
\mu_3\,H\,S_3\,S^\dagger_5\,+\,\text{h.c.}\,,
\end{aligned}
\label{eq:L2a}
\end{equation}
for which the neutrino mass matrix can be written as
\begin{equation}
\left(M^\text{2.a}_{\nu}\right)_{ij}= \frac{\mu_1 (Y^{ia}_1\,Y^{ab}_2\,Y^{bj}_3+Y^{ic}_3\,Y^{ba}_2\,Y^{aj}_1)}{(2\,\pi)^8}\,\sum_{\alpha\,=\,1}^3 R_{1\alpha}R_{3 \alpha} \,  \left(2 m_{1}\,m_{2}\,I_\chi^1\, -\, I_\chi^{k^2}\, -\, I_\chi^{q^2}\, +\, I_\chi^{(k+q)^2} \right)\,
\label{eq:M2a}
\end{equation}
with $\chi\equiv \{F_2,\,S_4,\,S_{356},\,F_1,\,S_7\}$ and $R_{i \alpha}$ are the components of the $3 \times 3$ rotation matrix.




\begin{table}[H]
	\centering
	\begin{tabular}{|c|c|c|c|c|}
		\hline 
		& \multicolumn{4}{c|}{A}\\
		\hline 
		\hline 
		& \multicolumn{2}{c|}{\textit{ii}} & \multicolumn{2}{c|}{\textit{iii}}\\
		\hline 
		$\alpha$ & 2 & -4 & 3 & 1\\
		\hline 
		$\beta$ & 3 & -3 & 4 & 2\\
		\hline 
		DM & $\boldsymbol \ast$ & $\boldsymbol \ast$ & $\boldsymbol \ast$ &  $\boldsymbol \ast$ \\
		\hline 
		\# & 5 & 6 & 6 & 4\\
		\hline 
		$\mathbf{2}^1_{S}/\mathbf{2}^3_{S}$  & 2/1 & 1/1 & 1/1 & 3/$\times$\\
		\hline 
		$F^{++}/S^{++}$ & \checkmark/\checkmark  & \checkmark/\checkmark  & \checkmark/\checkmark & $\times$/$\times$ \\
		\hline 
	\end{tabular}
	\caption{\label{tab:res2a} Realizations of diagram 2.a of Figure~\ref{fig:category2}. For each set of $\mathsf{SU(2)_L}$ quantum numbers (roman numerals) and $\mathsf{Z_2}$ assignments (capital letters A, B, C) we give the hypercharge in terms of $\alpha$ and $\beta$ parameters, the DM candidates, the number of BSM particles (\#), the number of $\mathbf{2}^1_{S}$ and $\mathbf{2}^3_{S}$ present as well as the existence (or not) of doubly-charged fermions ($F^{++}$) and doubly-charged scalars ($S^{++}$) particles. The symbol $\checkmark$($\times$) means present (absent) while the symbol $\boldsymbol{\ast}$, if present in the DM row, means mixing between $\mathbf{1}^0_{S}$ and $\mathbf{2}^1_{S}$. The $\mathsf{SU(2)_L}$, $\mathsf{Z_2}$ and $\mathsf{U(1)_Y}$ assignments are given in Table~\ref{tab:assignmentscategory2}.}
\end{table}

\paragraph{Class 2.b}

For this class, we found seven models three of them with $\mathbf{1}^0_{S}$ as a DM candidate. There are two similar sets of solutions for $\mathsf{SU(2)_L}$ assignment \textit{iii} but with different $\mathsf{Z_2}$ charges. See the results on Table~\ref{tab:res2b}.
The Lagrangian of this class is given by
\begin{equation}
\begin{aligned}
\mathcal{L}^{}_\text{2.b}&= 
Y^{ia}_1\,(\overline{L^\texttt{C}}_i\,P_L) {F^\texttt{C}_4}_a \,S_2\,+\,
Y^{ab}_2\,\overline{F^\texttt{C}_4}_a\, {F_6}_b\,S^\dagger_7 \, +\, 
Y^{bc}_3\,\overline{F_6}_b\, {F_5}_c\,H \, +\, 
Y^{cd}_4\,\overline{F_5}_c\, {F_3}_d\,H \\
&\,+\,Y^{de}_5\,\overline{F_3}_d\, {F_1}_e\,S^\dagger_2 \, +\, 
Y^{ej}_6\,\overline{F_1}_b\,(P_L\, L_j)\,S_7\,+\,\text{h.c.}\,,
\end{aligned}
\label{eq:L2b}
\end{equation}
the neutrino mass matrix is then
\begin{equation}
\begin{aligned}
(M^\text{2.b}_{\nu})_{ij} &= \frac{(Y^{ia}_1\,Y^{ab}_2\,Y^{bc}_5\,Y^{cj}_6+Y^{ic}_6\,Y^{cb}_5\,Y^{ba}_2\,Y^{aj}_1)}{(2 \pi)^8}  \\
&  \times \sum_{\alpha=1}^3 R_{1 \alpha}R_{2 \alpha} \left(2\,m_{1} \, m_{356 \alpha}\, m_{4} \, I_\chi^1\,+\,(-\, m_{1} -\, m_{356 \alpha}\, +\, m_{4}\,) \,I_\chi^{k^2}  \right. \\
& \hspace{20mm}\left. + \,(+\, m_{1} -\, m_{356 \alpha}\, -\, m_{4}\,)\, I_\chi^{q^2}\,+\,(\, m_{1} +\, m_{356 \alpha}\, +\, m_{4}\,)\,I_\chi^{(k+q)^2} \right)\,,
\end{aligned}
\label{eq:M2b}
\end{equation}
with $\chi\equiv \{ S_2,\, F_4,\, F_{345\alpha},\,F_1,S_7\}$.

\begin{table}[H]
	\centering
	\begin{tabular}{|c|c|c|c|c|c|c|c|}
		\hline 
		& \multicolumn{5}{c|}{A} & \multicolumn{2}{c|}{C}\\
		\hline 
		\hline 
		& \multicolumn{2}{c|}{\textit{ii}} & \multicolumn{3}{c|}{\textit{iii}} & \multicolumn{2}{c|}{\textit{iii}}\\
		\hline 
		$\alpha$ & -4 & 2 & 1 & 1 & 3 & -1 & 1\\
		\hline 
		$\beta$ & -3 & 3 & 0 & 2 & 4 & -2 & 0\\
		\hline 
		DM & $\mathbf{1}_{F}$ & $\mathbf{1}_{F}$ & $\mathbf{1}_{S}$ & $\mathbf{1}_{F}$ & $\mathbf{1}_{F}$ & $\mathbf{1}_{S}$ & $\mathbf{1}_{S}$\\
		\hline 
		\# & 6 & 5 & 5 & 4 & 6 & 5 & 4\\
		\hline 
		$\mathbf{2}^1_{S}/\mathbf{2}^3_{S}$  & $\times$/2 & $\times$/2 & $\times$/$\times$ & $\times$/$\times$ & $\times$/$\times$ & $\times$/$\times$ & $\times$/$\times$\\
		\hline 
		$F^{++}/S^{++}$ & \checkmark/\checkmark & $\times$/\checkmark & \checkmark/$\times$ & $\times$/$\times$ & \checkmark/\checkmark & \checkmark/$\times$ & \checkmark/$\times$\\
		\hline 
	\end{tabular}
	\caption{\label{tab:res2b}	The same as in Table~\ref{tab:res2a} for Class 2.b of Figure~\ref{fig:category2}.}
\end{table}

\paragraph{Class 2.c}

Diagram 2.c is possible doing with ten different sets among which three require only the existence of five BSM fields. The DM candidate in this class is the mixing of the neutral component of $\mathbf{1}^0_{S}$ and $\mathbf{2}^1_{S}$.

The interactions from diagram 2.c are given by
\begin{equation}
\begin{aligned}
\mathcal{L}^{}_\text{2.c}&= 
Y^{ia}_1\,(\overline{L^\texttt{C}}_i\,P_L) {F_3}_a \,S^\dagger_5\,+\,
Y^{ab}_2\,\overline{F_3}_a\, {F_1}_b\,S^\dagger_2 \, +\, 
Y^{bj}_3\,\overline{F_1}_b\,(P_L\, L_j)\,S_7 \,+\,
\mu_1\,S_4\,S_6\,S^\dagger_7\\
&\,+\,\mu_2\,H\,S_5\,S^\dagger_6\,+\,
\mu_3\,H\,S_2\,S^\dagger_4\,+\,\text{h.c.}\,,
\end{aligned}
\label{eq:L2c}
\end{equation}
for which one extracts
\begin{equation}
\label{eq:M2c}
\begin{aligned}
\left(M^\text{2.c}_{\nu}\right)_{ij}&=\frac{\mu_1\,(Y^{ia}_1\,Y^{ab}_2\,Y^{bj}_3+Y^{ib}_3\,Y^{ba}_2\,Y^{aj}_1)}{4(2 \pi)^8}\,\sin(2\theta_{24})\,\sin(2\theta_{56})\\
& \times \sum_{\alpha,\,\beta\,=\,1}^{2} (-1)^\alpha\,(-1)^\beta \,\left(2\,m_{1}\,m_{3}\,I_\rho^1\,+\,I_\rho^{k^2}\,-\,I_\rho^{q^2} \,+\,I_\rho^{(k+q)^2} \right)\,,
\end{aligned}
\end{equation}
as the neutrino mass matrix, $\rho \equiv \{{S_{24}}^{}_\alpha,\,F_3,\,{S_{56}}^{}_\beta,\,S_7,\,F_1\}$.

\begin{table}[H]
	\centering
	\begin{tabular}{|c|c|c|c|c|c|c|c|c|c|c|}
		\hline 
		& \multicolumn{5}{c|}{A} & B & \multicolumn{4}{c|}{C}\\
		\hline 
		\hline 
		& \textit{i} & \multicolumn{2}{c|}{\textit{ii}} & \multicolumn{2}{c|}{\textit{iv}} & \textit{ii} & \textit{i} & \textit{ii} & \multicolumn{2}{c|}{\textit{iv}}\\
		\hline 
		$\alpha$ & 2 & -4 & 2 & 1 & 3 & -2 & -2 & -4 & -3 & 3\\
		\hline 
		$\beta$ & 0 & -1 & -1 & -1 & -1 & -3 & 0 & -1 & -1 & -1\\
		\hline 
		DM & $\boldsymbol \ast$ & $\boldsymbol \ast$ & $\boldsymbol \ast$ & $\boldsymbol \ast$ & $\boldsymbol \ast$ & $\boldsymbol \ast$ & $\boldsymbol \ast$ & $\boldsymbol \ast$ & $\boldsymbol \ast$ & $\boldsymbol \ast$\\
		\hline 
		\# & 5 & 7 & 6 & 5 & 7 & 6 & 5 & 7 & 6 & 7\\
		\hline 
		$\mathbf{2}^1_{S}/\mathbf{2}^3_{S}$ & 2/1 & 1/2 & 1/2 & 2/$\times$ & 1/1 & 2/1 & 2/1 & 1/2 & 1/1 & 1/1\\
		\hline 
		$F^{++}$/$S^{++}$ & $\times$/\checkmark{} & \checkmark/\checkmark{} & \checkmark/\checkmark{} & $\times$/$\times$ & \checkmark/\checkmark{} & $\times$/\checkmark{} & $\times$/\checkmark{} & \checkmark/\checkmark & \checkmark/\checkmark{} & \checkmark/\checkmark{}\\
		\hline 
	\end{tabular}
	\caption{\label{tab:res2c}	The same as in Table~\ref{tab:res2a} for diagram 2.c of Figure~\ref{fig:category2}.}
\end{table}

\paragraph{Class 2.d}

In this class we have eleven models with both fermion and scalar singlets as DM candidates. The minimal scenario requires four BSM fields and has neither doubly-charged particles nor 
scalar doublets. The solutions are presented in Table~\ref{tab:res2d}. The Lagrangian for this diagram is 
\begin{equation}
\begin{aligned}
\mathcal{L}^{}_\text{2.d}&= 
Y^{ia}_1\,(\overline{L^\texttt{C}}_i\,P_L) {F^\texttt{C}_5}_a \,S_3\,+\,
Y^{ab}_2\,\overline{F^\texttt{C}_5}_a\, {F^\texttt{C}_6}_b\,H\, +\, 
Y^{bc}_3\,\overline{F^\texttt{C}_6}_b\, {F_4}_c\,S^\dagger_7 \, +\, 
Y^{cd}_4\,\overline{F_4}_c\, {F_2}_d\,H \\
&\,+\,Y^{de}_5\,\overline{F_2}_d\, {F_1}_e\,S^\dagger_3 \, +\, 
Y^{ej}_6\,\overline{F_1}_e\,(P_L\, L_j)\,S_7\,+\,\text{h.c.}\,,
\end{aligned}
\label{eq:L2d}
\end{equation}
that leads to the neutrino mass matrix given by
\begin{equation}
\label{eq:M2d}
\begin{aligned}
\left(M^\text{2.d}_{\nu}\right)_{ij}&=\frac{(Y^{ia}_1\,Y^{ab}_3\,Y^{bc}_5\,Y^{cj}_6+Y^{ic}_6\,Y^{cb}_5\,Y^{ba}_3\,Y^{aj}_1)}{4(2\pi)^8}\sin(2\theta_{24})\sin(2\theta_{56}) \sum_{\alpha,\,\beta\,=\,1}^{2} (-1)^\alpha\,(-1)^\beta\left(2\,m_{1}\,m_{56 \beta}\,m_{24\alpha}\,I_\rho^1 \right. \\
&\left. \,+\,(m_{24\alpha}\,-\, m_{56 \beta}\,-\,m_{1})\,I_\rho^{k^2}\, + \, (m_{1}\,-\,m_{56\beta}\,-\,m_{24\alpha})\,I_\rho^{q^2}\right. \\
&\left. \,+\,(m_{1}\,+\,m_{56\beta}\,+\,m_{24\alpha}) \,I_\rho^{(k+q)^2} \right)\,,
\end{aligned}
\end{equation}
with $\rho \equiv \{{F_{24}}^{}_\alpha,\,S_3,\,{S_{56}}^{}_\beta,\,S_7,\,F_1\}$.

\begin{table}[H]
	\begin{tabular}{|c|c|c|c|c|c|c|c|c|c|c|c|}
		\hline 
		& \multicolumn{5}{c|}{A} & B & \multicolumn{5}{c|}{C}\\
		\hline 
		\hline 
		& \textbf{\textit{i}} & \multicolumn{2}{c|}{\textit{ii}} & \multicolumn{2}{c|}{\textit{iv}} & \textit{iv} & \textit{i} & \textit{ii} & \multicolumn{3}{c|}{\textit{iv}}\\
		\hline 
		$\alpha$ & 2 & -4 & 2 & 1 & 3 & -1 & 2 & 2 & -1 & 1 & 3\\
		\hline 
		$\beta$ & 0 & -1 & -1 & -1 & -1 & -3 & 0 & -1 & -3 & -1 & -1\\
		\hline 
		DM & $\mathbf{1}_{F}$ & $\mathbf{1}_{F}$ & $\mathbf{1}_{F}$ & $\mathbf{1}_{F}$ & $\mathbf{1}_{F}$ & $\mathbf{1}_{S}$ & $\mathbf{1}_{F}$ & $\mathbf{1}_{F}$ & $\mathbf{1}_{S}$ & $\mathbf{1}_{F}$ &  $\mathbf{1}_{F}$\\
		\hline 
		\# & 5 & 7 & 6 & 5 & 7 & 5 & 5 & 6 & 5 & 4 & 7\\
		\hline 
		$\mathbf{2}^1_{S}/\mathbf{2}^3_{S}$ & $\times$/1 & $\times$/2 & $\times$/2 & $\times$/$\times$ & $\times$/$\times$ & $\times$/$\times$ & $\times$/1 & $\times$/2 & $\times$/$\times$ & $\times$/$\times$ & $\times$/$\times$\\
		\hline 
		$F^{++}$/$S^{++}$ & $\times$/\checkmark & \checkmark/\checkmark & \checkmark/\checkmark & $\times$/$\times$ & \checkmark/\checkmark & \checkmark/$\times$ & $\times$/\checkmark & \checkmark/\checkmark & \checkmark/$\times$ & $\times$/$\times$ & \checkmark/\checkmark\\
		\hline 
	\end{tabular}
	\caption{\label{tab:res2d} The same as in Table~\ref{tab:res2a} for diagram 2.d of Figure~\ref{fig:category2}.}
\end{table}

\paragraph{Class 2.e}

There are fourteen models leading to neutrino mass via diagram 2.e whose the Lagrangian is
\begin{equation}
\begin{aligned}
\mathcal{L}^{}_\text{2.e}&= 
Y^{ia}_1\,(\overline{L^\texttt{C}}_i\,P_L) {F_5}_a \,S^\dagger_6\,+\,
Y^{ab}_2\,\overline{F_5}_a\, {F_3}_b\,H \, +\, 
Y^{bc}_3\,\overline{F_3}_b\, {F_1}_c\,S^\dagger_2\,+\, 
Y^{cj}_4\,\overline{F_1}_c\,(P_L\,L_j)\,S_7 \\
&\,+\,\mu_1\,S_4\,S_6\,S^\dagger_7\,+\,
\mu_2\,H\,S_2\,S^\dagger_4\,+\,\text{h.c.}\,.
\end{aligned}
\label{eq:L2e}
\end{equation}

The matrix $M_{\nu}$ is given by
\begin{equation}
\label{eq:M2e}
\begin{aligned}
\left(M^\text{2.e}_{\nu}\right)_{ij}&=\frac{\mu_1\,(Y^{ia}_1\,Y^{ab}_3\,Y^{bj}_4+Y^{ib}_4\,Y^{ba}_3\,Y^{aj}_1)}{4(2\pi)^8}\sin(2\theta_{24})\,\sin(2\theta_{35}) \\
& \times \sum_{\alpha,\,\beta\,=\,1}^{2} (-1)^\alpha\,(-1)^\beta \left(2 m_{35\beta}\,m_{1}\,I_\rho^1\, +\, I_\rho^{k^2}\, -\, I_\rho^{q^2}\, +\, I_\rho^{(k+q)^2} \right)\,,
\end{aligned}
\end{equation}
for $\rho \equiv \{{S_{24}}^{}_\alpha,\,S_6,\,{F_{35}}^{}_\beta,\,S_7,\,F_1\}$,

Among the solutions, presented in Table~\ref{tab:res2e}, there are models with a scalar singlet as the DM candidate and solutions where the DM is the mixture between $\mathbf{1}^0_{S}$ and $\mathbf{2}^0_{S}$. All the models apart from two have at least one doubly-charged particle.
\begin{table}[H]
	\setlength{\tabcolsep}{2pt}
	\begin{tabular}{|c|c|c|c|c|c|c|c|c|c|c|c|c|c|c|}
		\hline 
		& \multicolumn{5}{c|}{A} & \multicolumn{3}{c|}{B} & \multicolumn{6}{c|}{C}\\
		\hline 
		\hline 
		& \multicolumn{2}{c|}{\textit{i}} & \multicolumn{2}{c|}{\textit{iii}} & \textit{iv} & \multicolumn{2}{c|}{\textit{iii}} & \textit{iv} & \multicolumn{2}{c|}{\textit{i}} & \multicolumn{2}{c|}{\textit{iii}} & \multicolumn{2}{c|}{\textit{iv}}\\
		\hline 
		$\alpha$ & -4 & 2 & 1 & 3 & 1 & -1 & -1 & -1 & -4 & -2 & -3 & -1 & -3 & -1\\
		\hline 
		$\beta$ & 0 & 0 & 0 & 0 & -1 & -4 & 2 & 3 & 0 & 0 & 0 & 2 & -1 & 3\\
		\hline 
		DM & $\boldsymbol \ast$ & $\boldsymbol \ast$ & $\boldsymbol \ast$ & $\boldsymbol \ast$ & $\boldsymbol \ast$ & $\mathbf{1}_{S}$ & $\mathbf{1}_{S}$ & $\mathbf{1}_{S}$ & $\boldsymbol \ast$ & $\boldsymbol \ast$ & $\boldsymbol \ast$ & $\mathbf{1}_{S}$ & $\boldsymbol \ast$ & $\mathbf{1}_{S}$\\
		\hline 
		\# & 7 & 6 & 5 & 7 & 6 & 7 & 7 & 7 & 7 & 4 & 6 & 6 & 5 & 7\\
		\hline 
		$\mathbf{2}^1_{S}/\mathbf{2}^3_{S}$ & 1/1 & 1/1 & 2/$\times$ & 1/1 & 1/$\times$ & $\times$/2 & $\times$/2 & $\times$/1 & 1/1 & 2/$\times$ & 1/1 & $\times$/2 & 1/$\times$ & $\times$/1\\
		\hline 
		$F^{++}$/$S^{++}$ & \checkmark/\checkmark & \checkmark/\checkmark & $\times$/$\times$ & \checkmark/\checkmark & \checkmark/$\times$ & \checkmark /\checkmark & \checkmark/\checkmark & \checkmark /\checkmark{} & \checkmark/\checkmark & $\times$/$\times$ & \checkmark/\checkmark & \checkmark/\checkmark & \checkmark/$\times$ & \checkmark/\checkmark\\
		\hline 
	\end{tabular}
	\caption{\label{tab:res2e} The same as in Table~\ref{tab:res2a} for diagram 2.e of Figure~\ref{fig:category2}.}
\end{table}

\paragraph{Class 2.f}
This class has sixteen models, summarized in Table~\ref{tab:res2f}. The interactions present in this diagram are given by
\begin{equation}
\begin{aligned}
\mathcal{L}^{}_\text{2.f}&= 
Y^{ia}_1\,(\overline{L^\texttt{C}}_i\,P_L) {F^\texttt{C}_6}_a \,S_5\,+\,
Y^{ab}_2\,\overline{F^\texttt{C}_6}_a\, {F_4}_b\,S^\dagger_7\,+\, 
Y^{bc}_3\,\overline{F_4}_b\, {F_2}_c\,H\,+\, 
Y^{cd}_4\,\overline{F_2}_c\, {F_1}_d\,S^\dagger_3\\
&\,+\,Y^{dj}_5\,\overline{F_1}_d\,(P_L\,L_j)\,S_7 \,+\,
\mu\,H\,S_3\,S^\dagger_5\,+\,\text{h.c.}\,,
\end{aligned}
\label{eq:L2f}
\end{equation}
where the neutrino mass $M_{\nu}$ is given by
\begin{equation}
\label{eq:M2f}
\begin{aligned}
\left(M^\text{2.f}_{\nu}\right)_{ij}&=\frac{(Y^{ia}_1\,Y^{ab}_2\,Y^{bc}_4\,Y^{cj}_5+Y^{ic}_5\,Y^{cb}_4\,Y^{ba}_2\,Y^{aj}_1)}{4(2\pi)^8}\sin(2\theta_{24})\,\sin(2\theta_{35})\, \\
& \times  \sum_{\alpha,\,\beta\,=\,1}^{2} (-1)^\alpha\,(-1)^\beta \left(2\,m_{1}\,m_{24\alpha}\,m_{6}\, I_\rho^1\, +\,(m_{24\alpha}\,-\, m_{1}\, -\,m_{6})\,I_\rho^{k^2} \right. \\
&\left. + \,(m_{1}\, -\, m_{24\alpha}\,-\,m_{6})\,I_\rho^{q^2}\, +\, (m_{1}\, +\, m_{24\alpha}\, +\,m_{6})\, I_\rho^{(k+q)^2}\right)
\end{aligned}
\end{equation}
with $\rho \equiv \{{F_{24}}^{}_\alpha,\,F_6,\,{S_{35}}^{}_\beta,\,S_7,\,F_1\}$.

\begin{table}[H]
	\setlength{\tabcolsep}{1pt}
	\begin{tabular}{|c|c|c|c|c|c|c|c|c|c|c|c|c|c|c|c|c|}
		\hline 
		& \multicolumn{5}{c|}{A} & \multicolumn{4}{c|}{B} & \multicolumn{7}{c|}{C}\\
		\hline 
		\hline 
		& \multicolumn{2}{c|}{\textit{i}} & \multicolumn{2}{c|}{\textit{iii}} & \textit{iv} & \multicolumn{2}{c|}{\textit{iii}} & \multicolumn{2}{c|}{\textit{iv}} & \multicolumn{2}{c|}{\textit{i}} & \multicolumn{2}{c|}{\textit{iii}} & \multicolumn{3}{c|}{\textit{iv}}\\
		\hline 
		$\alpha$ & -4 & 2 & 3 & 1 & 1 & -1 & -1 & -1 & -1 & -4 & 2 & -1 & 3 & -1 & 1 & -1\\
		\hline 
		$\beta$ & 0 & 0 & 0 & 2 & -1 & -4 & 2 & -3 & 3 & 0 & 0 & -4 & 0 & -3 & -1 & 3\\
		\hline 
		DM & $\mathbf{1}_{F}$ & $\mathbf{1}_{F}$ & $\mathbf{1}_{F}$ & $\boldsymbol \ast$ & $\mathbf{1}_{F}$ & $\mathbf{1}_{S}$ & $\mathbf{1}_{S}$ & $\mathbf{1}_{S}$ & $\mathbf{1}_{S}$ & $\mathbf{1}_{F}$ & $\mathbf{1}_{F}$ & $\mathbf{1}_{S}$ & $\mathbf{1}_{F}$ & $\mathbf{1}_{S}$ & $\mathbf{1}_{F}$ & $\mathbf{1}_{S}$\\
		\hline 
		\# & 7 & 6 & 7 & 6 & 6 & 7 & 6 & 6 & 7 & 7 & 6 & 7 & 7 & 6 & 5 & 7\\
		\hline 
		$\mathbf{2}^1_{S}/\mathbf{2}^3_{S}$ & $\times$/2 & $\times$/2 & $\times$/1 & 1/$\times$ & $\times$/1 & $\times$/1 & $\times$/1 & $\times$/1 & $\times$/1 & $\times$/2 & $\times$/2 & $\times$/1 & $\times$/1 & $\times$/1 & $\times$/1 & $\times$/1\\
		\hline 
		$F^{++}$/$S^{++}$ & \checkmark/\checkmark & $\times$/\checkmark & \checkmark/\checkmark & $\times$/\checkmark & $\times$/\checkmark & \checkmark/\checkmark & \checkmark/\checkmark & \checkmark /\checkmark & \checkmark/\checkmark & \checkmark/\checkmark & $\times$/\checkmark & \checkmark/\checkmark & \checkmark/\checkmark & \checkmark/\checkmark & $\times$/\checkmark & \checkmark /\checkmark\\
		\hline 
	\end{tabular}
	\caption{\label{tab:res2f} The same as in Table~\ref{tab:res2a} for diagram 2.f of Figure~\ref{fig:category2}.}
\end{table}

\paragraph{Class 2.g}
The last class of this section contains five models with either scalar singlets or fermion singlets as DM candidates. All the models have doubly-charged fermions and scalars. The solutions given in Table~\ref{tab:res2g} have the following Lagrangian,
\begin{equation}
\begin{aligned}
\mathcal{L}^{}_\text{2.g}&= 
Y^{ia}_1\,(\overline{L^\texttt{C}}_i\,P_L) {F_3}_a \,S^\dagger_5\,+\,
Y^{ab}_2\,\overline{F_3}_a\, {F_1}_b\,S^\dagger_2\,+\, 
Y^{bc}_3\,\overline{F_1}_b\, {F_7}_c\,H\,+\, 
Y^{cd}_4\,\overline{F_7}_c\, {F_6}_d\,S_4\\
&\,+\,Y^{dj}_5\,\overline{F_6}_d\,(P_L\,L_j)\,S_5 \,+\,
\mu\,H\,S_2\,S^\dagger_4\,+\,\text{h.c.}\,,
\end{aligned}
\label{eq:L2g}
\end{equation}
and the neutrino mass matrix is given by 
\begin{equation}
\label{eq:M2g}
\begin{aligned}
\left(M^\text{2.g}_{\nu}\right)_{ij}&=\frac{ (Y^{ia}_1\,Y^{ab}_2\,Y^{bc}_4\,Y^{cj}_5+Y^{ic}_5\,Y^{cb}_4\,Y^{ba}_2\,Y^{aj}_1)}{4(2 \pi)^8}\sin (2 \theta_{17})\,\sin (2 \theta_{24}) \\
& \times \sum_{\alpha,\,\beta\,=\,1}^{2} (-1)^\alpha \,(-1)^\beta \left(2 m_{3}\, m_{17\alpha}\, m_{6}\, I_\eta^1 + (m_{3} +\, m_{6} ) I_\eta^{k^2} \right. \\
&\left. + \,(-m_{3}\, -\,m_{6})\, I_\eta^{q^2}\, +\, (m_{3}\, +\, 2\,{m_{17}}_{\alpha}\,+\,m_{6})\,I_\eta^{(k+q)^2} \right)\,,
\end{aligned}
\end{equation}
for $\eta\equiv \{{S_{24}}^{}_\alpha,\,F_3,\,S_5,\,F_6,\,{F_{17}}^{}_\beta\}$.

\begin{table}[H]
	\centering
	\begin{tabular}{|c|c|c|c|}
		\hline 
		& \multicolumn{3}{c|}{B}\\
		\hline 
		\hline 
		&  \multicolumn{3}{c|}{\textit{ii}}\\
		\hline 
		$\alpha$ & 4 & 0 & -2\\
		\hline 
		$\beta$  & 3 & -3 & -3\\
		\hline 
		DM &  $\mathbf{1}_{S}$ & $\mathbf{1}_{F}$ & $\mathbf{1}_{S}$\\
		\hline 
		\# & 6 & 6 & 6\\
		\hline 
		$\mathbf{2}^1_{S}/\mathbf{2}^3_{S}$ & $\times$/1 & $\times$/1 & $\times$/1\\
		\hline 
		$F^{++}$/$S^{++}$ & \checkmark/\checkmark & \checkmark/\checkmark & \checkmark/\checkmark\\
		\hline 
	\end{tabular}
	\caption{\label{tab:res2g} The same as in Table~\ref{tab:res2a} for diagram 2.g of Figure~\ref{fig:category2}.}
\end{table}

\subsection{Six-particle models}
\label{sub:6particles}

In this section we separate the possible diagrams into two categories based on the quartic coupling they have: diagrams that have a coupling of $H\,S_i\,S_j\,S_k$ type (section~\ref{sub:HSSS}) and those with the coupling $H\,H\,S_i\,S_j$ (section~\ref{sub:HHSS}), where $S_{i,j,k}$ refer to BSM scalars that can be singlets or doublets.

\subsubsection{Category 3}
\label{sub:HSSS}

\begin{figure}[H]
	\centering
	\captionsetup[subfigure]{labelformat=empty}
	\begin{subfigure}[b]{0.25\textwidth}
		\includegraphics[scale=0.2]{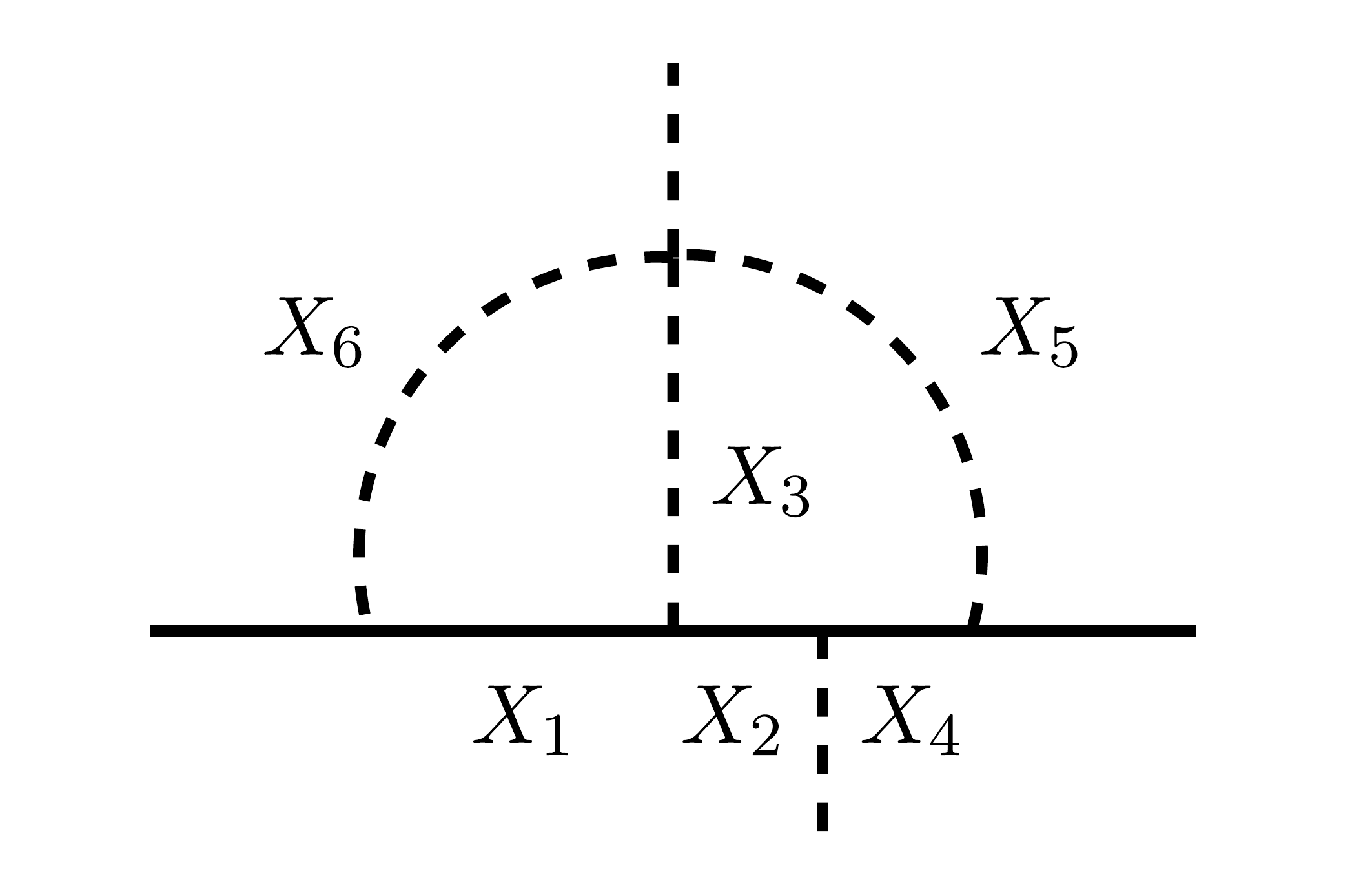}		
		\caption{\hspace{0.5cm}3.a}
	\end{subfigure} \hspace{0.8cm}
	\begin{subfigure}[b]{0.25\textwidth}
		\includegraphics[scale=0.2]{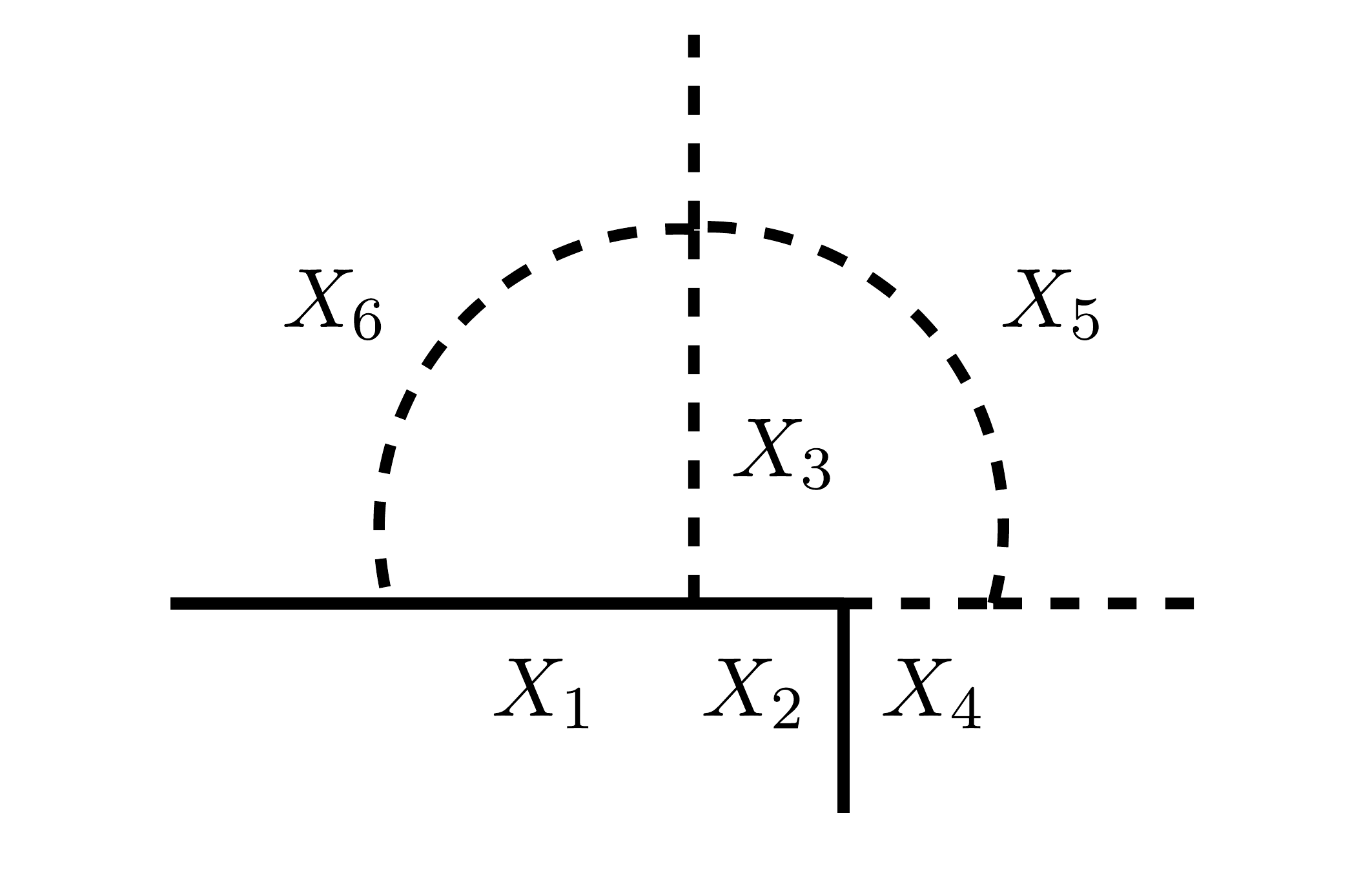}		
		\caption{\hspace{0.5cm}3.b}
	\end{subfigure} \hspace{0.8cm}                
	\begin{subfigure}[b]{0.25\textwidth}
		\includegraphics[scale=0.2]{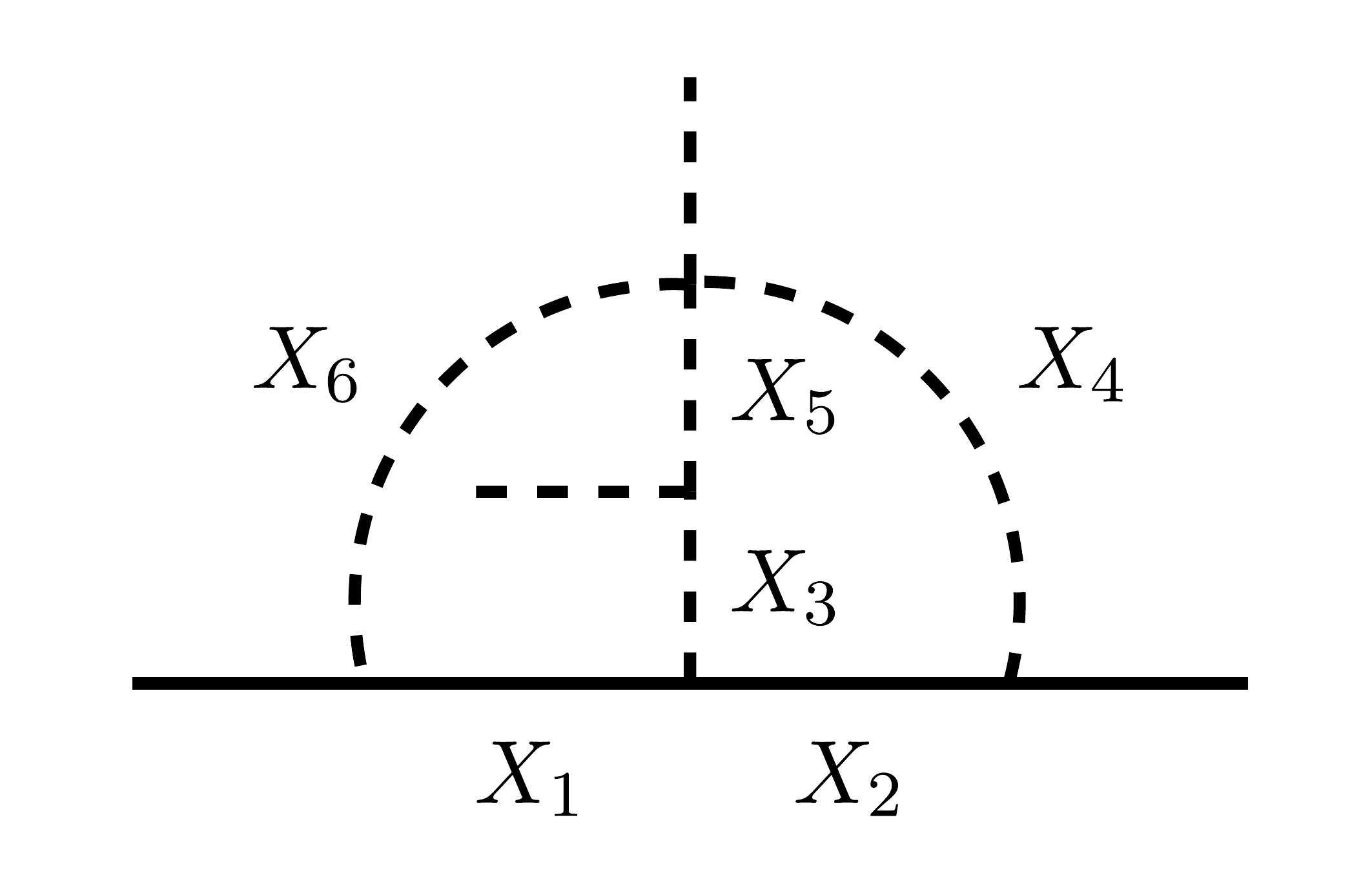}		
		\caption{\hspace{0.5cm}3.c}
	\end{subfigure}
	\caption{\label{fig:category3} Two-loop diagrams for Class 3.a - 3.c of Category 3.}
\end{figure}

This category is defined by the presence of a term of the form $H\,S_i\,S_j\,S_k$ in the scalar potential.
The diagrams within this category are shown in Figure~\ref{fig:category3} while the assignments for $\mathsf{SU(2)_L}$, $\mathsf{Z_2}$ and $\mathsf{U(1)_Y}$ are given, respectively, in the upper left, upper right and bottom charts of Table~\ref{tab:assignmentscategory3ab}
(Class 3.a and 3.b) and Table~\ref{tab:assignmentscategory3c} (Class 3.c).

\begin{table}[H]
	\centering
	\begin{tabular}{|c||c|c|c|c|c|c|}
		\hline 
		& $X_{1}$ & $X_{2}$ & $X_{3}$ & $X_{4}$ & $X_{5}$ & $X_{6}$\\
		\hline 
		\hline 
		\textit{i} & \textbf{1} & \textbf{1} & \textbf{1} & \textbf{2} & \textbf{1} & \textbf{2}\\
		\hline 
		\textit{ii} & \textbf{1} & \textbf{2} & \textbf{2} & \textbf{1} & \textbf{2} & \textbf{2}\\
		\hline 
		\textit{iii} & \textbf{2} & \textbf{1} & \textbf{2} & \textbf{2} & \textbf{1} & \textbf{1}\\
		\hline 
		\textit{iv} & \textbf{2} & \textbf{2} & \textbf{1} & \textbf{1} & \textbf{2} & \textbf{1}\\
		\hline 
	\end{tabular} \hspace{1cm}
	\begin{tabular}{|c||c|c|c|c|c|c|}
		\hline 
		& $X_{1}$ & $X_{2}$ & $X_{3}$ & $X_{4}$ & $X_{5}$ & $X_{6}$\\
		\hline 
		\hline 
		A & $\boldsymbol{+}$ & $\boldsymbol{-}$ & $\boldsymbol{-}$ & $\boldsymbol{-}$ & $\boldsymbol{-}$ & $\boldsymbol{+}$\\
		\hline 
		B & $\boldsymbol{-}$ & $\boldsymbol{+}$ & $\boldsymbol{-}$ & $\boldsymbol{+}$ & $\boldsymbol{+}$ & $\boldsymbol{-}$\\
		\hline 
		C & $\boldsymbol{-}$ & $\boldsymbol{-}$ & $\boldsymbol{+}$ & $\boldsymbol{-}$ & $\boldsymbol{-}$ & $\boldsymbol{-}$\\
		\hline 
	\end{tabular} \vspace{0.5cm}	
	
	\begin{tabular}{|c||c|c|c|c|c|c|}
		\hline 
		& $X_{1}$ & $X_{2}$ & $X_{3}$ & $X_{4}$ & $X_{5}$ & $X_{6}$\\
		\hline 
		\hline 
		a & $\alpha$ & $\beta$ & $\alpha-\beta$ & $\beta+1$ & $\beta$ & $\alpha+1$\\
		\hline 
		b & $\alpha$ & $\beta$ & $\alpha-\beta$ & $\beta-1$ & $\beta$ & $\alpha+1$\\
		\hline 
	\end{tabular}
	\caption{\label{tab:assignmentscategory3ab}  
		$\mathsf{SU(2)_L}$ (upper left), $\mathsf{Z_2}$ (upper right) and  $\mathsf{U(1)_Y}$ (bottom) assignments for particles $X_1$ to $X_6$ of diagrams 3.a and 3.b in Figure~\ref{fig:category3}.}
\end{table}

\begin{table}[H]
	\centering
	\begin{tabular}{|c||c|c|c|c|c|c|}
		\hline 
		& $X_{1}$ & $X_{2}$ & $X_{3}$ & $X_{4}$ & $X_{5}$ & $X_{6}$\\
		\hline 
		\hline 
		\textit{i}& \textbf{1} & \textbf{1} & \textbf{1} & \textbf{2} & \textbf{2} & \textbf{2}\\
		\hline 
		\textit{ii} & \textbf{1} & \textbf{2} & \textbf{2} & \textbf{1} & \textbf{1} & \textbf{2}\\
		\hline 
		\textit{iii} & \textbf{2} & \textbf{1} & \textbf{2} & \textbf{2} & \textbf{1} & \textbf{1}\\
		\hline 
		\textit{iv} & \textbf{2} & \textbf{2} & \textbf{1} & \textbf{1} & \textbf{2} & \textbf{1}\\
		\hline 
	\end{tabular}\hspace{1cm}
	\begin{tabular}{|c||c|c|c|c|c|c|}
		\hline 
		& $X_{1}$ & $X_{2}$ & $X_{3}$ & $X_{4}$ & $X_{5}$ & $X_{6}$\\
		\hline 
		\hline 
		A & $\boldsymbol{+}$ & $\boldsymbol{-}$ & $\boldsymbol{-}$ & $\boldsymbol{-}$ & $\boldsymbol{-}$ & $\boldsymbol{+}$\\
		\hline 
		B & $\boldsymbol{-}$ & $\boldsymbol{+}$ & $\boldsymbol{-}$ & $\boldsymbol{+}$ & $\boldsymbol{-}$ & $\boldsymbol{-}$\\
		\hline 
		C & $\boldsymbol{-}$ & $\boldsymbol{-}$ & $\boldsymbol{+}$ & $\boldsymbol{-}$ & $\boldsymbol{+}$ & $\boldsymbol{-}$\\
		\hline 
	\end{tabular}\vspace{0.5cm}
	
	\begin{tabular}{|c||c|c|c|c|c|c|}
		\hline 
		& $X_{1}$ & $X_{2}$ & $X_{3}$ & $X_{4}$ & $X_{5}$ & $X_{6}$\\
		\hline 
		\hline 
		a & $\alpha$ & $\beta$ & $\alpha-\beta$ & $\beta-1$ & $\alpha-\beta+1$ & $\alpha+1$\\
		\hline 
	\end{tabular}	
	\caption{\label{tab:assignmentscategory3c}  
		The same as Table~\ref{tab:assignmentscategory3ab} for diagram 3.c in Figure~\ref{fig:category3}.}
\end{table}

\paragraph{Class 3.a}

We compile in Table~\ref{tab:res3a} the fifteen models found for this class. Among the solutions seven have no doubly-charged particle. The DM candidate is, in all models, a scalar. The Lagrangian is given by
\begin{equation}
\begin{aligned}
\mathcal{L}^{}_\text{3.a}&= 
Y^{ia}_1\,(\overline{L^\texttt{C}}_i\,P_L) {F_4}_a \,S^\dagger_5\,+\,
Y^{ab}_2\,\overline{F_4}_a\, {F_2}_b\,H\,+\, 
Y^{bc}_3\,\overline{F_2}_b\, {F_1}_c\,S^\dagger_3\\
&\,+\,Y^{cj}_4\,\overline{F_1}_c\,(P_L\,L_j)\,S_6 \,+\,
\lambda\,H\,S_3\,S_5\,S^\dagger_6\,+\,\text{h.c.}\,,
\end{aligned}
\label{eq:L3a}
\end{equation}
leading to the neutrino mass matrix
\begin{equation}
\label{eq:M3a}
\begin{aligned}
\left(M^\text{3.a}_{\nu}\right)_{ij}&=\frac{\lambda \, v \, (Y^{ia}_1\,Y^{ab}_3\,Y^{bj}_4+Y^{ib}_4\,Y^{ba}_3\,Y^{aj}_1)}{2(2 \pi)^8}\,\sin (2 \theta_{24}) \, \times \\
& \sum_{\alpha\,=\,1}^{2}(-1)^{\alpha-1}\,\left(2\,(m_{24})_{\alpha}\,m_{1}\,I_\chi^1\,+\,I_\chi^{k^2}\,-\,I_\chi^{q^2}\,+\,I_\chi^{(k+q)^2}\right)\,,
\end{aligned}
\end{equation}
where $\chi\equiv \{S_5,\,{S_{24}}^{}_\alpha,\,S_3,\,F_1,\,S_6\}$.

\begin{table}[H]
	\setlength{\tabcolsep}{0pt}
	\centering
	\begin{tabular}{|c|c|c|c|c|c|c|c|c|c|c|c|c|c|c|c|}
		\hline 
		& \multicolumn{6}{c|}{A} & \multicolumn{4}{c|}{B} & \multicolumn{5}{c|}{C}\\
		\hline 
		& \multicolumn{3}{c|}{\textit{i}} & \textit{ii} & \textit{iii} & \textit{iv} & \textit{i} & \multicolumn{2}{c|}{\textit{iii}} & \textit{iv} & \multicolumn{2}{c|}{\textit{iii}} & \multicolumn{3}{c|}{\textit{iv}}\\
		\hline 
		$\alpha$ & -4 & {\color{red} -2} & 2 & 2 & 1 & 1 & -2 & -3 & -1 & {\color{red}-1} & -1 & {\color{red}-1} & -1 & -1 & -1\\
		\hline 
		$\beta$ & -4 & {\color{red}-2} & 2 & 1 & 2 & 1 & -2 & -2 & -2 & {\color{red}1} & -4 & {\color{red}-2} & -3 & 1 & 3\\
		\hline 
		DM & $\mathbf{1}_{S}$ & {$\color{red}\mathbf{1}_{S}$} & $\mathbf{1}_{S}$ & $\mathbf{2}_{S}$ & $\mathbf{2}_{S}$ & $\boldsymbol{\ast}$ & $\boldsymbol{\ast}$ & $\mathbf{2}_{S}$ & $\boldsymbol{\ast}$ & $\color{red}\mathbf{1}_{S}$ & $\mathbf{1}_{S}$ & $\color{red}\mathbf{1}_{S}$ & $\mathbf{1}_{S}$ & $\boldsymbol{\ast}$ & $\mathbf{1}_{S}$\\
		\hline 
		\# & 6 & {\color{red}5} & 5 & 4 & 5 & 5 & 4 & 4 & 4 & {\color{red}4} & 6 & {\color{red}5} & 6 & 5 & 6\\
		\hline 
		$\mathbf{2}^1_{S}/\mathbf{2}^3_{S}$ & $\times$/1 & {\color{red}1/$\times$} & $\times$/1 & 2/1 & 1/$\times$ & 1/$\times$ & 1/$\times$ & 1/$\times$ & 1/$\times$ & $\color{red}1/\times$ & $\times$/1 & $\color{red}1/\times$ & $\times$/1 & 1/$\times$ & $\times$/1\\
		\hline 
		$F^{++}/S^{++}$ & \checkmark/\checkmark & {\color{red}$\times$/$\times$} & \checkmark/\checkmark & $\times$/\checkmark & \checkmark/$\times$ & $\times$/$\times$ & $\times$/$\times$  & \checkmark/$\times$ & $\times$/$\times$ & $\color{red}\times/\times$ & \checkmark/\checkmark & $\color{red}\times/\times$ & \checkmark/\checkmark & $\times$/$\times$ & \checkmark/\checkmark\\
		\hline 
	\end{tabular}
	\caption{\label{tab:res3a} Realizations of diagram 3.a of Figure~\ref{fig:category3}. For each set of $\mathsf{SU(2)_L}$ quantum numbers (roman numerals) and $\mathsf{Z_2}$ assignments (capital letters A, B, C) we give the hypercharge in terms of $\alpha$ and $\beta$ parameters, the DM candidates, the number of beyond SM particles (\#), the number of $\mathbf{2}^1_{S}$ and $\mathbf{2}^3_{S}$ present as well as the existence (or not) of doubly-charged fermions ($F^{++}$) and doubly-charged scalars ($S^{++}$) particles. The symbol $\checkmark$($\times$) means present (absent) while the symbol $\boldsymbol{\ast}$, if present in the DM row, means mixing between $\mathbf{1}^0_{S}$ and $\mathbf{2}^1_{S}$. The $\mathsf{SU(2)_L}$, $\mathsf{Z_2}$ and $\mathsf{U(1)_Y}$ assignments are given in Table~\ref{tab:assignmentscategory3ab}. In red we marked models with at least one even-charged $\mathsf{Z}_2$ scalar doublet with hypercharge $Y=1$ that generically can create FCNCs.}
\end{table}

\paragraph{Class 3.b}

This class contains sixteen models all of them with a scalar as a DM candidate. The results presented in Table~\ref{tab:res3b} show that only two models have no doubly-charged particles. The interactions present in this class are given by
\begin{equation}
\begin{aligned}
\mathcal{L}^{}_\text{3.b}&= 
Y^{ia}_1\,(\overline{L^\texttt{C}}_i\,P_L) {F_2}_a \,S^\dagger_4\,+\,
Y^{ab}_2\,\overline{F_2}_a\, {F_1}_b\,S^\dagger_3\,+\,
Y^{bj}_3\,\overline{F_1}_b\,(P_L\,L_j)\,S_6 \,+\,
\mu\,H\,S_4\,S^\dagger_5\\
&\,+\,\lambda\,H\,S_3\,S_5\,S^\dagger_6\,+\,\text{h.c.}\,,
\end{aligned}
\label{eq:L3b}
\end{equation}
which have the following neutrino mass matrix,
\begin{equation}
\label{eq:M3b}
\begin{aligned}
\left(M^\text{3.b}_{\nu}\right)_{ij}&=\frac{\lambda \, v \, (Y^{ia}_1\,Y^{ab}_2\,Y^{bj}_3+Y^{ib}_3\,Y^{ba}_2\,Y^{aj}_1)}{2(2 \pi)^8}\sin(2 \theta_{45})\,\times \\
&\sum_{\alpha\,=\,1}^{2} (-1)^{\alpha-1}\,\left(2\,m_{1}\,m_{2}\,I_\chi^1\,+\, I_\chi^{k^2}\,-\,I_\chi^{q^2}\,+\,I_\chi^{(k+q)^2} \right)\,,
\end{aligned}
\end{equation}
where $\chi\equiv \{F_2,\,{F_{45}}^{}_\alpha,\,S_3,\,F_1,\,S_6\}$.

\begin{table}[H]
	\centering
	\setlength{\tabcolsep}{0pt}
	\begin{tabular}{|c|c|c|c|c|c|c|c|c|c|c|c|c|c|c|c|c|}
		\hline
		& \multicolumn{9}{c|}{A} & \multicolumn{5}{c|}{B} & \multicolumn{2}{c|}{C}\\
		\hline
		\hline
		& \multicolumn{2}{c|}{\textit{i}} & \multicolumn{5}{c|}{\textit{ii}} & \multicolumn{2}{c|}{\textit{iii}} & \textit{i} & \textit{ii} & \multicolumn{3}{c|}{\textit{iii}} & \textit{ii} & \textit{iii}\\
		\hline
		$\alpha$ & {\color{red}-2} & 2 & -4 & {\color{red}-2} & {\color{red}-2} & 2 & 2 & 1 & 3 & -2 & -4 & -3 & -1 & 3 & -2 & {\color{red}-1}\\
		\hline
		$\beta$ & {\color{red}-2} & 2 & -3 & {\color{red}-1} & {\color{red}1} & 1 & 3 & 2 & 4 & -2 & -3 & -2 & -2 & 4 & 1 & {\color{red}-2}\\
		\hline
		DM & {\color{red}$\mathbf{1}_{S}$} & $\boldsymbol{\ast}$ & $\mathbf{2}_{S}$ & {\color{red}$\mathbf{2}_{S}$} & {\color{red}$\boldsymbol{\ast}$} & $\boldsymbol{\ast}$ & $\mathbf{2}_{S}$ & $\mathbf{2}_{S}$ & $\mathbf{2}_{S}$ & $\boldsymbol{\ast}$ & $\mathbf{2}_{S}$ & $\mathbf{2}_{S}$ & $\boldsymbol{\ast}$ & $\mathbf{2}_{S}$ & $\boldsymbol{\ast}$ & {\color{red}$\mathbf{1}_{S}$}\\
		\hline
		\# & {\color{red}5} & 5 & 6 & {\color{red}4} & {\color{red}5} & 4 & 5 & 4 & 6 & 5 & 6 & 5 & 5 & 6 & 5 & {\color{red}6}\\
		\hline
		$\mathbf{2}^1_{S}/\mathbf{2}^3_{S}$ & {\color{red}1/1} & 1/1 & 1/2 & {\color{red}3/$\times$} & {\color{red}2/1} & 2/1 & 1/2 & 2/$\times$ & 1/1 & 1/1 & 1/2 & 1/1 & 1/1 & 1/1 & 2/1 & {\color{red}1/1}\\
		\hline
		$F^{++}/S^{++}$ & {\color{red}$\times$/\checkmark} & $\times$/\checkmark & \checkmark/\checkmark & {\color{red}$\times$/$\times$} & {\color{red}$\times$/\checkmark} & $\times$/\checkmark & \checkmark/\checkmark & $\times$/$\times$ & \checkmark/\checkmark & $\times$/\checkmark & \checkmark/\checkmark & \checkmark/\checkmark & $\times$/\checkmark & \checkmark/\checkmark & $\times$/\checkmark & {\color{red}$\times$/\checkmark}\\
		\hline
	\end{tabular}
	\caption{\label{tab:res3b} The same as in Table~\ref{tab:res3a} for diagram 3.b.}
\end{table}

\paragraph{Class 3.c}

From the seven models of Class 3.c only one has a pure scalar singlet as DM candidate being the remaining ones a mixture between $\mathbf{1}^0_S$ and $\mathbf{2}^1_S$. The result of this class with the Lagrangian given by
\begin{equation}
\begin{aligned}
\mathcal{L}^{}_\text{3.c}&= 
Y^{ia}_1\,(\overline{L^\texttt{C}}_i\,P_L) {F_2}_a \,S^\dagger_4\,+\,
Y^{ab}_2\,\overline{F_2}_a\, {F_1}_b\,S^\dagger_3\,+\,
Y^{bj}_3\,\overline{F_1}_b\,(P_L\,L_j)\,S_6 \,+\,
\mu\,H\,S_3\,S^\dagger_5\\
&\,+\,\lambda\,H\,S_4\,S_5\,S^\dagger_6\,+\,\text{h.c.}\, ,
\end{aligned}
\label{eq:L3c}
\end{equation}
are presented in Table~\ref{tab:res3c}. 

The neutrino mass matrix is 
\begin{equation}
\label{eq:M3c}
\begin{aligned}
\left(M^\text{3.c}_{\nu}\right)_{ij}&=\frac{\lambda \, v \, (Y^{ia}_1\,Y^{ab}_2\,Y^{bj}_3+Y^{ib}_3\,Y^{ba}_2\,Y^{aj}_1)}{2(2 \pi)^8}\sin (2 \theta_{35})  \, \times \\
& \sum_{\alpha\,=\,1}^{2} (-1)^{\alpha-1} \,\left(2\, m_{1} \, m_{2}\, I_\chi^1\, +\, I_\chi^{k^2}\, -\, I_\chi^{q^2}\, +\, I_\chi^{(k+q)^2} \right)\,,
\end{aligned}
\end{equation}
where $\chi\equiv \{F_2\,S_4,\,{F_{35}}^{}_\alpha,\,F_1\,S_6\}$.

\begin{table}[H]
	\centering
	\begin{tabular}{|c|c|c|c|c|c|c|c|}
		\hline
		& \multicolumn{6}{c|}{A} & C\\
		\hline
		\hline
		& \textit{i} & \multicolumn{3}{c|}{\textit{ii}} & \multicolumn{2}{c|}{\textit{iii}} & \textit{ii}\\
		\hline
		$\alpha$ & 2 & 2 &  {\color{red}-2} & -4 & 3 & 1 & -2\\
		\hline
		$\beta$ & 2 & 3 &  {\color{red}1} & -3 & 4 & 2 & 1\\
		\hline
		DM & $\boldsymbol{\ast}$ & $\boldsymbol{\ast}$ &  {\color{red}$\mathbf{1}_{S}$} & $\boldsymbol{\ast}$ & $\boldsymbol{\ast}$ & $\boldsymbol{\ast}$ & $\boldsymbol{\ast}$\\
		\hline
		\# & 4 & 5 &  {\color{red}5} & 6 & 6 & 4 & 6\\
		\hline
		$\mathbf{2}^1_{S}/\mathbf{2}^3_{S}$ & 2/1 & 1/1 &  {\color{red}1/1} & 1/1 & 1/1 & 3/$\times$ & 1/1\\
		\hline
		$F^{++}/S^{++}$ & $\times$/\checkmark{} & \checkmark/\checkmark &  {\color{red}$\times$/\checkmark} & \checkmark/\checkmark & \checkmark/\checkmark & $\times$/$\times$ & $\times$/ \checkmark\tabularnewline
		\hline
	\end{tabular}
	\caption{\label{tab:res3c} The same as in Table~\ref{tab:res3a} for diagram 3.c. The $\mathsf{SU(2)_L}$, $\mathsf{Z_2}$ and $\mathsf{U(1)_Y}$ assignments for this model are given in Table~\ref{tab:assignmentscategory3c}.}
\end{table}

\subsubsection{Category 4}
\label{sub:HHSS}

This category contains two different classes for which a scalar term $H\,H\,S_i\,S_j$ is present. The $\mathsf{SU(2)_L}$ and $\mathsf{U(1)_Y}$ assignments are given in Table~\ref{tab:assignmentscategory4} while the $\mathsf{Z_2}$ charges are given in Table~\ref{tab:assignmentscategory4Z2} for both 4.a (left) and 4.b (right).

\begin{figure}[H]
	\centering
	\captionsetup[subfigure]{labelformat=empty}
	\begin{subfigure}[b]{0.25\textwidth}
		\includegraphics[scale=0.2]{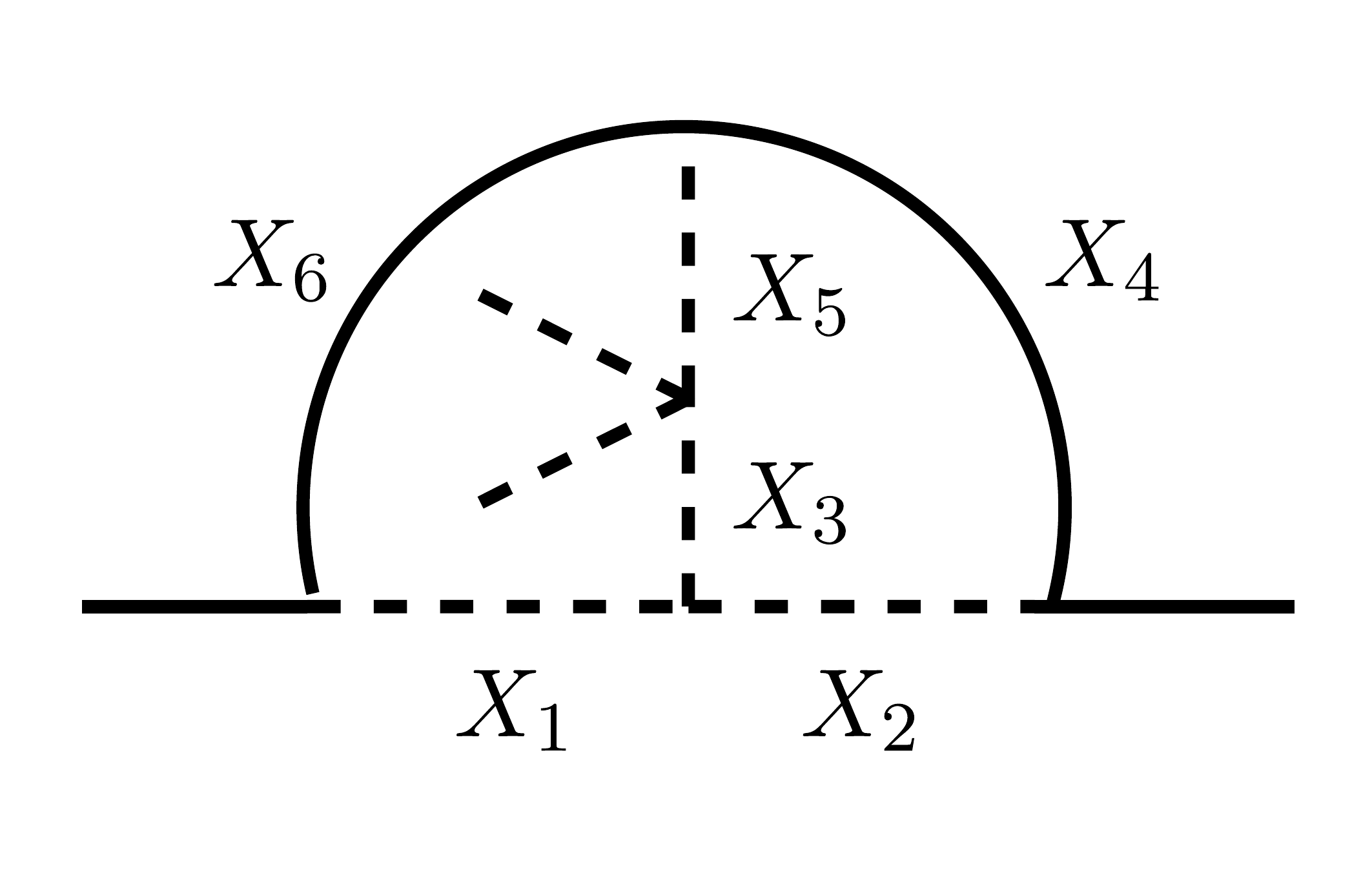}
		\caption{\hspace{0.5cm}4.a}
	\end{subfigure} 
	\hspace{0.8cm}
	\begin{subfigure}[b]{0.25\textwidth}
		\includegraphics[scale=0.2]{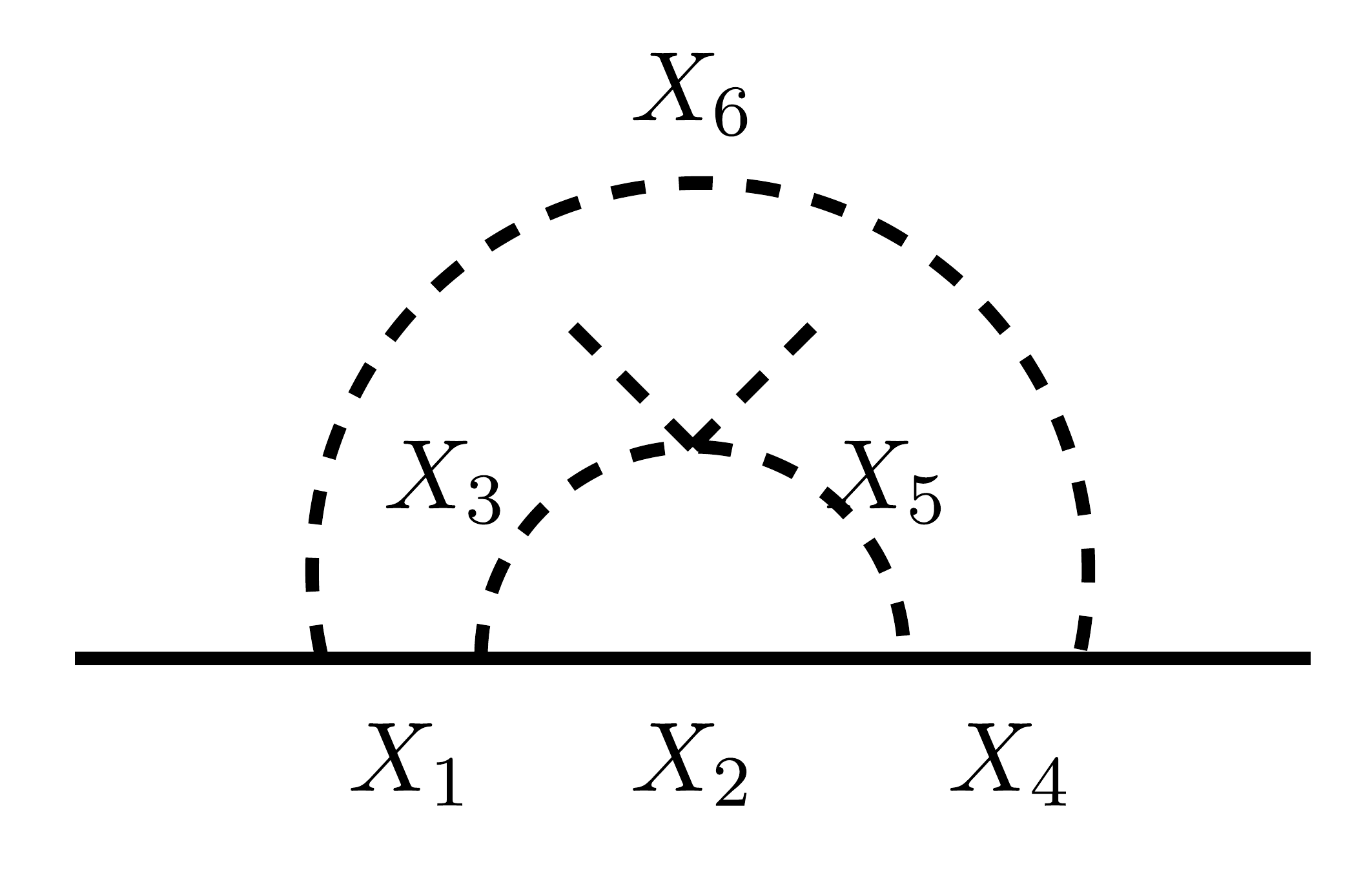}
		\caption{\hspace{0.5cm}4.b}
	\end{subfigure} 
	\caption{\label{fig:category4} Two-loop diagrams for Class 4.a and 4.b of Category 4.}
\end{figure}

\begin{table}[H]
	\setlength{\tabcolsep}{3pt}
	\centering
	\begin{tabular}{|c||c|c|c|c|c|c|}
		\hline
		& $X_{1}$ & $X_{2}$ & $X_{3}$ & $X_{4}$ & $X_{5}$ & $X_{6}$\\
		\hline
		\hline
		\textit{ii} & \textbf{1} & \textbf{2} & \textbf{2} & \textbf{1} & \textbf{2} & \textbf{2}\\
		\hline
		\textit{iii} & \textbf{2} & \textbf{1} & \textbf{2} & \textbf{2} & \textbf{2} & \textbf{1}\\
		\hline
	\end{tabular}\hspace{0.5cm}
	\begin{tabular}{|c||c|c|c|c|c|c|}
		\hline
		& $X_{1}$ & $X_{2}$ & $X_{3}$ & $X_{4}$ & $X_{5}$ & $X_{6}$\\
		\hline
		\hline
		a & $\alpha$ & $\beta$ & $\alpha-\beta$ & $\beta-1$ & $\alpha-\beta+2$ & $\alpha+1$\\
		\hline
		b & $\alpha$ & $\beta$ & $\alpha-\beta$ & $\alpha+2$ & $\alpha-\beta+2$ & $\alpha+1$\\
		\hline
	\end{tabular} 
	\caption{\label{tab:assignmentscategory4}  
		$\mathsf{SU(2)_L}$ (left) and $\mathsf{U(1)_Y}$ (right) assignments for particles $X_1$ to $X_6$ of diagrams in Figure~\ref{fig:category4}.}
\end{table}

\begin{table}[H]
	\centering
	\begin{tabular}{|c||c|c|c|c|c|c|}
		\hline
		& $X_{1}$ & $X_{2}$ & $X_{3}$ & $X_{4}$ & $X_{5}$ & $X_{6}$\\
		\hline
		\hline
		A & $\boldsymbol{+}$ & $\boldsymbol{-}$ & $\boldsymbol{-}$ & $\boldsymbol{-}$ & $\boldsymbol{-}$ & $\boldsymbol{+}$\\
		\hline
		B & $\boldsymbol{-}$ & $\boldsymbol{+}$ & $\boldsymbol{-}$ & $\boldsymbol{+}$ & $\boldsymbol{-}$ & $\boldsymbol{-}$\\
		\hline
		C & $\boldsymbol{-}$ & $\boldsymbol{-}$ & $\boldsymbol{+}$ & $\boldsymbol{-}$ & $\boldsymbol{+}$ & $\boldsymbol{-}$\\
		\hline
	\end{tabular}\hspace{0.5cm}
	\begin{tabular}{|c||c|c|c|c|c|c|}
		\hline
		& $X_{1}$ & $X_{2}$ & $X_{3}$ & $X_{4}$ & $X_{5}$ & $X_{6}$\\
		\hline
		\hline
		A & $\boldsymbol{+}$ & $\boldsymbol{-}$ & $\boldsymbol{-}$ & $\boldsymbol{+}$ & $\boldsymbol{-}$ & $\boldsymbol{+}$\\
		\hline
		B & $\boldsymbol{-}$ & $\boldsymbol{+}$ & $\boldsymbol{-}$ & $\boldsymbol{-}$ & $\boldsymbol{-}$ & $\boldsymbol{-}$\\
		\hline
		C & $\boldsymbol{-}$ & $\boldsymbol{-}$ & $\boldsymbol{+}$ & $\boldsymbol{-}$ & $\boldsymbol{+}$ & $\boldsymbol{-}$\\
		\hline  
	\end{tabular}
	\caption{\label{tab:assignmentscategory4Z2}  
		$\mathsf{Z_2}$ assignments for particles $X_1$ to $X_6$ of diagrams 4.a~(left) and  4.b~(right)  of Figure~\ref{fig:category4}.}
\end{table}

\paragraph{Class 4.a}
The seven models found for this class are summarized in Table~\ref{tab:res4a}. Two of the models have no doubly-charged particles and for the first time we find a model that requires only three BSM fields to be complete. The interactions in this class are given by
\begin{equation}
\begin{aligned}
\mathcal{L}^{}_\text{4.a}&= 
Y^{ia}_1\,(\overline{L^\texttt{C}}_i\,P_L) {F^\texttt{C}_4}_a \,S_2\,+\,
Y^{ab}_2\,\overline{F^\texttt{C}_4}_a\, {F^\texttt{C}_6}_b\,S_5\,+\,
Y^{bj}_3\,\overline{F^\texttt{C}_6}_b\,(P_L\,L_j)\,S^\dagger_1 \,+\,
\mu\,S_1\,S^\dagger_2\,S^\dagger_3\\
&\,+\,\lambda\,H\,H\,S_3\,S^\dagger_5\,+\,\text{h.c.}\,,
\end{aligned}
\label{eq:L4a}
\end{equation}
which lead to the following neutrino mass matrix
\begin{equation}
\label{eq:M4a}
\begin{aligned}
\left(M^\text{4.a}_{\nu}\right)^{ij}&=\frac{\mu \, (Y^{ia}_1\,Y^{ab}_2\,Y^{bj}_3+Y^{ib}_3\,Y^{ba}_2\,Y^{aj}_1)}{2(2 \pi)^8}\sin (2 \theta_{35}) \, \times  \\
& \sum_{\alpha\,=\,1}^{2} (-1)^{\alpha-1}\,\left(2\,m_{4}\, m_{6}\, I_\chi^1\, +\, I_\chi\, -\, I_\chi^{q^2}\, +\, I_\chi^{(k+q)^2} \right)\,,
\end{aligned}
\end{equation}
where $\chi\equiv \{S_2,\,F_4,\,{F_{35}}^{}_\alpha,\,S_1,\,S_6\}$.

\begin{table}[H]
	\centering
	\begin{tabular}{|c|c|c|c|c|c|c|c|}
		\hline 
		& \multicolumn{7}{c|}{A}\\
		\hline 
		\hline 
		& \multicolumn{2}{c|}{\textit{ii}} & \multicolumn{5}{c|}{\textit{iii}}\\
		\hline 
		$\alpha$ & -2 & -4 & 3 &  {\color{red}1} &  {\color{red}1}& -3 & -3\\
		\hline 
		$\beta$ & -1 & -3 & 4 &  {\color{red}2} &  {\color{red}0} & 0 & -2\\
		\hline 
		DM & $\mathbf{2}_{S}$ & $\mathbf{2}_{S}$ & $\mathbf{2}_{S}$ &  {\color{red}$\mathbf{2}_{S}$} &  {\color{red}$\boldsymbol{\ast}$} & $\boldsymbol{\ast}$ & $\mathbf{2}_{S}$\\
		\hline 
		\# & 3 & 5 & 5 &  {\color{red}4} &  {\color{red}5} & 5 & 4\\
		\hline 
		$\mathbf{2}^1_{S}$/$\mathbf{2}^3_{S}$ & 3/$\times$ & 2/1 & 2/1 &  {\color{red}3/$\times$} &  {\color{red}2/1} & 1/2 & 2/1\\
		\hline 
		$F^{++}/S^{++}$ & $\times/\times$ & \checkmark/\checkmark & \checkmark/\checkmark &  {\color{red}$\times/\times$} &  {\color{red}$\times$/\checkmark} & $\times$/\checkmark & \checkmark/\checkmark \\
		\hline 
	\end{tabular}
	\caption{\label{tab:res4a} Realizations of diagram 4.a of Figure~\ref{fig:category4}. For each set of $\mathsf{SU(2)_L}$ quantum numbers (roman numerals) and $\mathsf{Z_2}$ assignments (capital letters A, B, C) we give the hypercharge in terms of $\alpha$ and $\beta$ parameters, the DM candidates, the number of beyond SM particles (\#), the number of $\mathbf{2}^1_{S}$ and $\mathbf{2}^3_{S}$ present as well as the existence (or not) of doubly-charged fermions ($F^{++}$) and doubly-charged scalars ($S^{++}$) particles. The symbol $\checkmark$($\times$) means present (absent) while the symbol $\boldsymbol{\ast}$, if present in the DM row, means mixing between $\mathbf{1}^0_{S}$ and $\mathbf{2}^1_{S}$. The $\mathsf{SU(2)_L}$, $\mathsf{Z_2}$ and $\mathsf{U(1)_Y}$ assignments are given in Table~\ref{tab:assignmentscategory4}. In red we marked models with at least one even-charged $\mathsf{Z}_2$ scalar doublet with hypercharge $Y=1$ that generically can create FCNCs.}
\end{table}

\paragraph{Class 4.b}
This class has only two models requiring four BSM fields each. They both have  doubly-charged particles and two scalar doublets in addition to the SM Higgs. The details of these two solutions are given in Table~\ref{tab:res4b}, and their interactions are given by the following Lagrangian
\begin{equation}
\begin{aligned}
\mathcal{L}^{}_\text{4.b}&= 
Y^{ia}_1\,(\overline{L^\texttt{C}}_i\,P_L) {F_4}_a \,S^\dagger_6\,+\,
Y^{ab}_2\,\overline{F_4}_a\,{F_2}_b\,S_5\,+\,
Y^{bc}_3\,\overline{F_2}_b\, {F_1}_c\,S^\dagger_3\,+\,
Y^{cj}_4\,\overline{F_1}_c\,(P_L\,L_j)\,S_6 \\
&\,+\,\lambda\,H\,H\,S_3\,S^\dagger_5\,+\,\text{h.c.}\,.
\end{aligned}
\label{eq:L4b}
\end{equation}

The mass matrix is given by
\begin{equation}
\label{eq:M4b}
\begin{aligned}
(M^\text{4.b}_{\nu})^{ij}&=\frac{( Y^{ia}_1\,Y^{ab}_2\,Y^{bc}_3\,Y^{cj}_4+Y^{ic}_4\,Y^{cb}_3\,Y^{ba}_2\,Y^{aj}_1)}{2(2 \pi)^8}\sin (2 \theta_{35}) \\
& \times \sum_{\alpha\,=\,1}^{2} (-1)^{\alpha-1}\, \left(2\,m_{1}\,m_{2}\,m_{4}\, I_\gamma^1\, +\, (m_{1}\, + \, 2\,m_{2}\,+\,m_{4})\, I_\gamma^{k^2} \right. \\
& \left. + (-m_{1}-m_{4})\, I_\gamma^{q^2}\, +\, (m_{1}+m_{4})\,I_\gamma^{(k+q)^2} \right)\,,
\end{aligned}
\end{equation}
where $\gamma\equiv \{{F_{35}}^{}_\alpha,\,F_2,\,F_1,\,F_4,\,S_6\}$.

\begin{table}[H]
	\centering
	\begin{tabular}{|c|c|c|}
		\hline
		& \multicolumn{2}{c|}{A}\\
		\hline
		\hline
		& \multicolumn{2}{c|}{\textit{iii}}\\
		\hline
		$\alpha$ & -1 & 1\\
		\hline
		$\beta$ & -2 & 2\\
		\hline
		DM & $\boldsymbol{\ast}$ & $\mathbf{2}_{S}$\\
		\hline
		\# & 4 & 4\\
		\hline
		$\mathbf{2}^1_{S}/\mathbf{2}^3_{S}$ & 1/1 & 2/$\times$\\
		\hline
		$F^{++}/S^{++}$ & $\times$/\checkmark & \checkmark/$\times$\\
		\hline
	\end{tabular}
	\caption{\label{tab:res4b} The same as in Table~\ref{tab:res4a} for diagram 4.b.}
\end{table}

\section{Some Remarks on the Models Phenomenology}
\label{sec:pheno}
In this section we briefly comment on some phenomenological aspects of the models presented above.

\subsection{The Scalar Sector}
\label{sub:MHDM}

Depending on the specific diagram, two-loop radiative neutrino masses require six or seven internal messengers to close the loop. More concretely, the number of additional scalars that need to be added are controlled by the number of Yukawa interactions and cubic scalar terms of each diagram. 
Four diagrams (1.e, 1.g, 2.b and 2.d) need the addition of only two scalars, six diagrams (1.a, 1.c, 2.g, 2.f, 3.a and 4.b) require three extra scalars, seven models (1.d, 1.f, 1.h, 2.e, 3.b, 3.c and 4.a) use four extra scalars and finally three diagrams (1.b, 2.a and 2.c) employ five additional scalars. By inspection of the models one can see that the simplest scalar sector is for a model for which one has the SM Higgs plus two singlets, while the more complicated scalar sector will be for a model with the SM Higgs, plus three doublets and two singlets. Although each model has a specific scalar potential, it will be a combination of the SM Higgs potential plus the contribution coming from singlets with hypercharges $Y=0,\,2,\,4$ and/or doublets with hypercharges $Y=1,\,3$.

The scalar potential for the pure SM is given by
\begin{equation}
V_\text{SM}=(m_{H})^2\,( H^{\dagger}H)\,+\,\frac{\lambda_H}{2}\,( H^{\dagger}H)^2\,,
\label{eq:VSM}
\end{equation}
while the generic scalar potential for SM plus scalar doublets is given as
\begin{eqnarray}
\begin{aligned}
\label{eq:V1}
V_{1}&= \Gamma_1^{ab}(\,\phi_a^\dagger\,\phi_b\,)+ \Gamma_2^{ab}(\,\chi_a^\dagger\,\chi_b\,) \\ &+\Lambda_1^{abcd}\,(\phi_a^\dagger\,\phi_b)\,(\phi_c^\dagger\ \phi_d)\,+ \Lambda_2^{abcd}\,(\chi_a^\dagger\,\phi_b)\,(\phi_c\cdot \phi_d)+ \text{h.c.}, \,
\end{aligned}
\end{eqnarray}
where $a$ and $b$ represent different fields.  $\phi$ refers to a BSM scalar $\mathsf{SU(2)_L}$ doublet with $Y=1$ and $\chi$ is a $\mathsf{SU(2)_L}$ doublet with $Y=3$. For $a\text{ or }b=1$, $\phi_1=H$, is the SM Higgs.

In the case of doublets plus singlets ($S_i$), the generic scalar potential has the extra terms, 
\begin{equation}
\label{eq:V2}
\begin{aligned}
V_{2}&=\lambda_1(S_i^\ast\,S_j)\,(\phi_a^\dagger\,\phi_b) 
\,+\,\lambda_2\,(S_{Y=0}\,S_{Y=2}^*)\,(\phi_a^ \cdot\,\phi_b)
\,+\,\lambda_3\,(S_{Y=2}\,S_{Y=4}^\ast)(\phi_a^ \cdot\,\phi_b) \\
&+\,\lambda_4\,(S_{Y=2}^\ast)^2\,(H\cdot \chi)
\,+\,\lambda_5\,(S_{Y=0}\,S_{Y=4}^\ast)\,(H\cdot \chi_i)+\text{h.c.},
\end{aligned}
\end{equation}
where, in addition to the labels in the last equation, we have the indices $i$ and $j$ that represent different singlets with the same hypercharge.

Finally, the potential with the cubic terms is given by
\begin{equation}
\label{eq:V3}
\begin{aligned}
V_{3}=\mu_1\,S_{Y=0}\,\phi_a^\dagger\,\phi_b\,+\,\mu_2\, S_{Y=2}^\ast\, (\phi_a\cdot\phi_b)\,+\,\mu_3\, S_{Y=2}^\ast\, \phi_a^\dagger\,\chi_b\,+\,\mu_4\,S_{Y=4}^\ast\,(\phi_a\cdot\chi_b)\,+\text{h.c.}.
\end{aligned}
\end{equation}

Note that the scalar potential given above is very generic and  each specific model contains different terms. This is not only because of the difference between the particle content of each model but also due to the presence of the $\mathsf{Z_2}$ symmetry that forbids certain terms from appearing. The expressions given above in eqs.~\eqref{eq:VSM}--\eqref{eq:V3} are generic and need to be taken together with the $\mathsf{Z_2}$ charge assignments.

When writing the potential we have considered $H$ as the only scalar that will acquire a VEV, i.e., the BSM scalars are all inert~\footnote{This is so to ensure the two-loop diagrams are genuine and cannot be broken into lower order diagrams.}. Although in most cases $H$ is, in fact, the SM Higgs, this is not the case for all models, the reason for this has to do with the fact that in some models it mixes with other scalars.  It is very common in the literature to add an \textit{ad hoc} $\mathsf{Z_2}$ symmetry which forbids the terms that mix the SM Higgs and other doublets ($\lambda_4^{Hi}$ and $m_{Hi}$ terms). In our case the $\mathsf{Z_2}$ charges are fixed by the two-loop diagrams, this allows some models to have a scalar doublet $\phi$ with the same $\mathsf{Z_2}$ charge as the SM Higgs and hence mixing them. This implies that there could be cases where the SM Higgs is the lightest of a two-eigenstate system $\{H^0,\phi^0\}$.

Note that one could also think of models where $\mu_1$ mixes a hypercharge $Y$=1 scalar doublet, $\phi$, with a neutral scalar singlet, $S_0$. In this generic case, the SM Higgs is now  the lightest of the system $\{H^0,\phi^0,S_0\}$. Given the $\mathsf{Z_2}$ assignments, none of the models in this article have these three eigenstates at once and we do not need to be concerned about these cases.

In section \ref{sec:results} we showed for each model the number of scalar doublets needed. In twenty four of the models the scalar sector only consists of the SM Higgs plus singlets. Also, it can be seen that some of the models that were presented have a scalar potential that has been studied in multi-Higgs-doublet-model~(MHDM), more exactly there are fifteen 2HDM, nine 3HDM and five 4HDM. These models are a simple extension of the SM obtained by adding scalar doublets and they have been thoroughly studied in the literature, see e.g.~\cite{Branco:2011iw}. The most common MHDM contains only $\mathsf{SU(2)_L}$ doublets with hypercharge $Y=1$,  nevertheless, we can see that most models exhibit more of a modified structure since they include also doublets with hypercharge $Y=3$, that have been studied a little in literature~\cite{Huitu:1996su,Aoki:2011yk,Chiang:2012dk,Babu:2013ega,Okada:2015hia} and still require more analysis. The addition of inert singlets does not affect the phenomenology~\cite{Drozd:2014yla} of MHDM models.

\subsection{Flavor Changing Neutral Currents}
\label{sub:SDY1}

In the simplest of these cases, when a second Higgs scalar is added to the SM,  then the two Higgs doublets will allow SM fermions to couple to up-type, down-type fermions and leptons. Therefore, generic Yukawa couplings are not allowed without inducing dangerous contributions to flavor-changing neutral currents~(FCNCs). In other words, in the NHDM generic case the are terms in the Yukawa Lagrangian that will create unwanted FCNCs~\cite{Branco:2011iw,Ginzburg:2004vp,Haber:2006ue,Altmannshofer:2012ar}. 

This will be translated into our models, that unless some other physics is involved,  there should be a restriction for the models to contain an even-charged $\mathsf{Z}_2$ scalar doublets with hypercharge $Y=1$ (other than the SM Higgs)~\cite{Celis:2013rcs}. For this reason we kept all such models marked in red in the results shown before. Luckily models in Category~1 and~2 never have this field. The reason is that whenever this field appears it is always  accompanied by fields that allow for one-loop realizations. Nevertheless, this is not the case for Category~3 and 4 where only a limited number of such models appear. 

There are many possibilities of how to avoid FCNCs in these multi-Higgs models, the most common is to chose the $\mathsf{Z_2}$ of the scalars to forbid in the Yukawa terms the dangerous contribution~\cite{Inoue:2014nva}.

However, the models in this paper lack this freedom, since the $\mathsf{Z_2}$ charges are assign uniquely based on what the two loops diagrams permit.
Other option is to chose the VEV alignment of the scalars, but once again in all the models presented only the SM Higgs gets a VEV, and therefore this is not a solution to eliminate FCNCs. Nevertheless, there are options of how to eliminate FCNCs, for example, hierachical, Yukawa-alinged, VEV-aligned, minimal flavor violation models~\cite{Pich:2009sp,Jung:2010ik,Dev:2014yca,Buras:2003jf,Agashe:2005hk,Gavela:2009cd,Kagan:2009bn}.

\subsection{Addition on Triplets}
\label{sub:triplets}

Interestingly, none of the models presented have both fermion and scalar as possible DM candidates at the same time. This statement changes if we relax the condition of having just singlets and doublets and introduce $\mathsf{SU(2)_L}$ triplets. In fact, all the diagrams presented can be extended to include $\mathsf{SU(2)_L}$ triplets, to do so, one needs to trade in each vertex of the type doublet-doublet-singlet, the singlet for a triplet and in the singlet-singlet-singlet vertices, two singlets for two triplets or all singles for triplets. Such an extension allows us to find solutions that were previously not viable due to our restrictions, particularly those that can create neutrino masses but have no DM particle inside the loop or those for which the external legs were not all neutral. However, the introduction of triplets escalates the number of allowed models. As an example, we use diagram 1.d; without triplets there are 14 possible models, with them we found 162. Even if other diagrams do not behave as dramatically as 1.d, it is clear that the number of models, if one wishes to include triplets, would be too great to be presented in a concise manner.
One needs to be aware that the addition of $\mathsf{SU(2)_L}$ triplets to the models needs to be made carefully since they might induce neutrino masses at tree level via type-II and/or type-III seesaw mechanism.

\subsection{Doubly-Charged Particles}
\label{sub:doubly}

In section~\ref{sec:results} we showed all models that can generate genuine two-loop neutrino masses when considering the condition of adding just singlets and doublets to the SM. This condition was implemented, mainly, to keep the amount of presented solutions to a manageable number but also to restrain how ``exotic'' the new fields are. Nevertheless, most of these models do contain at least one peculiar particle characterized by having an electric charge $Q=2$. 

The total number of models presented is 141, out of those 118 have the presence of at least one doubly-charged particle: 15 have doubly-charged fermions, 33 doubly-charged scalars and 70 have both. Given our construction these particles can only be singlets with hypercharge $Y=4$ or the component of a doublet with $Y=3$. 

The connection between neutrino masses and doubly-charged particles is not new and it has been studied since the Babu-Zee model ~\cite{Babu:2002uu} which includes a doubly-charged scalar singlet.

The possibility of detecting such a particle at colliders has been discussed in the literature~\cite{Gunion:1996pq,Babu:2002uu,AristizabalSierra:2006gb,Nebot:2007bc,Ohlsson:2009vk,Babu:2016rcr} and several searches are being done at the CMS~\cite{Chatrchyan:2012ya} and ATLAS experiments~\cite{ATLAS:2012hi}. Because they have not been found, these experiments have set bounds on their masses. For example, a thorough study has been done in Ref.~\cite{Alloul:2013raa}  of possible LHC discovery of doubly-charged particles, considering spins 0, 1/2 and 1 together with different $\mathsf{SU(2)_L}$ assignments, singlets, doublets and triplets. One of the interesting characteristic for these searches is that the SM background is suppressed making these signatures easy to analyze.  Any positive signature would be excellent to discriminate between many of the models.

Doubly-charged particles coming from singlets or doublets will have different decays and therefore different signatures. There are multiple channels for production of doubly-charged particles, but the production via s-channel gauge-boson exchange in the one with the biggest branching ratio: 
\begin{subequations}
	\begin{align}
	\label{eq:reactions1}
	&p p \rightarrow {W^{+}}^\ast \rightarrow X^{++}\,Y^{-}\,, \\
	\label{eq:reactions2}
	&p p \rightarrow (Z/\gamma^\ast) \rightarrow X^{++}\,X^{--}\,. 
	\end{align}
\end{subequations}
In eq.~\eqref{eq:reactions1} $X$ and $Y$ represent two different particles from the same doublet while in eq.~\eqref{eq:reactions2} $X$ represents either a singlet or a doublet.

One of the possible phenomenological differences between singlets and doublets is that singlets can only decay through the second channel, eq.~\eqref{eq:reactions2}, given they do not have $W$-boson interactions. 

Once a doubly-charged particle is produced its subsequent decay depends  on the specificity of the model, namely on the nature of the particle, fermion or scalar. Therefore the final collider signatures and their corresponding branching ratios would be able to distinguish between singlets and doublets, and fermions or scalars ~\cite{Huitu:1996su,Aoki:2011yk,Chiang:2012dk}. In particular, the shape of the invariant mass distribution $m_{\ell \ell}$ for the final states will allow to distinguish between doubly-charged fermion or scalars in the initial state, while the scalar will exhibit a sharp peak around its original mass, the fermion will not display this peak \cite{Babu:2013ega}. 


\subsection{Parameter Space and Scale}

We will show that the atmospheric neutrino scale $m_{atm}=0.05\text{ eV}$ can be obtained for a benchmark point where we keep the numerical value of the Yukawas perturbative, $Y_{\nu}< \sqrt{4\,\pi}$. Furthermore the value of the cubic parameters,  $\mu$, is restricted by the one-loop induced quartic coupling ($\lambda_\text{eff}<1$) which is about the mass of the Higgs times a constant that depends on the masses hierarchy. For benchmark purposes we will use  $\mu \sim 4 m_h$~\cite{Babu:2002uu} where we consider the mass of all fields heavier than $m_h$. In this exercise we will determine the approximate scale of the messenger particles inside the loops.

After the Higgs acquires a VEV, $v=174$ GeV, the internal propagators connected to the Higgs leg will mix and that will reduce the number of internal messengers in the loops. All diagrams will have five messengers. Diagrams 1.g, 1.h, 2.a  and 2.b are special in the sense that the mixture creates three-state eigenvalues. Diagrams of Category~3 have an $H\,S_i\,S_j\,S_k$ vertex that does not create mixing, leaving only one vertex that does so (and reducing the number of messengers from 6 to 5). Diagrams of Category~4 have an $H\,H\,S_i\,S_j$ vertex that creates the mixing but since it is the only one that does so (the same as in Category~3) the internal messengers only get reduced by one unit. This has the consequence that most of these diagrams have the same generic integrals, with the exception of 1.g, 1.h, 2.g and 4.b (one can see that this is due to the position and composition of the external legs, compared to the other diagrams of the same category), although the integrals of these diagrams are still similar to the aforementioned.

We will use the results from Ref.~\cite{Sierra:2014rxa}, in it, all of the two-loop integrals needed for the masses were calculated in a model-independent fashion.

When calculating the neutrino masses we will always use the mass-eigenstate basis, in this basis these models will always have particles that get mixed by the Higgs in the external legs. Depending on the model there are two types of mixing angles, those that are proportional to a Yukawa, $Y$ (these angles mix two fermions), and the angles that mix two scalars (these are proportional to a cubic coupling $\mu$),
\begin{eqnarray}
\tan 2\theta^Y_{12} = \frac{2\; Y v}{M_{F_1}-M_{F_2}}, \; \;\;\;\;\; \tan 2\theta^\mu_{12} = \frac{2\; \mu\; v}{M_{S_1}^2-M_{S_2}^2}\,, 
\label{eq:mixangle}
\end{eqnarray}
where the indices $1,2$ refer to the two different mass eigenstates.

It is convenient to introduce a variable that parametrizes the degeneracy between two mass parameters, $M_2 = deg\times M_1$. Also, the scale of the new fields added is such that $M\gg\sqrt{2 \mu v}\gg 2 Y v$, generically speaking the value of the mixing angles will be very small, unless we consider that the mass eigenstates are quasidegenerate.

\begin{figure}[H]
	\centering
	\captionsetup[subfigure]{labelformat=empty}
	\includegraphics[scale=.5]{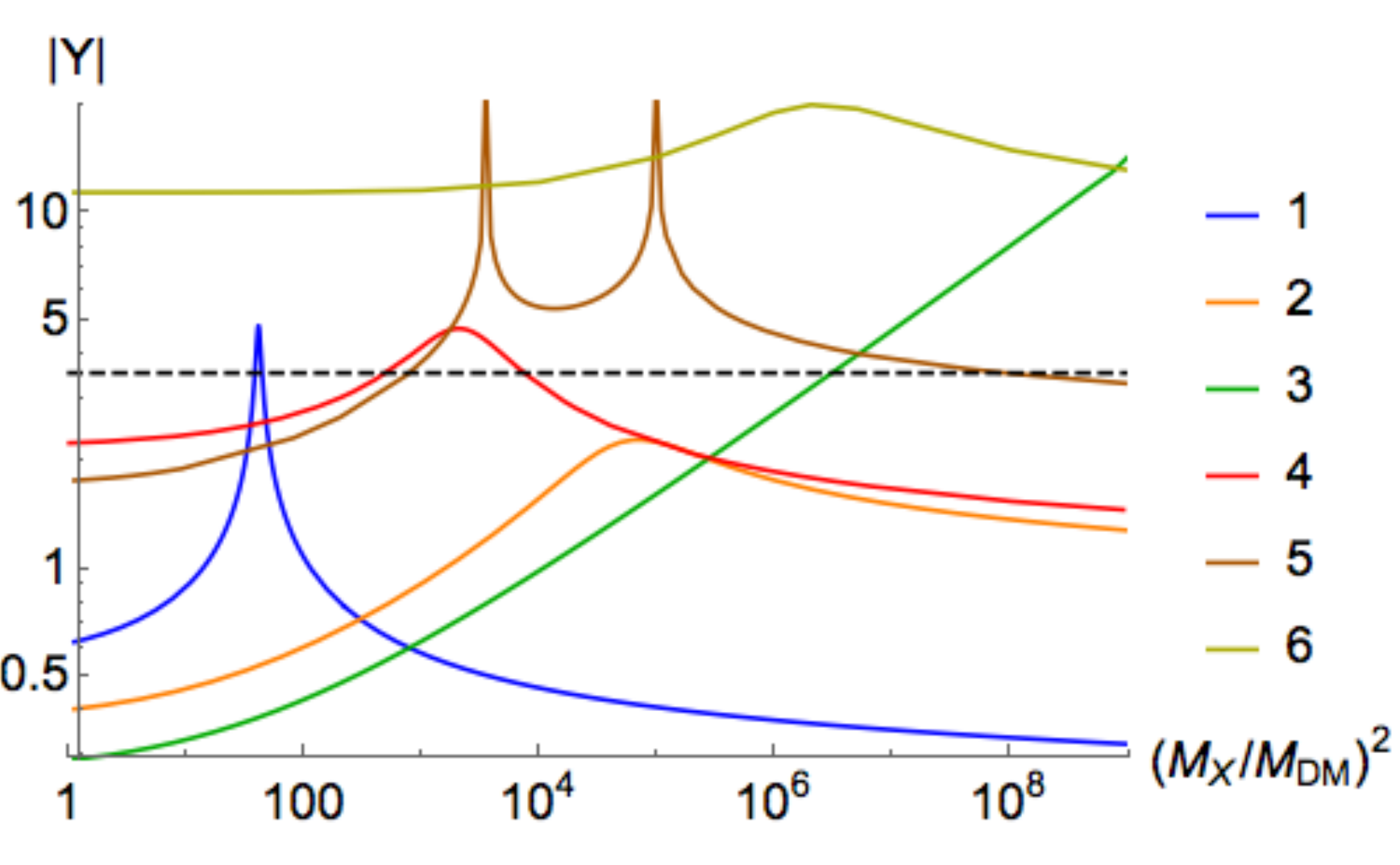}
	\caption{\label{fig:parspace} Absolute value of the generalized Yukawas as function of the ratio $M_X/M_{DM}$ squared. 
		The benchmark values are: $\textbf{1. } \left(M_{S_2}/M_{F_3}\right)^2=1$, $\left(M_{F_{45}}/M_{F_3}\right)^2=1.1$, $\left(M_{F_{17}}/M_{F_3}\right)^2=1.5$, $deg=5$,
		\textbf{2. }$\left(M_{S_2}/M_{F_3}\right)^2=10^4$, $\left(M_{F_{45}}/M_{F_3}\right)^2=1.1$, $\left(M_{F_{17}}/M_{F_3}\right)^2=1.5$, $deg=5$,
		\textbf{3. }$\left(M_{S_2}/M_{F_3}\right)^2=10^{12}$, $\left(M_{F_{45}}/M_{F_3}\right)^2=1.1$, $\left(M_{F_{17}}/M_{F_3}\right)^2=1.5$, $deg=5$,
		\textbf{4. }$\left(M_{S_2}/M_{F_3}\right)^2=1$, $\left(M_{F_{45}}/M_{F_3}\right)^2=1.1$, $\left(M_{F_{17}}/M_{F_3}\right)^2=1.5$, $deg=50$,
		\textbf{5. }$\left(M_{S_2}/M_{F_3}\right)^2=1$, $\left(M_{F_{45}}/M_{F_3}\right)^2=1.1\times 10^4$, $\left(M_{F_{17}}/M_{F_3}\right)^2=1.5$, $deg=5$,
		\textbf{6. }$\left(M_{S_2}/M_{F_3}\right)^2=10^4$, $\left(M_{F_{45}}/M_{F_3}\right)^2=1.1\times 10^4$, $\left(M_{F_{17}}/M_{F_3}\right)^2=1.5\times 10^4$, $deg=5$.}
\end{figure}

To illustrate the scale of the new physics required in two-loop neutrino mass models, we use class 1.e as an example. To keep the analysis manageable, we will use a simplified parameter space, where all Yukawas are set equal to each other, the cubic couplings are all equal and have a value of  $\mu=500$~GeV (this being a limit case, as explained above). The mass of the particle in the middle of the loop, i.e., the one with no interactions with charged leptons (which in most cases is the DM candidate) will be set to $M_\text{DM}=5000$~GeV, all other masses will be scaled to this one. Also, we will set both parameters $deg$ equal to each other and use values away from the degenerate and quasidegenerate cases.
Figure~\ref{fig:parspace} shows the value of the generalized Yukawa for different values of the messenger masses, parametrized by the ratio squared of the mass divided by the DM mass. Values above the dashed line will require Yukawas that are non perturbative, making this region of the parameter space unacceptable. All of the lines have peaks that are caused in the calculation by accidental cancellations of parameters. One can see from the general trend of the plots that one or two of the masses can have a very high scale (higher than PeV), as long as all other are kept at a TeV scale.

\subsection{Number of Generations}

After symmetry breaking there are two types of models: the ones where $M_{\nu}\sim Y^{ia}_{\alpha}Y^{ab}_{\beta}Y^{bj}_{\gamma} + Y^{ib}_{\gamma}Y^{ba}_{\beta}Y^{aj}_{\alpha}$ (1.a, 1.b, 1.d, 2.a, 2.c, 2.e, 3.a, 3.b, 3.c, 4.a), and the ones where $M_{\nu}\sim Y^{ia}_{\alpha}Y^{ab}_{\beta}Y^{bc}_{\gamma}Y^{cj}_{\delta} + Y^{ic}_{\delta}Y^{cb}_{\gamma}Y^{ba}_{\beta}Y^{aj}_{\alpha}$ (1.c, 1.e, 1.f, 1.g, 1.h, 2.b, 2.d, 2.f, 2.g, 4.b), where $i,j=1,2,3$ are the indices of the SM families, $a$, $b$ and $c$ depend on the number of families for the BSM fields and the Greek characters are just to mark that the Yukawas might correspond to different interactions.

For the first case, if there is one generation per BSM fields then $Y_{ab}$ is a constant $C$ and hence the neutrino matrix is given by $\left(M_{\nu}\right)^{ij} \sim C\,(Y^{iT}_{\alpha}\, Y^j_{\gamma}+ Y^{jT}_{\gamma}\, Y^i_{\alpha})$, in this scenario there are two possibilities, either $\alpha=\gamma$, which implies that $M_{\nu}$ only has one nonzero eigenvalue, or $\alpha \neq \gamma$ which means that $M_{\nu}$ has two nonzero eigenvalues.
In the second case the Yukawas $Y_{ab}$ and $Y_{bc}$ always connect three BSM fields and they must have the same dimensions, which translates into the fields having the same number of generations. If it is only one, then $Y_{ab}Y_{bc}$ is a constant, and similarly to  the previous case the neutrino mass is given by $\left(M_{\nu}\right)^{ij} \sim C\,(Y^{iT}_{\alpha}\, Y^j_{\delta}+ Y^{jT}_{\delta}\, Y^i_{\alpha})$, which has one or two non zero eigenvalues depending if $\alpha=\delta$ or not.

This can be resumed by the fact that if both Yukawa interactions for the external legs connecting the SM charge lepton with BSM fields are the same, then one needs at least two generations of each BSM field to be able to reproduce neutrino data, or three generations if one wishes to reproduce three nonzero neutrino masses. If these Yukawa interactions are different, then one BSM generation suffices (but two generations are necessary to have all three nonzero neutrino masses).

Higher loop diagrams could be use to generate the smallest of neutrino masses and therefore reduce the number of BSM generations, but this calculation is beyond the scope of this paper.

\section{Conclusions}
\label{sec:conclusion}
In this work we provided a list of genuine two-loop models for neutrino mass generation where one of the internal messengers is a dark matter candidate. We obtained these two-loop realizations of the Weinberg operator by adding only singlets or doublets of $\mathsf{SU(2)_L}$ to the SM. 
Given the bounds set by direct detection experiments, DM will be either fermionic or scalar singlets with null hypercharge, or a scalar doublet with $Y=1$ (or a mixture of the latter two). Moreover, we ensured the stability of the DM candidate by the introduction of a $\mathsf{Z_2}$ symmetry. 

As it was shown, there are 20 different diagrams that generate genuine two-loop neutrino masses via a Weinberg operator, one of these diagrams can not accommodate simultaneously the restrictions on the fields and the existence of a DM candidate as a messenger leaving 19 different diagrams with results. Therefore we found in total 141 different sets which fill our constrains, these models are generated via a total of 19 different diagrams which we classify into two types, those with seven particles and those with six-particles, with two subsections each, called categories. The results are compiled in Tables~\ref{tab:res1a}--\ref{tab:res1g} for Category~1, Tables~\ref{tab:res2a}--\ref{tab:res2g} for Category~2, Tables~\ref{tab:res3a}--\ref{tab:res3c} for Category~3 and finally in Tables~\ref{tab:res4a} and~\ref{tab:res4b} for Category~4. 

A short discussion on some phenomenological aspects of the models has been presented in section~\ref{sec:pheno}, it was mentioned the possibility of these models to have FCNCs if a scalar $(2,1,+)$, under $\mathsf{SU(2)_L \times U(1)_Y \times Z_2}$, is present. We also mentioned how the models in this paper can be discriminated by the use of doubly-charged particle signatures. Furthermore, we included a short comment regarding the possibility of extending these models with $\mathsf{SU(2)_L}$ triplets; this is simple to do but increases the number of models presented to an unpractical amount. We noted the fact that most models only need one generation of BSM fields to recreate neutrino data, but two generations might be necessary in some specific cases or if one wants to have three non-zero eigenvalues for the neutrino masses.

Finally, we showed that for a particular set of parameters all the models fit the atmospheric neutrino mass whose latest experimental result is placed around $m_{atm}=0.05$~eV~\cite{Forero:2014bxa}. It was not our intention to show all the parameter space, nevertheless, given the multiple free parameters of the models as well as the solutions we did find, we expect a reasonable parameter space that allows for realistic neutrino mass spectra and a viable DM candidate.

\appendix
\section{Integrals}
\label{sec:app}

In this appendix we explicitly write the integrals appearing in section~\ref{sec:results}. The integrals are specified by the indices $\chi$, $\rho$, $\eta$ and $\gamma$. 

For diagrams 1.a-1.f, 2.a-2.b, 3.a-3.b and 4.a the integral is of the form
\begin{equation}
\label{eq:integralT1}
I^{\zeta}_\chi=\int d^4k \int d^4q\, \frac{\zeta}{\left(q^2-m^2_x\right)\left(q^2-m^2_y\right)(\left(q+k\right)^2-m^2_z)\left(k^2-m^2_r\right)\left(k^2-m^2_s\right)}\,,
\end{equation}
with $\chi\equiv\{x,\,y,\,z,\,r,\,s\}$.

For diagrams 2.c-2.f the integral is given by
\begin{equation}
\label{eq:integralT2}
I^{\zeta}_\rho=\int d^4k \int d^4q\, \frac{\zeta}{\left(q^2-m^2_x\right)(\left(q+k\right)^2-m^2_y)(\left(q+k\right)^2-m^2_z)\left(k^2-m^2_r\right)\left(k^2-m^2_s\right)}\,,
\end{equation}
with $\rho\equiv \{x,\,y,\,z,\,r,\,s\}$ while for diagram 2.g the integral reads as
\begin{equation}
\label{eq:integralT3}
I^{\zeta}_\eta=\int d^4k \int d^4q\, \frac{\zeta}{\left(q^2-m^2_x\right)(\left(q+k\right)^2-m^2_y)(\left(q+k\right)^2-m^2_z)(\left(q+k\right)^2-m^2_r)\left(k^2-m^2_s\right)}\,,
\end{equation}
with $\eta\equiv \{x,\,y,\,z,\,r,\,s\}$. Finally for diagram 1.g, 1.h and 4.b it is given by
\begin{equation}
\label{eq:integralT4}
I^{\zeta}_\gamma=\int d^4k \int d^4q\, \frac{\zeta}{\left(q^2-m^2_x\right)(\left(q+k\right)^2-m^2_y)\left(k^2-m^2_z\right)\left(k^2-m^2_r\right)\left(k^2-m^2_s\right)}\,,
\end{equation}
with $\gamma\equiv \{x,\,y,\,z,\,r,\,s\}$, where $\zeta$ refers to 1, $k^2$, $q^2$ and $(q+k)^2$.

The solution to these integrals is computed in a model-independent fashion explicitly in~\cite{Sierra:2014rxa}.

\section{Examples}
\label{sec:appB}
\subsection{$\mathsf{Z_2}$ symmetry to forbid one-loop contributions}

Diagrams 1.a, 1.b and 4.b have bilateral symmetry, i.e., if one draws a vertical line right through the middle of the diagram there will be two mirror images on both sides of the lines.  In these three cases it is not necessary to consider the $\mathsf{Z_2}$  assignments A and B of Table~\ref{tab:assignmentscategory1}, and only one suffices. Diagrams 1.c, 1.e and 1.f also have a bilateral symmetry over an inclined line that also reduce the number of possible models. Nevertheless, that is not the case for all other models, and the specific $\mathsf{Z_2}$ assignment is not arbitrary and have real physical consequences for the models. 

To explain this in more detail we use an example, with no particular reason, we choose a model from diagram 1.d, with $\alpha=-2$ and $\beta=-1$, as well as  $ii$ under $\mathsf{SU(2)_L}$ assignment as given by Table \ref{fig:category1}.  With these quantum numbers we look at two different models: In one we assign $\mathsf{Z_2}$ charge to the fields in the left loop ($\mathsf{Z_2}$ assignment A), and for a second model the $\mathsf{Z_2}$ charges are for the right loop ($\mathsf{Z_2}$ assignment B).  In Table \ref{tab:example} we show the particle content for these models.

\begin{table}[H]
	\centering
	\begin{tabular}{|c|c|c|c|c|}
		\hline 
		& $\mathsf{SU(2)_L}$ & $\mathsf{U(1)_Y}$ & $\mathsf{Z_2}$ (model 1) & $\mathsf{Z_2}$ (model 2)\\
		\hline 
		\hline 
		\multirow{3}{*}{Fermions} & 1 & -2 & $\boldsymbol{-}$ & $\boldsymbol{+}$\\
		\cline{2-5} 
		& 2 & -1 & $\boldsymbol{+}$ & $\boldsymbol{-}$\\
		\cline{2-5} 
		& 1 & -2 & $\boldsymbol{+}$ & $\boldsymbol{-}$\\
		\hline 
		\multirow{3}{*}{Scalars}  & 2 & -3 & $\boldsymbol{-} $& $\boldsymbol{+}$ \tabularnewline
		\cline{2-5} 
		& 1 & -2 & $\boldsymbol{-}$ & $\boldsymbol{+}$\\
		\cline{2-5} 
		& 2 & -1 & $\boldsymbol{-}$ & $\boldsymbol{-}$\\
		\cline{2-5} 
		& 2 & -1 & $\boldsymbol{+}$ & $\boldsymbol{-}$\\
		\hline 
	\end{tabular}
	\caption{The quantum numbers for a set of particles to exemplify the importance of $\mathsf{Z_2}$ charge assignment: model~1 corresponds to assignment A while model~2 corresponds to B.}
	\label{tab:example}
\end{table}

\begin{figure}[H]
	\captionsetup[subfigure]{labelformat=empty}
	\hspace{0cm}
	\begin{subfigure}{0.2\textwidth}	
		\includegraphics[scale=0.3]{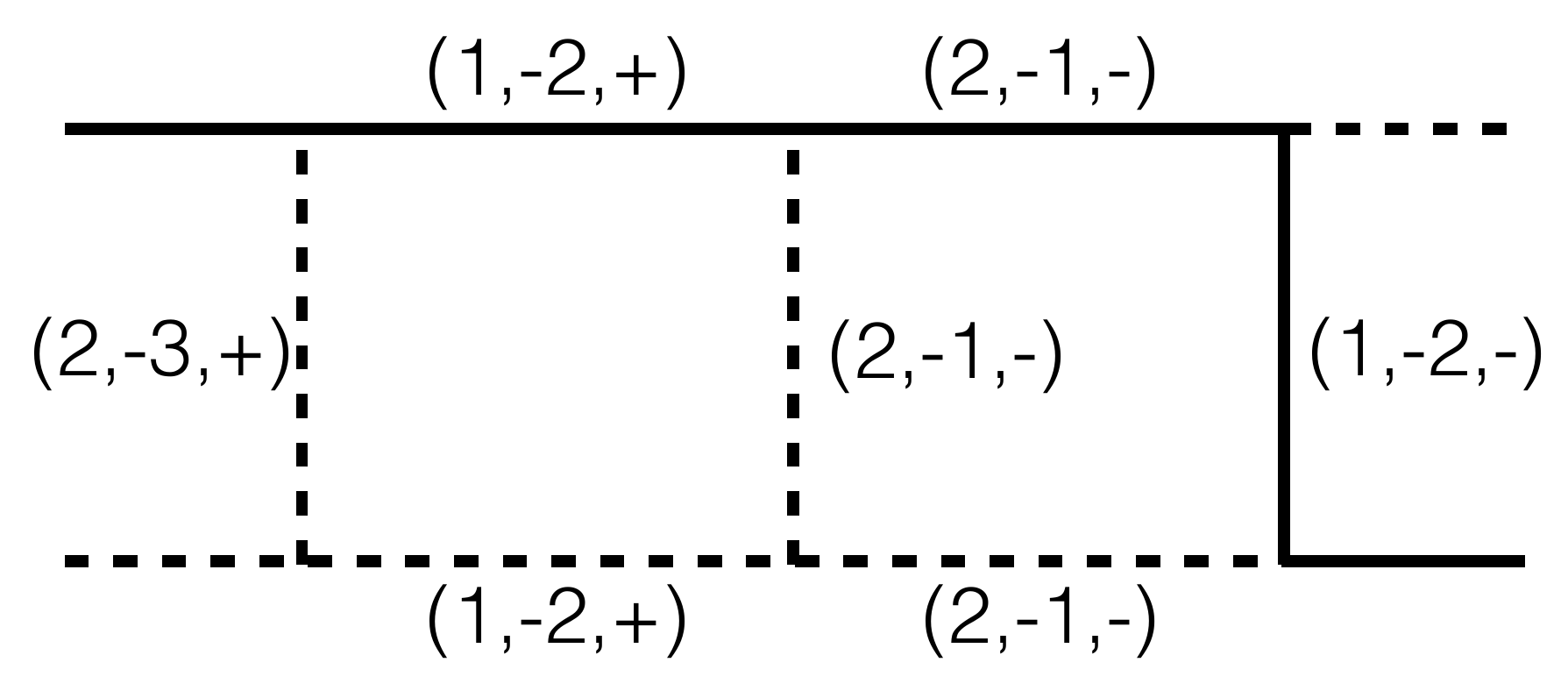}		
		\caption{\label{fig:appb}\hspace{1.9cm}  (a) }
	\end{subfigure}   \hspace{0.9cm}
	\hspace{1.5cm}
	\vspace{.1cm}
	\begin{subfigure}{0.2\textwidth}
		\includegraphics[scale=0.3]{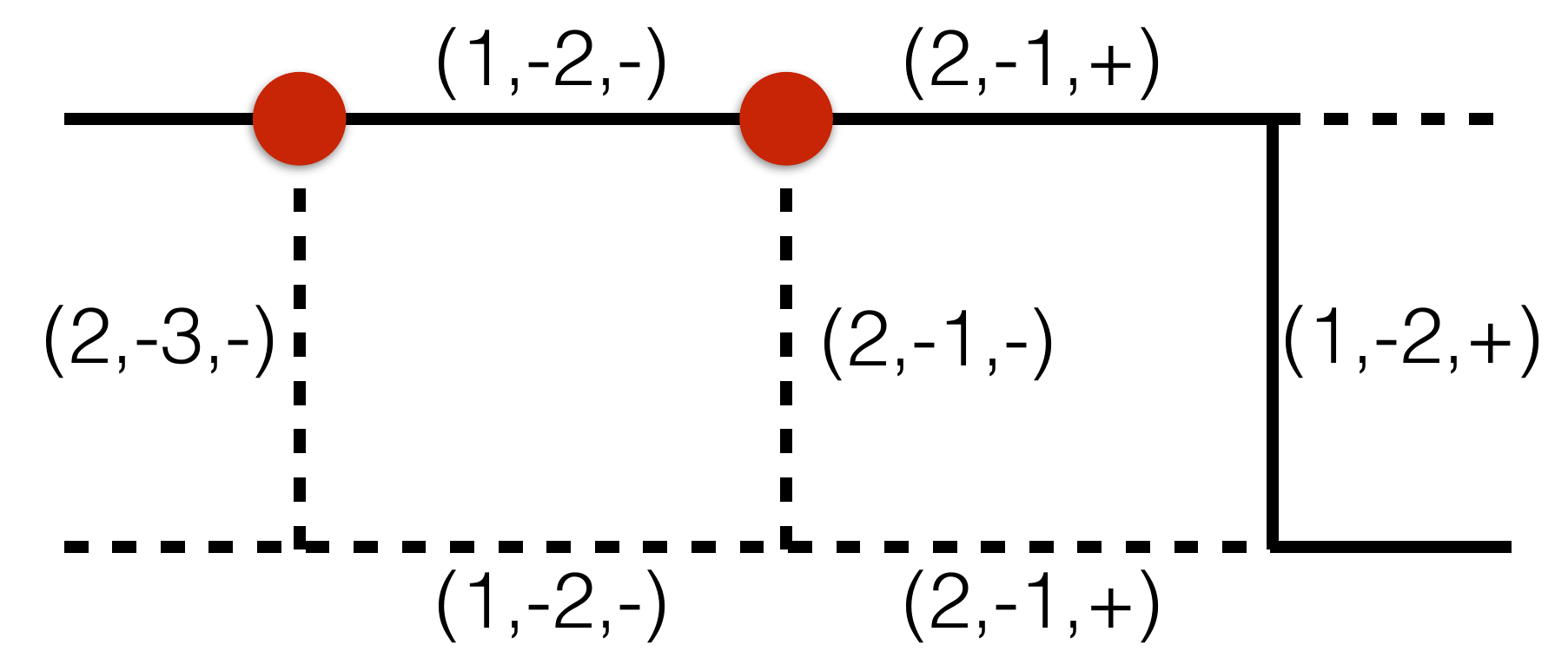}		
		\caption{\hspace{1.9cm} (b) }
	\end{subfigure} 
	\hspace{2.25cm}  
	\vspace{.5cm}
	\begin{subfigure}{0.2\textwidth}
		\includegraphics[scale=0.27]{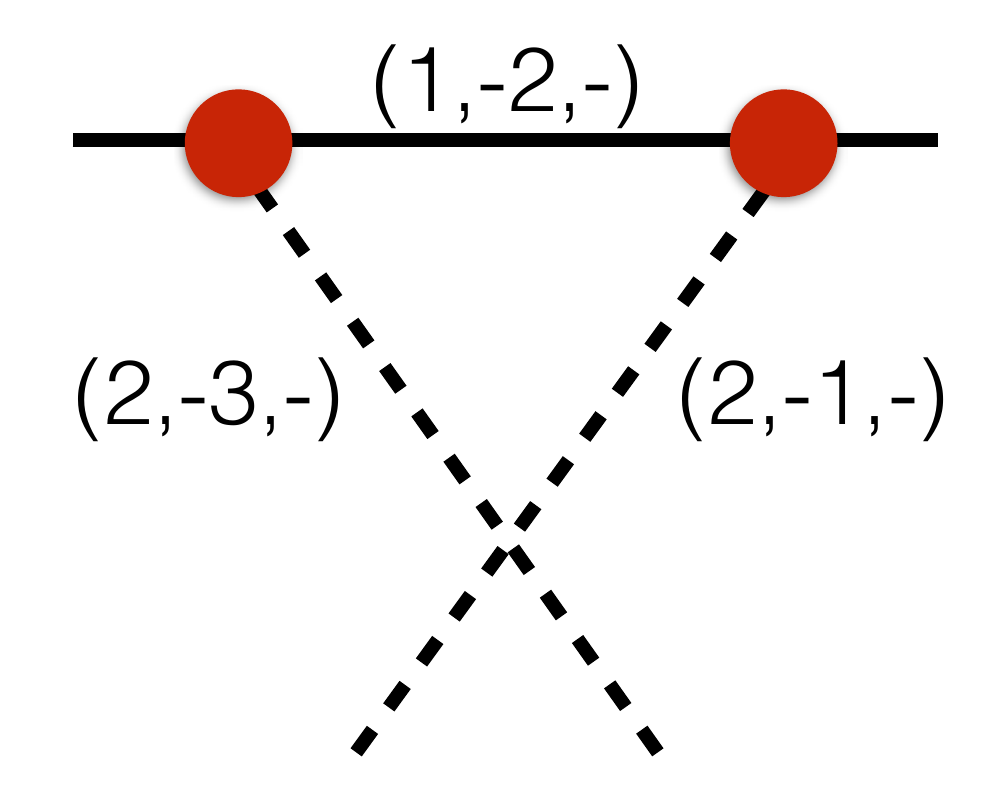}		
		\caption{\hspace{0cm} (c)}
	\end{subfigure} \hspace{0.8cm}
	\caption{\label{fig:example1} Example of two different models to explain the importance of $\mathsf{Z_2}$ to forbid one-loop contributions: (a)  two-loop contribution from model~1 of Table~\ref{tab:example} ($\mathsf{Z_2}$ assignment A), (b) two-loop contribution from model~2 of Table~\ref{tab:example} ($\mathsf{Z_2}$ assignment B), (c) one-loop contribution obtained from model~2 ($\mathsf{Z_2}$ assignment B).}
\end{figure}

In Figure \ref{fig:example1} we show the Feynman diagrams for the Majorana neutrino masses. In the upper panel we show the main contribution given by a two-loop diagram 1.d, while in the bottom panel we show the two main contributions, in red we marked the interactions in the two-loop contribution that can be used to generate a one-loop. One can clearly see that if one wishes to search for all possible models then the $\mathsf{Z_2}$ assignment for the fields is important and one can not just assign charges to one-loop.

Hence, as can be seen by this example the $\mathsf{Z_2}$ not only stabilizes a DM candidate but also helps to forbid contributions from one-loop or tree level.

\subsection{A model with DM that does not contribute to neutrino mass}

\begin{figure}[H]
	\includegraphics[scale=0.44]{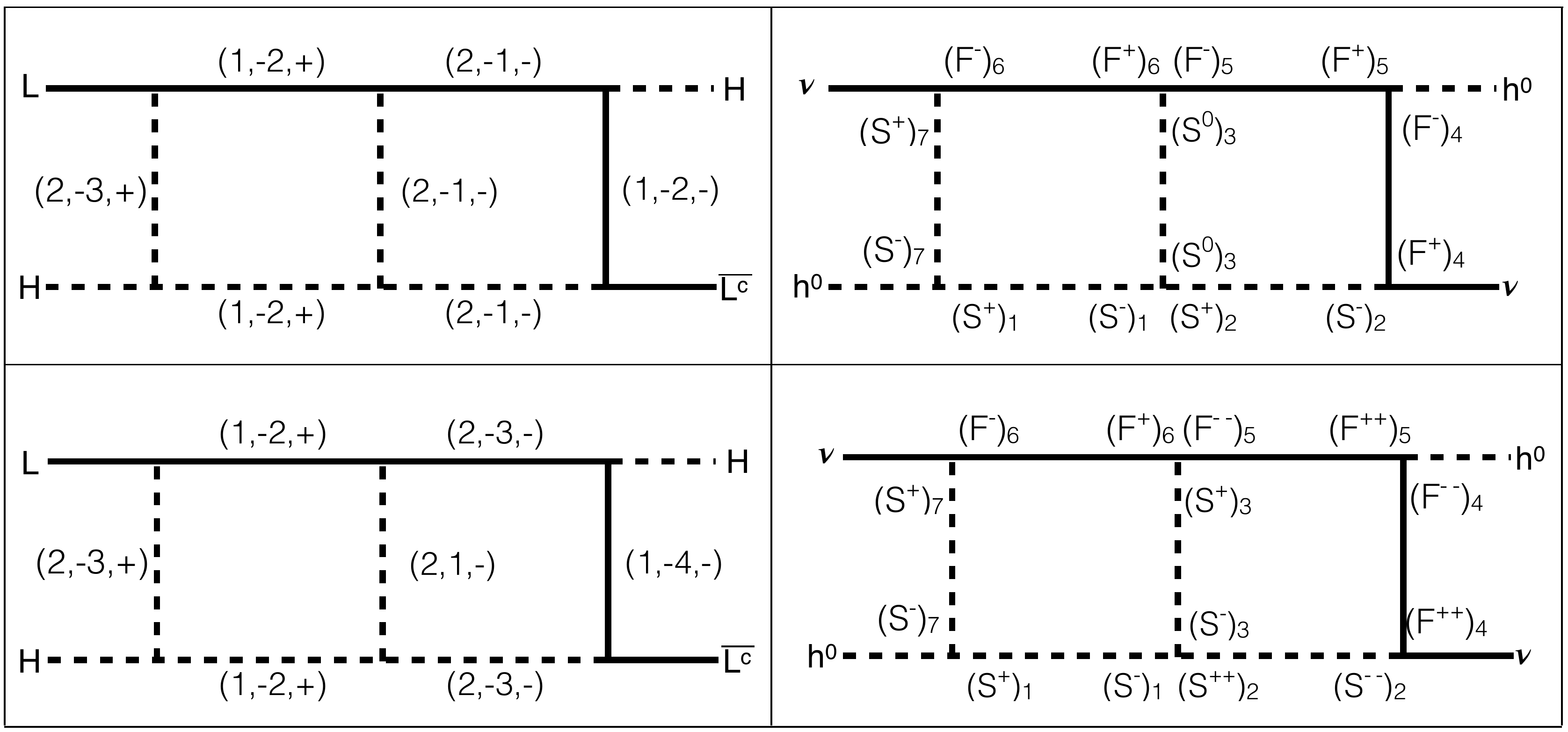}		
	\caption{\label{fig:example2} Two models for two-loop diagram of Class 1.d: on the left-hand side we present the two-loop filled with the generic particles ($Y=-1$ upper panel, $Y=1$ lower panel) while on right-hand side we specify in terms of their components.}
\end{figure}

In section \ref{sec:results} it was mentioned that:  "[...]we have excluded all of the models that, while theoretically allow for a  genuine two-loop neutrino mass and have a field that contains a DM candidate, they do not have an electric neutral particle in the loop[...]". Since this idea is difficult to grasp, we will explain it by using an example. In Figure \ref{fig:example2} we show two models both generate a Majorana neutrino mass at a genuine two-loop level with a diagram of the type 1.d, the difference between them is that for the second one we have rescaled the hypercharge of one of the loops by 2. On the left we show the diagrams constructed with the fields and their $\mathsf{SU(2)_L \times U(1)_Y} \times \mathsf{Z_2}$ numbers. On the right we show the same model and diagram but with the explicit field components for each vertex of the diagram. As can be seen in both models there is a DM candidate given by a doublet scalar with hypercharge $Y=1$, nevertheless in the first model the neutral component of the scalar doublet is the one that participates in the loop, while in the second model is the charged component of this doublet. Therefore, the second model does not have an explicit DM candidate contributing to the neutrino mass.

\acknowledgments

We thank Jean-Rene Cudell, Atri Bhattacharya and Diego Aristizabal for interesting discussions and comments.
This work was supported by the ``Fonds de la Recherche Scientifique-FNRS'' under 
grant number 4.4501.15. The work of C.S. was supported by the ``Universit\'e de Li\`ege'' 
and the EU in the context of the MSCA-COFUND-BeIPD project.


\bibliographystyle{JHEP}
\bibliography{refs}

\providecommand{\href}[2]{#2}\begingroup\raggedright\begin{thebibliography}{100}

\bibitem{Forero:2014bxa}
D.~V. Forero, M.~T\'ortola and J.~W.~F. Valle, \emph{{Neutrino oscillations
  refitted}}, \href{http://dx.doi.org/10.1103/PhysRevD.90.093006}{\emph{Phys.
  Rev. D} {\bf 90} (2014) 093006}, [\href{http://arxiv.org/abs/1405.7540}{{\tt
  1405.7540}}].

\bibitem{Ahmed:2009zw}
{\scshape CDMS-II} collaboration, Z.~Ahmed et~al., \emph{{Dark Matter Search
  Results from the CDMS II Experiment}},
  \href{http://dx.doi.org/10.1126/science.1186112}{\emph{Science} {\bf 327}
  (2010) 1619--1621}, [\href{http://arxiv.org/abs/0912.3592}{{\tt 0912.3592}}].

\bibitem{Aprile:2012nq}
{\scshape XENON100} collaboration, E.~Aprile et~al., \emph{{Dark Matter Results
  from 225 Live Days of XENON100 Data}},
  \href{http://dx.doi.org/10.1103/PhysRevLett.109.181301}{\emph{Phys. Rev.
  Lett.} {\bf 109} (2012) 181301}, [\href{http://arxiv.org/abs/1207.5988}{{\tt
  1207.5988}}].

\bibitem{Akerib:2015rjg}
{\scshape LUX} collaboration, D.~S. Akerib et~al., \emph{{Improved Limits on
  Scattering of Weakly Interacting Massive Particles from Reanalysis of 2013
  LUX Data}},
  \href{http://dx.doi.org/10.1103/PhysRevLett.116.161301}{\emph{Phys. Rev.
  Lett.} {\bf 116} (2016) 161301}, [\href{http://arxiv.org/abs/1512.03506}{{\tt
  1512.03506}}].

\bibitem{Ade:2013zuv}
{\scshape Planck} collaboration, P.~A.~R. Ade et~al., \emph{{Planck 2013
  results. XVI. Cosmological parameters}},
  \href{http://dx.doi.org/10.1051/0004-6361/201321591}{\emph{Astron.
  Astrophys.} {\bf 571} (2014) A16},
  [\href{http://arxiv.org/abs/1303.5076}{{\tt 1303.5076}}].

\bibitem{Krauss:2002px}
L.~M. Krauss, S.~Nasri and M.~Trodden, \emph{{A Model for neutrino masses and
  dark matter}},
  \href{http://dx.doi.org/10.1103/PhysRevD.67.085002}{\emph{Phys. Rev.} {\bf
  D67} (2003) 085002}, [\href{http://arxiv.org/abs/hep-ph/0210389}{{\tt
  hep-ph/0210389}}].

\bibitem{Hirsch:2004he}
M.~Hirsch and J.~W.~F. Valle, \emph{{Supersymmetric origin of neutrino mass}},
  \href{http://dx.doi.org/10.1088/1367-2630/6/1/076}{\emph{New J. Phys.} {\bf
  6} (2004) 76}, [\href{http://arxiv.org/abs/hep-ph/0405015}{{\tt
  hep-ph/0405015}}].

\bibitem{Aoki:2008av}
M.~Aoki, S.~Kanemura and O.~Seto, \emph{{Neutrino mass, Dark Matter and Baryon
  Asymmetry via TeV-Scale Physics without Fine-Tuning}},
  \href{http://dx.doi.org/10.1103/PhysRevLett.102.051805}{\emph{Phys. Rev.
  Lett.} {\bf 102} (2009) 051805}, [\href{http://arxiv.org/abs/0807.0361}{{\tt
  0807.0361}}].

\bibitem{Sierra:2008wj}
D.~Aristizabal~Sierra, J.~Kubo, D.~Restrepo, D.~Suematsu and O.~Zapata,
  \emph{{Radiative seesaw: Warm dark matter, collider and lepton flavour
  violating signals}},
  \href{http://dx.doi.org/10.1103/PhysRevD.79.013011}{\emph{Phys. Rev.} {\bf
  D79} (2009) 013011}, [\href{http://arxiv.org/abs/0808.3340}{{\tt
  0808.3340}}].

\bibitem{Asaka:2005pn}
T.~Asaka and M.~Shaposhnikov, \emph{{The nuMSM, dark matter and baryon
  asymmetry of the universe}},
  \href{http://dx.doi.org/10.1016/j.physletb.2005.06.020}{\emph{Phys. Lett.}
  {\bf B620} (2005) 17--26}, [\href{http://arxiv.org/abs/hep-ph/0505013}{{\tt
  hep-ph/0505013}}].

\bibitem{Restrepo:2015ura}
D.~Restrepo, A.~Rivera, M.~Sánchez-Peláez, O.~Zapata and W.~Tangarife,
  \emph{{Radiative Neutrino Masses in the Singlet-Doublet Fermion Dark Matter
  Model with Scalar Singlets}},
  \href{http://dx.doi.org/10.1103/PhysRevD.92.013005}{\emph{Phys. Rev.} {\bf
  D92} (2015) 013005}, [\href{http://arxiv.org/abs/1504.07892}{{\tt
  1504.07892}}].

\bibitem{Klasen:2013jpa}
M.~Klasen, C.~E. Yaguna, J.~D. Ruiz-Alvarez, D.~Restrepo and O.~Zapata,
  \emph{{Scalar dark matter and fermion coannihilations in the radiative seesaw
  model}}, \href{http://dx.doi.org/10.1088/1475-7516/2013/04/044}{\emph{JCAP}
  {\bf 1304} (2013) 044}, [\href{http://arxiv.org/abs/1302.5298}{{\tt
  1302.5298}}].

\bibitem{Wang:2015saa}
W.~Wang and Z.-L. Han, \emph{{Radiative linear seesaw model, dark matter, and
  $U(1)_{B-L}$}},
  \href{http://dx.doi.org/10.1103/PhysRevD.92.095001}{\emph{Phys. Rev.} {\bf
  D92} (2015) 095001}, [\href{http://arxiv.org/abs/1508.00706}{{\tt
  1508.00706}}].

\bibitem{Merle:2016scw}
A.~Merle, M.~Platscher, N.~Rojas, J.~W.~F. Valle and A.~Vicente,
  \emph{{Consistency of WIMP Dark Matter as radiative neutrino mass
  messenger}}, \href{http://dx.doi.org/10.1007/JHEP07(2016)013}{\emph{JHEP}
  {\bf 07} (2016) 013}, [\href{http://arxiv.org/abs/1603.05685}{{\tt
  1603.05685}}].

\bibitem{Ma:2006km}
E.~Ma, \emph{{Verifiable radiative seesaw mechanism of neutrino mass and dark
  matter}}, \href{http://dx.doi.org/10.1103/PhysRevD.73.077301}{\emph{Phys.
  Rev.} {\bf D73} (2006) 077301},
  [\href{http://arxiv.org/abs/hep-ph/0601225}{{\tt hep-ph/0601225}}].

\bibitem{Ma:1998dn}
E.~Ma, \emph{{Pathways to naturally small neutrino masses}},
  \href{http://dx.doi.org/10.1103/PhysRevLett.81.1171}{\emph{Phys. Rev. Lett.}
  {\bf 81} (1998) 1171--1174}, [\href{http://arxiv.org/abs/hep-ph/9805219}{{\tt
  hep-ph/9805219}}].

\bibitem{Angel:2012ug}
P.~W. Angel, N.~L. Rodd and R.~R. Volkas, \emph{{Origin of neutrino masses at
  the LHC: $\Delta L = 2$ effective operators and their ultraviolet
  completions}},
  \href{http://dx.doi.org/10.1103/PhysRevD.87.073007}{\emph{Phys. Rev.} {\bf
  D87} (2013) 073007}, [\href{http://arxiv.org/abs/1212.6111}{{\tt
  1212.6111}}].

\bibitem{Weinberg:1979sa}
S.~Weinberg, \emph{{Baryon and Lepton Nonconserving Processes}},
  \href{http://dx.doi.org/10.1103/PhysRevLett.43.1566}{\emph{Phys. Rev. Lett.}
  {\bf 43} (1979) 1566--1570}.

\bibitem{Lehman:2014jma}
L.~Lehman, \emph{{Extending the Standard Model Effective Field Theory with the
  Complete Set of Dimension-7 Operators}},
  \href{http://dx.doi.org/10.1103/PhysRevD.90.125023}{\emph{Phys. Rev.} {\bf
  D90} (2014) 125023}, [\href{http://arxiv.org/abs/1410.4193}{{\tt
  1410.4193}}].

\bibitem{Babu:2001ex}
K.~S. Babu and C.~N. Leung, \emph{{Classification of effective neutrino mass
  operators}},
  \href{http://dx.doi.org/10.1016/S0550-3213(01)00504-1}{\emph{Nucl. Phys.}
  {\bf B619} (2001) 667--689}, [\href{http://arxiv.org/abs/hep-ph/0106054}{{\tt
  hep-ph/0106054}}].

\bibitem{Bonnet:2009ej}
F.~Bonnet, D.~Hernandez, T.~Ota and W.~Winter, \emph{{Neutrino masses from
  higher than d=5 effective operators}},
  \href{http://dx.doi.org/10.1088/1126-6708/2009/10/076}{\emph{JHEP} {\bf 10}
  (2009) 076}, [\href{http://arxiv.org/abs/0907.3143}{{\tt 0907.3143}}].

\bibitem{Liao:2016qyd}
Y.~Liao and X.-D. Ma, \emph{{Operators up to Dimension Seven in Standard Model
  Effective Field Theory Extended with Sterile Neutrinos}},
  \href{http://arxiv.org/abs/1612.04527}{{\tt 1612.04527}}.

\bibitem{delAguila:2008cj}
F.~del Aguila and J.~A. Aguilar-Saavedra, \emph{{Distinguishing seesaw models
  at LHC with multi-lepton signals}},
  \href{http://dx.doi.org/10.1016/j.nuclphysb.2008.12.029}{\emph{Nucl. Phys.}
  {\bf B813} (2009) 22--90}, [\href{http://arxiv.org/abs/0808.2468}{{\tt
  0808.2468}}].

\bibitem{Minkowski:1977sc}
P.~Minkowski, \emph{{$\mu \to e\gamma$ at a Rate of One Out of $10^{9}$ Muon
  Decays?}}, \href{http://dx.doi.org/10.1016/0370-2693(77)90435-X}{\emph{Phys.
  Lett.} {\bf B67} (1977) 421--428}.

\bibitem{Yanagida:1979as}
T.~Yanagida, \emph{{Horizontal Symmetry and Masses of Neutrinos}}, {\emph{Conf.
  Proc.} {\bf C7902131} (1979) 95--99}.

\bibitem{GellMann:1980vs}
M.~Gell-Mann, P.~Ramond and R.~Slansky, \emph{{Complex Spinors and Unified
  Theories}}, {\emph{Conf. Proc.} {\bf C790927} (1979) 315--321},
  [\href{http://arxiv.org/abs/1306.4669}{{\tt 1306.4669}}].

\bibitem{Mohapatra:1979ia}
R.~N. Mohapatra and G.~Senjanovic, \emph{{Neutrino Mass and Spontaneous Parity
  Violation}}, \href{http://dx.doi.org/10.1103/PhysRevLett.44.912}{\emph{Phys.
  Rev. Lett.} {\bf 44} (1980) 912}.

\bibitem{Schechter:1980gr}
J.~Schechter and J.~W.~F. Valle, \emph{{Neutrino Masses in $SU(2)\times U(1)$
  Theories}}, \href{http://dx.doi.org/10.1103/PhysRevD.22.2227}{\emph{Phys.
  Rev.} {\bf D22} (1980) 2227}.

\bibitem{Magg:1980ut}
M.~Magg and C.~Wetterich, \emph{{Neutrino Mass Problem and Gauge Hierarchy}},
  \href{http://dx.doi.org/10.1016/0370-2693(80)90825-4}{\emph{Phys. Lett.} {\bf
  B94} (1980) 61--64}.

\bibitem{Mohapatra:1980yp}
R.~N. Mohapatra and G.~Senjanovic, \emph{{Neutrino Masses and Mixings in Gauge
  Models with Spontaneous Parity Violation}},
  \href{http://dx.doi.org/10.1103/PhysRevD.23.165}{\emph{Phys. Rev.} {\bf D23}
  (1981) 165}.

\bibitem{Cheng:1980qt}
T.~P. Cheng and L.-F. Li, \emph{{Neutrino Masses, Mixings and Oscillations in
  $SU(2)\times U(1)$ Models of Electroweak Interactions}},
  \href{http://dx.doi.org/10.1103/PhysRevD.22.2860}{\emph{Phys. Rev.} {\bf D22}
  (1980) 2860}.

\bibitem{Perez:2008ha}
P.~Fileviez~Perez, T.~Han, G.-y. Huang, T.~Li and K.~Wang, \emph{{Neutrino
  Masses and the CERN LHC: Testing Type II Seesaw}},
  \href{http://dx.doi.org/10.1103/PhysRevD.78.015018}{\emph{Phys. Rev.} {\bf
  D78} (2008) 015018}, [\href{http://arxiv.org/abs/0805.3536}{{\tt
  0805.3536}}].

\bibitem{Foot:1988aq}
R.~Foot, H.~Lew, X.~G. He and G.~C. Joshi, \emph{{Seesaw Neutrino Masses
  Induced by a Triplet of Leptons}},
  \href{http://dx.doi.org/10.1007/BF01415558}{\emph{Z. Phys.} {\bf C44} (1989)
  441}.

\bibitem{Franceschini:2008pz}
R.~Franceschini, T.~Hambye and A.~Strumia, \emph{{Type-III see-saw at LHC}},
  \href{http://dx.doi.org/10.1103/PhysRevD.78.033002}{\emph{Phys. Rev.} {\bf
  D78} (2008) 033002}, [\href{http://arxiv.org/abs/0805.1613}{{\tt
  0805.1613}}].

\bibitem{Bonnet:2012kz}
F.~Bonnet, M.~Hirsch, T.~Ota and W.~Winter, \emph{{Systematic study of the d=5
  Weinberg operator at one-loop order}},
  \href{http://dx.doi.org/10.1007/JHEP07(2012)153}{\emph{JHEP} {\bf 07} (2012)
  153}, [\href{http://arxiv.org/abs/1204.5862}{{\tt 1204.5862}}].

\bibitem{Restrepo:2013aga}
D.~Restrepo, O.~Zapata and C.~E. Yaguna, \emph{{Models with radiative neutrino
  masses and viable dark matter candidates}},
  \href{http://dx.doi.org/10.1007/JHEP11(2013)011}{\emph{JHEP} {\bf 11} (2013)
  011}, [\href{http://arxiv.org/abs/1308.3655}{{\tt 1308.3655}}].

\bibitem{Sierra:2014rxa}
D.~Aristizabal~Sierra, A.~Degee, L.~Dorame and M.~Hirsch, \emph{{Systematic
  classification of two-loop realizations of the Weinberg operator}},
  \href{http://dx.doi.org/10.1007/JHEP03(2015)040}{\emph{JHEP} {\bf 03} (2015)
  040}, [\href{http://arxiv.org/abs/1411.7038}{{\tt 1411.7038}}].

\bibitem{Dolgov:2013una}
A.~D. Dolgov, S.~L. Dubovsky, G.~I. Rubtsov and I.~I. Tkachev,
  \emph{{Constraints on millicharged particles from Planck data}},
  \href{http://dx.doi.org/10.1103/PhysRevD.88.117701}{\emph{Phys. Rev.} {\bf
  D88} (2013) 117701}, [\href{http://arxiv.org/abs/1310.2376}{{\tt
  1310.2376}}].

\bibitem{DelNobile:2015bqo}
E.~Del~Nobile, M.~Nardecchia and P.~Panci, \emph{{Millicharge or Decay: A
  Critical Take on Minimal Dark Matter}},
  \href{http://dx.doi.org/10.1088/1475-7516/2016/04/048}{\emph{JCAP} {\bf 1604}
  (2016) 048}, [\href{http://arxiv.org/abs/1512.05353}{{\tt 1512.05353}}].

\bibitem{Agrawal:2016quu}
P.~Agrawal, F.-Y. Cyr-Racine, L.~Randall and J.~Scholtz, \emph{{Make Dark
  Matter Charged Again}},  \href{http://arxiv.org/abs/1610.04611}{{\tt
  1610.04611}}.

\bibitem{Foot:2014uba}
R.~Foot and S.~Vagnozzi, \emph{{Dissipative hidden sector dark matter}},
  \href{http://dx.doi.org/10.1103/PhysRevD.91.023512}{\emph{Phys. Rev.} {\bf
  D91} (2015) 023512}, [\href{http://arxiv.org/abs/1409.7174}{{\tt
  1409.7174}}].

\bibitem{Abe:2014gua}
T.~Abe, R.~Kitano and R.~Sato, \emph{{Discrimination of dark matter models in
  future experiments}},
  \href{http://dx.doi.org/10.1103/PhysRevD.91.095004}{\emph{Phys. Rev.} {\bf
  D91} (2015) 095004}, [\href{http://arxiv.org/abs/1411.1335}{{\tt
  1411.1335}}].

\bibitem{Ibarra:2013zia}
A.~Ibarra, A.~S. Lamperstorfer and J.~Silk, \emph{{Dark matter annihilations
  and decays after the AMS-02 positron measurements}},
  \href{http://dx.doi.org/10.1103/PhysRevD.89.063539}{\emph{Phys. Rev.} {\bf
  D89} (2014) 063539}, [\href{http://arxiv.org/abs/1309.2570}{{\tt
  1309.2570}}].

\bibitem{Rott:2014kfa}
C.~Rott, K.~Kohri and S.~C. Park, \emph{{Superheavy dark matter and IceCube
  neutrino signals: Bounds on decaying dark matter}},
  \href{http://dx.doi.org/10.1103/PhysRevD.92.023529}{\emph{Phys. Rev.} {\bf
  D92} (2015) 023529}, [\href{http://arxiv.org/abs/1408.4575}{{\tt
  1408.4575}}].

\bibitem{Ando:2015qda}
S.~Ando and K.~Ishiwata, \emph{{Constraints on decaying dark matter from the
  extragalactic gamma-ray background}},
  \href{http://dx.doi.org/10.1088/1475-7516/2015/05/024}{\emph{JCAP} {\bf 1505}
  (2015) 024}, [\href{http://arxiv.org/abs/1502.02007}{{\tt 1502.02007}}].

\bibitem{Giesen:2015ufa}
G.~Giesen, M.~Boudaud, Y.~Génolini, V.~Poulin, M.~Cirelli, P.~Salati et~al.,
  \emph{{AMS-02 antiprotons, at last! Secondary astrophysical component and
  immediate implications for Dark Matter}},
  \href{http://dx.doi.org/10.1088/1475-7516/2015/09/023,
  10.1088/1475-7516/2015/9/023}{\emph{JCAP} {\bf 1509} (2015) 023},
  [\href{http://arxiv.org/abs/1504.04276}{{\tt 1504.04276}}].

\bibitem{Agrawal:2010fh}
P.~Agrawal, Z.~Chacko, C.~Kilic and R.~K. Mishra, \emph{{A Classification of
  Dark Matter Candidates with Primarily Spin-Dependent Interactions with
  Matter}},  \href{http://arxiv.org/abs/1003.1912}{{\tt 1003.1912}}.

\bibitem{Burgess:2000yq}
C.~P. Burgess, M.~Pospelov and T.~ter Veldhuis, \emph{{The Minimal model of
  nonbaryonic dark matter: A Singlet scalar}},
  \href{http://dx.doi.org/10.1016/S0550-3213(01)00513-2}{\emph{Nucl. Phys.}
  {\bf B619} (2001) 709--728}, [\href{http://arxiv.org/abs/hep-ph/0011335}{{\tt
  hep-ph/0011335}}].

\bibitem{Barger:2007im}
V.~Barger, P.~Langacker, M.~McCaskey, M.~J. Ramsey-Musolf and G.~Shaughnessy,
  \emph{{LHC Phenomenology of an Extended Standard Model with a Real Scalar
  Singlet}}, \href{http://dx.doi.org/10.1103/PhysRevD.77.035005}{\emph{Phys.
  Rev.} {\bf D77} (2008) 035005}, [\href{http://arxiv.org/abs/0706.4311}{{\tt
  0706.4311}}].

\bibitem{Hambye:2009pw}
T.~Hambye, F.~S. Ling, L.~Lopez~Honorez and J.~Rocher, \emph{{Scalar Multiplet
  Dark Matter}}, \href{http://dx.doi.org/10.1007/JHEP05(2010)066,
  10.1088/1126-6708/2009/07/090}{\emph{JHEP} {\bf 07} (2009) 090},
  [\href{http://arxiv.org/abs/0903.4010}{{\tt 0903.4010}}].

\bibitem{Cline:2013gha}
J.~M. Cline, K.~Kainulainen, P.~Scott and C.~Weniger, \emph{{Update on scalar
  singlet dark matter}}, \href{http://dx.doi.org/10.1103/PhysRevD.92.039906,
  10.1103/PhysRevD.88.055025}{\emph{Phys. Rev.} {\bf D88} (2013) 055025},
  [\href{http://arxiv.org/abs/1306.4710}{{\tt 1306.4710}}].

\bibitem{Brdar:2013iea}
V.~Brdar, I.~Picek and B.~Radovcic, \emph{{Radiative Neutrino Mass with
  Scotogenic Scalar Triplet}},
  \href{http://dx.doi.org/10.1016/j.physletb.2013.11.045}{\emph{Phys. Lett.}
  {\bf B728} (2014) 198--201}, [\href{http://arxiv.org/abs/1310.3183}{{\tt
  1310.3183}}].

\bibitem{Aoki:2014cja}
M.~Aoki and T.~Toma, \emph{{Impact of semi-annihilation of $\mathbb{Z}_3$
  symmetric dark matter with radiative neutrino masses}},
  \href{http://dx.doi.org/10.1088/1475-7516/2014/09/016}{\emph{JCAP} {\bf 1409}
  (2014) 016}, [\href{http://arxiv.org/abs/1405.5870}{{\tt 1405.5870}}].

\bibitem{Kim:2006af}
Y.~G. Kim and K.~Y. Lee, \emph{{The Minimal model of fermionic dark matter}},
  \href{http://dx.doi.org/10.1103/PhysRevD.75.115012}{\emph{Phys. Rev.} {\bf
  D75} (2007) 115012}, [\href{http://arxiv.org/abs/hep-ph/0611069}{{\tt
  hep-ph/0611069}}].

\bibitem{Kusenko:2009up}
A.~Kusenko, \emph{{Sterile neutrinos: The Dark side of the light fermions}},
  \href{http://dx.doi.org/10.1016/j.physrep.2009.07.004}{\emph{Phys. Rept.}
  {\bf 481} (2009) 1--28}, [\href{http://arxiv.org/abs/0906.2968}{{\tt
  0906.2968}}].

\bibitem{Kumericki:2012bh}
K.~Kumericki, I.~Picek and B.~Radovcic, \emph{{TeV-scale Seesaw with Quintuplet
  Fermions}}, \href{http://dx.doi.org/10.1103/PhysRevD.86.013006}{\emph{Phys.
  Rev.} {\bf D86} (2012) 013006}, [\href{http://arxiv.org/abs/1204.6599}{{\tt
  1204.6599}}].

\bibitem{McDonald:1993ex}
J.~McDonald, \emph{{Gauge singlet scalars as cold dark matter}},
  \href{http://dx.doi.org/10.1103/PhysRevD.50.3637}{\emph{Phys. Rev.} {\bf D50}
  (1994) 3637--3649}, [\href{http://arxiv.org/abs/hep-ph/0702143}{{\tt
  hep-ph/0702143}}].

\bibitem{Birkedal:2006fz}
A.~Birkedal, A.~Noble, M.~Perelstein and A.~Spray, \emph{{Little Higgs dark
  matter}}, \href{http://dx.doi.org/10.1103/PhysRevD.74.035002}{\emph{Phys.
  Rev.} {\bf D74} (2006) 035002},
  [\href{http://arxiv.org/abs/hep-ph/0603077}{{\tt hep-ph/0603077}}].

\bibitem{Bhattacharya:2011tr}
S.~Bhattacharya, J.~L. Diaz-Cruz, E.~Ma and D.~Wegman, \emph{{Dark
  Vector-Gauge-Boson Model}},
  \href{http://dx.doi.org/10.1103/PhysRevD.85.055008}{\emph{Phys. Rev.} {\bf
  D85} (2012) 055008}, [\href{http://arxiv.org/abs/1107.2093}{{\tt
  1107.2093}}].

\bibitem{Baek:2012se}
S.~Baek, P.~Ko, W.-I. Park and E.~Senaha, \emph{{Higgs Portal Vector Dark
  Matter : Revisited}},
  \href{http://dx.doi.org/10.1007/JHEP05(2013)036}{\emph{JHEP} {\bf 05} (2013)
  036}, [\href{http://arxiv.org/abs/1212.2131}{{\tt 1212.2131}}].

\bibitem{Boehm:2003ha}
C.~Boehm, P.~Fayet and J.~Silk, \emph{{Light and heavy dark matter particles}},
  \href{http://dx.doi.org/10.1103/PhysRevD.69.101302}{\emph{Phys. Rev.} {\bf
  D69} (2004) 101302}, [\href{http://arxiv.org/abs/hep-ph/0311143}{{\tt
  hep-ph/0311143}}].

\bibitem{Cao:2007fy}
Q.-H. Cao, E.~Ma, J.~Wudka and C.~P. Yuan, \emph{{Multipartite dark matter}},
  \href{http://arxiv.org/abs/0711.3881}{{\tt 0711.3881}}.

\bibitem{Profumo:2009tb}
S.~Profumo, K.~Sigurdson and L.~Ubaldi, \emph{{Can we discover multi-component
  WIMP dark matter?}},
  \href{http://dx.doi.org/10.1088/1475-7516/2009/12/016}{\emph{JCAP} {\bf 0912}
  (2009) 016}, [\href{http://arxiv.org/abs/0907.4374}{{\tt 0907.4374}}].

\bibitem{Bhattacharya:2013hva}
S.~Bhattacharya, A.~Drozd, B.~Grzadkowski and J.~Wudka, \emph{{Two-Component
  Dark Matter}}, \href{http://dx.doi.org/10.1007/JHEP10(2013)158}{\emph{JHEP}
  {\bf 10} (2013) 158}, [\href{http://arxiv.org/abs/1309.2986}{{\tt
  1309.2986}}].

\bibitem{Bhattacharya:2013nya}
S.~Bhattacharya, E.~Ma and D.~Wegman, \emph{{Supersymmetric left-right model
  with radiative neutrino mass and multipartite dark matter}},
  \href{http://dx.doi.org/10.1140/epjc/s10052-014-2902-7}{\emph{Eur. Phys. J.}
  {\bf C74} (2014) 2902}, [\href{http://arxiv.org/abs/1308.4177}{{\tt
  1308.4177}}].

\bibitem{Kajiyama:2013rla}
Y.~Kajiyama, H.~Okada and T.~Toma, \emph{{Multicomponent dark matter particles
  in a two-loop neutrino model}},
  \href{http://dx.doi.org/10.1103/PhysRevD.88.015029}{\emph{Phys. Rev.} {\bf
  D88} (2013) 015029}, [\href{http://arxiv.org/abs/1303.7356}{{\tt
  1303.7356}}].

\bibitem{Cirelli:2005uq}
M.~Cirelli, N.~Fornengo and A.~Strumia, \emph{{Minimal dark matter}},
  \href{http://dx.doi.org/10.1016/j.nuclphysb.2006.07.012}{\emph{Nucl. Phys. B}
  {\bf 753} (2006) 178--194}, [\href{http://arxiv.org/abs/hep-ph/0512090}{{\tt
  hep-ph/0512090}}].

\bibitem{Cirelli:2007xd}
M.~Cirelli, A.~Strumia and M.~Tamburini, \emph{{Cosmology and Astrophysics of
  Minimal Dark Matter}},
  \href{http://dx.doi.org/10.1016/j.nuclphysb.2007.07.023}{\emph{Nucl. Phys.}
  {\bf B787} (2007) 152--175}, [\href{http://arxiv.org/abs/0706.4071}{{\tt
  0706.4071}}].

\bibitem{Cirelli:2009uv}
M.~Cirelli and A.~Strumia, \emph{{Minimal Dark Matter: Model and results}},
  \href{http://dx.doi.org/10.1088/1367-2630/11/10/105005}{\emph{New J. Phys.}
  {\bf 11} (2009) 105005}, [\href{http://arxiv.org/abs/0903.3381}{{\tt
  0903.3381}}].

\bibitem{Ahriche:2015wha}
A.~Ahriche, K.~L. McDonald, S.~Nasri and T.~Toma, \emph{{A Model of Neutrino
  Mass and Dark Matter with an Accidental Symmetry}},
  \href{http://dx.doi.org/10.1016/j.physletb.2015.05.031}{\emph{Phys. Lett.}
  {\bf B746} (2015) 430--435}, [\href{http://arxiv.org/abs/1504.05755}{{\tt
  1504.05755}}].

\bibitem{Sierra:2016qfa}
D.~Aristizabal~Sierra, C.~Sim\~oes and D.~Wegman, \emph{{Closing in on minimal
  dark matter and radiative neutrino masses}},
  \href{http://dx.doi.org/10.1007/JHEP06(2016)108}{\emph{JHEP} {\bf 06} (2016)
  108}, [\href{http://arxiv.org/abs/1603.04723}{{\tt 1603.04723}}].

\bibitem{Sierra:2016rcz}
D.~A. Sierra, C.~Sim\~oes and D.~Wegman, \emph{{Radiative accidental matter}},
  \href{http://arxiv.org/abs/1605.08267}{{\tt 1605.08267}}.

\bibitem{Essig:2007az}
R.~Essig, \emph{{Direct Detection of Non-Chiral Dark Matter}},
  \href{http://dx.doi.org/10.1103/PhysRevD.78.015004}{\emph{Phys. Rev.} {\bf
  D78} (2008) 015004}, [\href{http://arxiv.org/abs/0710.1668}{{\tt
  0710.1668}}].

\bibitem{Farzan:2012ev}
Y.~Farzan, S.~Pascoli and M.~A. Schmidt, \emph{{Recipes and Ingredients for
  Neutrino Mass at Loop Level}},
  \href{http://dx.doi.org/10.1007/JHEP03(2013)107}{\emph{JHEP} {\bf 03} (2013)
  107}, [\href{http://arxiv.org/abs/1208.2732}{{\tt 1208.2732}}].

\bibitem{Casas:2001sr}
J.~A. Casas and A.~Ibarra, \emph{{Oscillating neutrinos and muon ---> e,
  gamma}}, \href{http://dx.doi.org/10.1016/S0550-3213(01)00475-8}{\emph{Nucl.
  Phys.} {\bf B618} (2001) 171--204},
  [\href{http://arxiv.org/abs/hep-ph/0103065}{{\tt hep-ph/0103065}}].

\bibitem{Branco:2011iw}
G.~C. Branco, P.~M. Ferreira, L.~Lavoura, M.~N. Rebelo, M.~Sher and J.~P.
  Silva, \emph{{Theory and phenomenology of two-Higgs-doublet models}},
  \href{http://dx.doi.org/10.1016/j.physrep.2012.02.002}{\emph{Phys. Rept.}
  {\bf 516} (2012) 1--102}, [\href{http://arxiv.org/abs/1106.0034}{{\tt
  1106.0034}}].

\bibitem{Huitu:1996su}
K.~Huitu, J.~Maalampi, A.~Pietila and M.~Raidal, \emph{{Doubly charged Higgs at
  LHC}}, \href{http://dx.doi.org/10.1016/S0550-3213(97)87466-4}{\emph{Nucl.
  Phys. B} {\bf 487} (1997) 27--42},
  [\href{http://arxiv.org/abs/hep-ph/9606311}{{\tt hep-ph/9606311}}].

\bibitem{Aoki:2011yk}
M.~Aoki, S.~Kanemura and K.~Yagyu, \emph{{Doubly-charged scalar bosons from the
  doublet}}, \href{http://dx.doi.org/10.1016/j.physletb.2011.11.043,
  10.1016/j.physletb.2011.07.017}{\emph{Phys. Lett.} {\bf B702} (2011)
  355--358}, [\href{http://arxiv.org/abs/1105.2075}{{\tt 1105.2075}}].

\bibitem{Chiang:2012dk}
C.-W. Chiang, T.~Nomura and K.~Tsumura, \emph{{Search for doubly charged Higgs
  bosons using the same-sign diboson mode at the LHC}},
  \href{http://dx.doi.org/10.1103/PhysRevD.85.095023}{\emph{Phys. Rev.} {\bf
  D85} (2012) 095023}, [\href{http://arxiv.org/abs/1202.2014}{{\tt
  1202.2014}}].

\bibitem{Babu:2013ega}
K.~S. Babu, A.~Patra and S.~K. Rai, \emph{{New Signals for Doubly-Charged
  Scalars and Fermions at the Large Hadron Collider}},
  \href{http://dx.doi.org/10.1103/PhysRevD.88.055006}{\emph{Phys. Rev.} {\bf
  D88} (2013) 055006}, [\href{http://arxiv.org/abs/1306.2066}{{\tt
  1306.2066}}].

\bibitem{Okada:2015hia}
H.~Okada and K.~Yagyu, \emph{{Three-loop neutrino mass model with doubly
  charged particles from isodoublets}},
  \href{http://dx.doi.org/10.1103/PhysRevD.93.013004}{\emph{Phys. Rev. D} {\bf
  93} (2016) 013004}, [\href{http://arxiv.org/abs/1508.01046}{{\tt
  1508.01046}}].

\bibitem{Drozd:2014yla}
A.~Drozd, B.~Grzadkowski, J.~F. Gunion and Y.~Jiang, \emph{{Extending
  two-Higgs-doublet models by a singlet scalar field - the Case for Dark
  Matter}}, \href{http://dx.doi.org/10.1007/JHEP11(2014)105}{\emph{JHEP} {\bf
  11} (2014) 105}, [\href{http://arxiv.org/abs/1408.2106}{{\tt 1408.2106}}].

\bibitem{Ginzburg:2004vp}
I.~F. Ginzburg and M.~Krawczyk, \emph{{Symmetries of two Higgs doublet model
  and CP violation}},
  \href{http://dx.doi.org/10.1103/PhysRevD.72.115013}{\emph{Phys. Rev.} {\bf
  D72} (2005) 115013}, [\href{http://arxiv.org/abs/hep-ph/0408011}{{\tt
  hep-ph/0408011}}].

\bibitem{Haber:2006ue}
H.~E. Haber and D.~O'Neil, \emph{{Basis-independent methods for the
  two-Higgs-doublet model. II. The Significance of tan$\beta$}},
  \href{http://dx.doi.org/10.1103/PhysRevD.74.015018,
  10.1103/PhysRevD.74.059905}{\emph{Phys. Rev.} {\bf D74} (2006) 015018},
  [\href{http://arxiv.org/abs/hep-ph/0602242}{{\tt hep-ph/0602242}}].

\bibitem{Altmannshofer:2012ar}
W.~Altmannshofer, S.~Gori and G.~D. Kribs, \emph{{A Minimal Flavor Violating
  2HDM at the LHC}},
  \href{http://dx.doi.org/10.1103/PhysRevD.86.115009}{\emph{Phys. Rev.} {\bf
  D86} (2012) 115009}, [\href{http://arxiv.org/abs/1210.2465}{{\tt
  1210.2465}}].

\bibitem{Celis:2013rcs}
A.~Celis, V.~Ilisie and A.~Pich, \emph{{LHC constraints on two-Higgs doublet
  models}}, \href{http://dx.doi.org/10.1007/JHEP07(2013)053}{\emph{JHEP} {\bf
  07} (2013) 053}, [\href{http://arxiv.org/abs/1302.4022}{{\tt 1302.4022}}].

\bibitem{Inoue:2014nva}
S.~Inoue, M.~J. Ramsey-Musolf and Y.~Zhang, \emph{{CP-violating phenomenology
  of flavor conserving two Higgs doublet models}},
  \href{http://dx.doi.org/10.1103/PhysRevD.89.115023}{\emph{Phys. Rev.} {\bf
  D89} (2014) 115023}, [\href{http://arxiv.org/abs/1403.4257}{{\tt
  1403.4257}}].

\bibitem{Pich:2009sp}
A.~Pich and P.~Tuzon, \emph{{Yukawa Alignment in the Two-Higgs-Doublet Model}},
  \href{http://dx.doi.org/10.1103/PhysRevD.80.091702}{\emph{Phys. Rev.} {\bf
  D80} (2009) 091702}, [\href{http://arxiv.org/abs/0908.1554}{{\tt
  0908.1554}}].

\bibitem{Jung:2010ik}
M.~Jung, A.~Pich and P.~Tuzon, \emph{{Charged-Higgs phenomenology in the
  Aligned two-Higgs-doublet model}},
  \href{http://dx.doi.org/10.1007/JHEP11(2010)003}{\emph{JHEP} {\bf 11} (2010)
  003}, [\href{http://arxiv.org/abs/1006.0470}{{\tt 1006.0470}}].

\bibitem{Dev:2014yca}
P.~S. Bhupal~Dev and A.~Pilaftsis, \emph{{Maximally Symmetric Two Higgs Doublet
  Model with Natural Standard Model Alignment}},
  \href{http://dx.doi.org/10.1007/JHEP11(2015)147,
  10.1007/JHEP12(2014)024}{\emph{JHEP} {\bf 12} (2014) 024},
  [\href{http://arxiv.org/abs/1408.3405}{{\tt 1408.3405}}].

\bibitem{Buras:2003jf}
A.~J. Buras, \emph{{Minimal flavor violation}}, {\emph{Acta Phys. Polon.} {\bf
  B34} (2003) 5615--5668}, [\href{http://arxiv.org/abs/hep-ph/0310208}{{\tt
  hep-ph/0310208}}].

\bibitem{Agashe:2005hk}
K.~Agashe, M.~Papucci, G.~Perez and D.~Pirjol, \emph{{Next to minimal flavor
  violation}},  \href{http://arxiv.org/abs/hep-ph/0509117}{{\tt
  hep-ph/0509117}}.

\bibitem{Gavela:2009cd}
M.~B. Gavela, T.~Hambye, D.~Hernandez and P.~Hernandez, \emph{{Minimal Flavour
  Seesaw Models}},
  \href{http://dx.doi.org/10.1088/1126-6708/2009/09/038}{\emph{JHEP} {\bf 09}
  (2009) 038}, [\href{http://arxiv.org/abs/0906.1461}{{\tt 0906.1461}}].

\bibitem{Kagan:2009bn}
A.~L. Kagan, G.~Perez, T.~Volansky and J.~Zupan, \emph{{General Minimal Flavor
  Violation}}, \href{http://dx.doi.org/10.1103/PhysRevD.80.076002}{\emph{Phys.
  Rev.} {\bf D80} (2009) 076002}, [\href{http://arxiv.org/abs/0903.1794}{{\tt
  0903.1794}}].

\bibitem{Babu:2002uu}
K.~S. Babu and C.~Macesanu, \emph{{Two loop neutrino mass generation and its
  experimental consequences}},
  \href{http://dx.doi.org/10.1103/PhysRevD.67.073010}{\emph{Phys. Rev.} {\bf
  D67} (2003) 073010}, [\href{http://arxiv.org/abs/hep-ph/0212058}{{\tt
  hep-ph/0212058}}].

\bibitem{Gunion:1996pq}
J.~F. Gunion, C.~Loomis and K.~T. Pitts, \emph{{Searching for doubly charged
  Higgs bosons at future colliders}}, {\emph{eConf} {\bf C960625} (1996)
  LTH096}, [\href{http://arxiv.org/abs/hep-ph/9610237}{{\tt hep-ph/9610237}}].

\bibitem{AristizabalSierra:2006gb}
D.~Aristizabal~Sierra and M.~Hirsch, \emph{{Experimental tests for the Babu-Zee
  two-loop model of Majorana neutrino masses}},
  \href{http://dx.doi.org/10.1088/1126-6708/2006/12/052}{\emph{JHEP} {\bf 12}
  (2006) 052}, [\href{http://arxiv.org/abs/hep-ph/0609307}{{\tt
  hep-ph/0609307}}].

\bibitem{Nebot:2007bc}
M.~Nebot, J.~F. Oliver, D.~Palao and A.~Santamaria, \emph{{Prospects for the
  Zee-Babu Model at the CERN LHC and low energy experiments}},
  \href{http://dx.doi.org/10.1103/PhysRevD.77.093013}{\emph{Phys. Rev.} {\bf
  D77} (2008) 093013}, [\href{http://arxiv.org/abs/0711.0483}{{\tt
  0711.0483}}].

\bibitem{Ohlsson:2009vk}
T.~Ohlsson, T.~Schwetz and H.~Zhang, \emph{{Non-standard neutrino interactions
  in the Zee-Babu model}},
  \href{http://dx.doi.org/10.1016/j.physletb.2009.10.025}{\emph{Phys. Lett.}
  {\bf B681} (2009) 269--275}, [\href{http://arxiv.org/abs/0909.0455}{{\tt
  0909.0455}}].

\bibitem{Babu:2016rcr}
K.~S. Babu and S.~Jana, \emph{{Probing Doubly Charged Higgs Bosons at the LHC
  through Photon Initiated Processes}},
  \href{http://arxiv.org/abs/1612.09224}{{\tt 1612.09224}}.

\bibitem{Chatrchyan:2012ya}
{\scshape CMS} collaboration, S.~Chatrchyan et~al., \emph{{A search for a
  doubly-charged Higgs boson in $pp$ collisions at $\sqrt{s}=7$ TeV}},
  \href{http://dx.doi.org/10.1140/epjc/s10052-012-2189-5}{\emph{Eur. Phys. J.}
  {\bf C72} (2012) 2189}, [\href{http://arxiv.org/abs/1207.2666}{{\tt
  1207.2666}}].

\bibitem{ATLAS:2012hi}
{\scshape ATLAS} collaboration, G.~Aad et~al., \emph{{Search for doubly-charged
  Higgs bosons in like-sign dilepton final states at $\sqrt{s}=7$ TeV with the
  ATLAS detector}},
  \href{http://dx.doi.org/10.1140/epjc/s10052-012-2244-2}{\emph{Eur. Phys. J.}
  {\bf C72} (2012) 2244}, [\href{http://arxiv.org/abs/1210.5070}{{\tt
  1210.5070}}].

\bibitem{Alloul:2013raa}
A.~Alloul, M.~Frank, B.~Fuks and M.~Rausch~de Traubenberg,
  \emph{{Doubly-charged particles at the Large Hadron Collider}},
  \href{http://dx.doi.org/10.1103/PhysRevD.88.075004}{\emph{Phys. Rev. D} {\bf
  88} (2013) 075004}, [\href{http://arxiv.org/abs/1307.1711}{{\tt 1307.1711}}].

\end{thebibliography}\endgroup

\end{document}